\documentclass[reprint,amsmath,amssymb,aps,prb,groupedaddress,nofootinbib,superscriptaddress]{revtex4-1}
\usepackage{graphicx,xcolor}
\usepackage{amsthm,amssymb,amsmath,dsfont,braket,mathdots,tensor}
\usepackage{bm}
\usepackage[hidelinks,pagebackref=false,pdfnewwindow=true]{hyperref}
\usepackage{relsize,amsbsy}
\usepackage[export]{adjustbox}

\DeclareMathOperator{\sgn}{sgn}
\DeclareMathOperator{\tr}{Tr}

\DeclareMathOperator{\im}{Im}

\newcommand{\1}{\mathds{1}}

\newcommand{\be}{\begin{equation}}
\newcommand{\ee}{\end{equation}}
\newcommand{\ms}{Majorana spinon }
\newcommand{\mss}{Majorana spinons }
\newcommand{\mssns}{Majorana spinons}

\newcommand{\bit}{\begin{enumerate}}
	\newcommand{\eit}{\end{enumerate}}

\definecolor{bananayellow}{rgb}{1.0, 0.88, 0.21}
\definecolor{bronze}{rgb}{0.62, 0.28, 0.1}

\definecolor{palatinatepurple}{rgb}{0.49, 0.24, 0.46}

\definecolor{darkblue}{rgb}{0.0, 0.0, 0.55}

\definecolor{darkgreen}{rgb}{0.0, 0.5, 0.0}

\begin{document}	
	
	\title{Resonant Raman scattering theory for Kitaev models \\ and their Majorana fermion boundary modes}
		\author{Brent Perreault}
		\affiliation{\small School of Physics and Astronomy, University of Minnesota, Minneapolis, Minnesota 55455, USA}
		%
	\author{Johannes Knolle}
	\affiliation{\small Department of Physics, Cavendish Laboratory, JJ Thomson Avenue, Cambridge CB3 0HE, U.K.}
		\author{Natalia B. Perkins}
		\author{F. J. Burnell}
		\affiliation{\small School of Physics and Astronomy, University of Minnesota, Minneapolis, Minnesota 55455, USA}
		
		\date{\today} 
				
		\begin{abstract}
		 	We study the inelastic light scattering response in two- (2D) and three-dimensional (3D) Kitaev spin-liquid models with \ms band structures in the symmetry classes BDI and D leading to protected gapless surface modes. We present a detailed calculation of the resonant Raman/Brillouin scattering vertex relevant to iridate and ruthenate compounds whose low-energy physics is believed to be proximate to these spin-liquid phases. In the symmetry class BDI, we find that while the resonant scattering on thin films can detect the gapless boundary modes of spin liquids, the non-resonant processes do not couple to them. For the symmetry class D, however, we find that the coupling between both types of light-scattering processes and the low-energy surface states is strongly suppressed. Additionally, we describe the effect of weak time-reversal symmetry breaking perturbations on the bulk Raman response of these systems. 
		\end{abstract}
		
		\maketitle
			
	
	\section{Introduction}
	
Borders and boundaries can have much more drastic effects than just separating different regions in space: their presence can give rise to fundamentally new types of excitations. In condensed matter physics, this phenomenon has been firmly established since the discovery of the quantum Hall effect in two-dimensional electron gases.~\cite{Klitzing1980} The boundary separating systems with different topological properties (quantified by differing bulk topological invariants~\cite{Thouless1982}) harbors special ``protected" edge states, whose presence is guaranteed by properties of the bulk. This celebrated {\it bulk-boundary correspondence} underlies the robustness of such edge states to local perturbations. 

In addition to quantum Hall systems, a variety of other symmetry protected topological phases (SPTs) can arise in weakly-interacting electronic systems. The resulting new materials, such as topological insulators (TIs)~\cite{Hasan2010,Qi2011} and Dirac and Weyl semimetals,~\cite{wan11,Vafek2014} have drawn considerable attention from the condensed matter community. This excitement has been driven by both the identification and the synthesis of such materials,\cite{BernevigHughesZhang,Teo08,DaiBernevig15} and the direct experimental observation of their protected surface states.~\cite{Hsieh08,Xu15,Lv} For example, the helical edge states of 2D topological insulators lead to a quantized spin Hall transport\cite{koenig07}, while Weyl semimetals harbor exotic Fermi arc surface states which can be probed by high-resolution angle-resolved photoemission spectroscopy (ARPES).~\cite{Xu15,Lv}

Considerable progress has also been made in understanding the possibilities for protected boundary modes in strongly-interacting systems where a band-theory type description of electron-like quasiparticles is not appropriate. Of particular interest in this regard are strongly-interacting systems whose bulk exhibits topological order (TO),~\cite{Wen1990} leading to a particularly rich set of possible boundary states.\cite{Levin2012,Swingle11,Levin11,Maciejko12,LuVishwanath,MesarosRan,LongQPaper}
Relative to their weakly interacting counterparts, however, these phases pose significant experimental challenges: besides the difficulty in identifying materials that might host such phases, the low-energy quasiparticles in TO systems are generically {\it fractionalized}, meaning that they contain only fractional parts of the quantum numbers of the constituent electrons. Consequently, these quasiparticles do not couple directly to standard experimental probes, making a definitive identification of the surface states challenging. 

Quantum spin liquids (QSLs)~\cite{Anderson1973} are TO phases that have drawn particular interest in recent years as evidence that several highly frustrated magnetic materials may harbor these topological phases has accumulated.~\cite{Lacroix2011,Balents,savary16}
In these phases fractionalization is exhibited by spin-charge separation, with the charge degree of freedom being pinned, while emergent dispersing quasiparticles known as {\it spinons} carry the electron's spin. These spinons can in principle participate in protected surface or edge states, similar to those realized in topological superconductors.~\cite{Pesin} However, the chargeless spinons would not show up in charge transport experiments or ARPES. This raises an interesting question of how best to detect the resulting TO, and in particular of how to probe any protected boundary modes that such a system may exhibit.

In this paper we investigate resonant inelastic light scattering as a tool to probe bulk fractionalized excitations\cite{Knolle14-2,Perreault15, Perreault16,Halasz16} and, in particular, surface states, characteristic of the topological orders realized in quantum spin liquids. Because the literature on inelastic scattering of near-visible light has been primarily focused on Raman scattering, we use `Raman' to refer to the intensity and operators in general, but the experimental reader should keep in mind that Brillouin scattering is expected to be better suited to the energy scales of the proposed low-energy signatures.\cite{Perreault16} 

We focus on the QSLs realized by a particular class of spin-exchange Hamiltonians inspired by the Kitaev honeycomb model,\cite{Kitaev} which we refer to as Kitaev QSLs. The Kitaev QSL models can be formulated on any 2D or 3D tri-coordinated lattice,~\cite{Mandal,Obrien} and have the advantage that they are exactly solvable, with excitations naturally described by an exact fractionalization of the spin degrees of freedom (d.o.f.) into dispersing Majorana fermions and static Z$_2$ gauge fluxes. By studying a range of different trivalent lattices, the Kitaev QSL models enable us to consider a rich variety of QSL phases with distinct topological boundary modes. For example, the 2D decorated honeycomb lattice realizes a QSL with non-Abelian excitations and chiral edge modes\cite{YaoKivelson}, while trivalent 3D lattices can for example exhibit Majorana band structures mimicking Weyl semi-metals with protected surface Fermi arcs.~\cite{Hermanns}

Kitaev QSLs have the further advantage that certain class of materials, which we refer to here as Jackeli-Khaliullin Kitaev (JKK) systems, are believed to be proximate to the Kitaev QSL phases. In these materials, first described in the seminal work of Jackeli and Khaliullin,\cite{Jackeli} edge-sharing oxygen octahedra enclosing transition-metal ions with partially filled t$_{2g}$ levels and strong spin-orbit coupling exhibit dominant Kitaev interactions between $j_{eff} = 1/2$ magnetic moments. Besides the layered iridates\cite{Chaloupka10} of the A$_2$IrO$_3$ (A=Na,Li) family, this general scenario also applies to $\alpha$-RuCl$_3$ \cite{Plumb14,Kim15-1} and the three-dimensional harmonic-honeycomb iridates\cite{Modic,Kimchi14,Takayama} $\beta$-Li$_2$IrO$_3$ and $\gamma$-Li$_2$IrO$_3$. Though current JKK materials at ambient pressure show residual long-ranged magnetism at low temperatures, growing evidence suggests that Kitaev interactions are dominant,~\cite{Chun2015,Banerjee16} and Kitaev QSL phases may be achievable in closely related materials systems.

In Kitaev QSL systems, Raman response is a particularly useful probe since, in contrast to the dynamical spin structure factor,\cite{Knolle14-1} it does not couple to the flux d.o.f. and, consequently, gives a more direct probe of the Majorana fermion density of states (DOS).\cite{Knolle14-2} In addition, information can be deduced from the polarization dependence, for which scattering matrix elements determine which aspects of the DOS are observable in practice. In particular, the symmetry properties of the Kitaev QSL systems are responsible for the fact that resonant Raman vertices can couple to the protected surface modes, while non-resonant Raman channels cannot.\cite{Perreault16} Finally, as shown in Refs. \cite{Knolle14-2,Perreault15,Perreault16}, the energy and polarization dependence of the Raman response in these systems contains signatures characteristic of the QSL states both in the 2D honeycomb 3D hyperhoneycomb lattices. Here, we extend these results by presenting a unified description of the Raman response of Majorana modes in a variety Kitaev QSLs,~\cite{Obrien} whose low energy bulk excitations consist of Dirac points, nodal lines or Weyl nodes. We also investigate the potential of Raman scattering (or Brillouin scattering) on thin films to provide evidence for the various types of protected boundary modes, such as surface flat bands or Fermi arcs, that arise in these systems. 

Overall, we establish inelastic light scattering as a powerful experimental tool for measuring the bulk-boundary correspondence in QSLs, which we corroborate by concrete calculations for a variety of Kitaev QSL phases in different dimensions, each representing different phenomenologies of its fractionalized bulk and boundary modes. Our derivation of the resonant Raman/Brillouin scattering vertex is presented in a way such that it is easily generalized to other systems beyond the integrable Kitaev limits studied here, including the more experimentally realistic situations where sub-dominant exchange interactions are present.\cite{Sizyuk}
 
The paper is organized as follows. In section~\ref{ModelSec}, we first introduce the Kitaev model and its exact solution on the honeycomb, hyperhoneycomb and (8,3)b lattices, and then 
review the relation between bulk topological invariants of their Majorana band structures and properties of the associated edge states in slab geometries. In section~\ref{RamanSec}, we discuss the derivation of the Raman vertex in Mott insulators in the Loudon-Fleury approach and generalize it to effective $j_{eff}=1/2$ systems in spin-orbit coupled magnets with dominant Kitaev interactions. In particular, we develop a formalism to calculate the chiral three-spin terms that appear in the next-to leading order in a perturbative expansion which becomes relevant in the resonant Raman scattering regime. In section~\ref{ResultsSec}, we present a microscopic calculation of the resonant and off-resonant Raman response in the different QSL phases. There we argue that the observability of different types of surface modes in Raman scattering of thin films is controlled by simple selection rules connecting symmetry properties of the Raman vertex to those of the surface states. Since Raman scattering is directly related to the Majorana DOS, we confirm that the bulk response can be used to detect different QSLs -- for example, nodal line and Weyl QSL show different asymptotic low energy behaviors in the Raman response. Using group theory, we show that detailed examination of the Raman polarization dependence presents another handle with which to diagnose QSLs. Finally, in section~\ref{DiscussionSec}, we summarize our main results and asses their applicability to future experiments. We close with a discussion. 
		

\section{Kitaev QSLs and topological band structures} \label{ModelSec}

\subsection{The model}
The Kitaev model, originally conceived on the honeycomb lattice,\cite{Kitaev} is an exactly solvable spin-exchange Hamiltonian with a spin-liquid ground state that can be formulated on any 
tri-coordinated lattice.\cite{Mandal,YaoKivelson,Modic}
The Hamiltonian has the general form:
\begin{align}\label{H}
{H}_K &= \sum_{\left<ij\right>^\alpha} J^\alpha \sigma^\alpha_i \sigma^\alpha_j \ \ ,
\end{align}
where one bond of each type $\alpha = x,y$ and $z$ emanates from every vertex, as shown in Figs. \ref{fig:HC_primitive}, \ref{fig:3DKitaev-basicunitcell-Hermanns} and \ref{fig:ETB_primitive}, and $\left<ij\right>$ are nearest-neighbor (NN) pairs. 

The model is solved exactly by replacing the spin variable at each site $j$ with four Majorana fermions (denoted $c_j$ and $b_j^\alpha$), via
\begin{align} \label{Spin2Mar}
\sigma^\alpha_j =i b_j^\alpha c_j  \ \ .
\end{align}
The four Majorana fermions are mutually anticommuting and self-conjugate, obeying $c_j^2 = 1 = (b_j^\alpha)^2$ so that $c^\dagger = c$. In terms of the Majoranas the Hamiltonian can be expressed
\begin{align}\label{H2}
{H}_K &= \sum_{\left<ij\right>^\alpha} J^\alpha b^\alpha_i b^\alpha_j c_i c_j \ \ .
\end{align}
Since the bond operators $u_{\left<ij\right>^\alpha} = i b_i^\alpha b_j^\alpha$ are conserved on each link, the Hamiltonian (\ref{H2}) decouples into orthogonal flux sectors described by sets of $\{u_{\left<ij\right>^\alpha}\}$ variables.~\cite{Kitaev} Each flux sector now can be considered individually and in each of them the Hamiltonian can be solved exactly, as it reduces to a bilinear form in $c$, describing Majorana fermions with hopping matrix elements determined by the underlying flux configuration. 

The ground state of the Hamiltonian (\ref{H}) is the Kitaev QSL. The elementary excitations in this state are of two kinds: first, there are dispersing fermionic excitations associated with exciting the $c$-type Majorana fermions. Second, there are flux defects, associated with exciting the $b$-type Majorana fermions. The representation (\ref{Spin2Mar}) is redundant, such that the number of independent physical $b$-type excitations is equal to the number of plaquettes $P$, rather than to the number of edges or the number of $b$ Majorana fermions, with $W_P = \prod_{\left<ij\right> \in P} u_{\left<ij\right>^\alpha}$ representing the only conserved quantities in the {\it physical} Hilbert space.

\subsection{The tri-coordinated lattices}

\begin{figure}
	\centering
	\includegraphics[width=.4\linewidth]{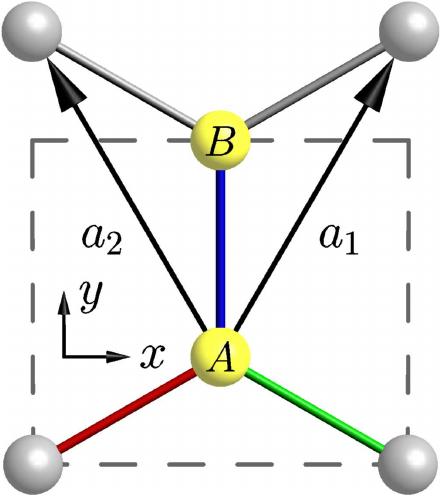}
	\caption{(Color Online) The primitive unit cell of the honeycomb lattice. A and B denote the two sublattices of the honeycomb lattice. The bond vectors are given by $d^z = (0,1)$ and $d^{x/y}=(\pm \sqrt{3},-1)/2$ with the unit vectors $a_{1/2} = (\pm \sqrt{3},3)/2$.}
	\label{fig:HC_primitive}
\end{figure}

\begin{figure}
	\centering
	\includegraphics[width=.87\linewidth]{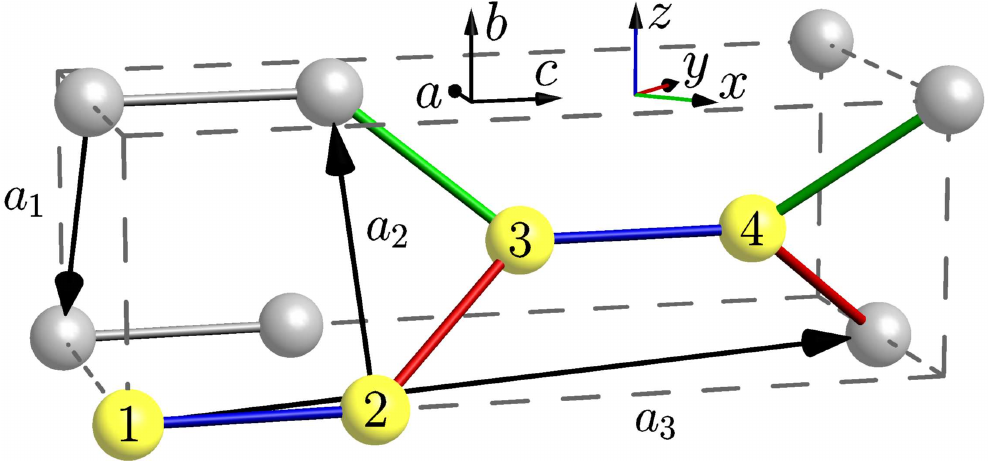}
	\caption{(Color Online) The primitive unit cell of the hyperhoneycomb lattice. There are bonds along five different directions given by $d^z = (0,0,1)$, 
		$d^{x/y}_{32} = \frac{1}{2}(\pm 1,\sqrt{2},-1)$, and $ 
		d^{x/y}_{14} = \frac{1}{2}(\pm 1,-\sqrt{2},-1)$. The unit vectors are $a_3 = (-1,0,3)$ and $a_{1/2} = (-1,\mp \sqrt{2}, 0)$.}
	\label{fig:3DKitaev-basicunitcell-Hermanns}
\end{figure}

\begin{figure}
	\centering
	\includegraphics[width=.82\linewidth]{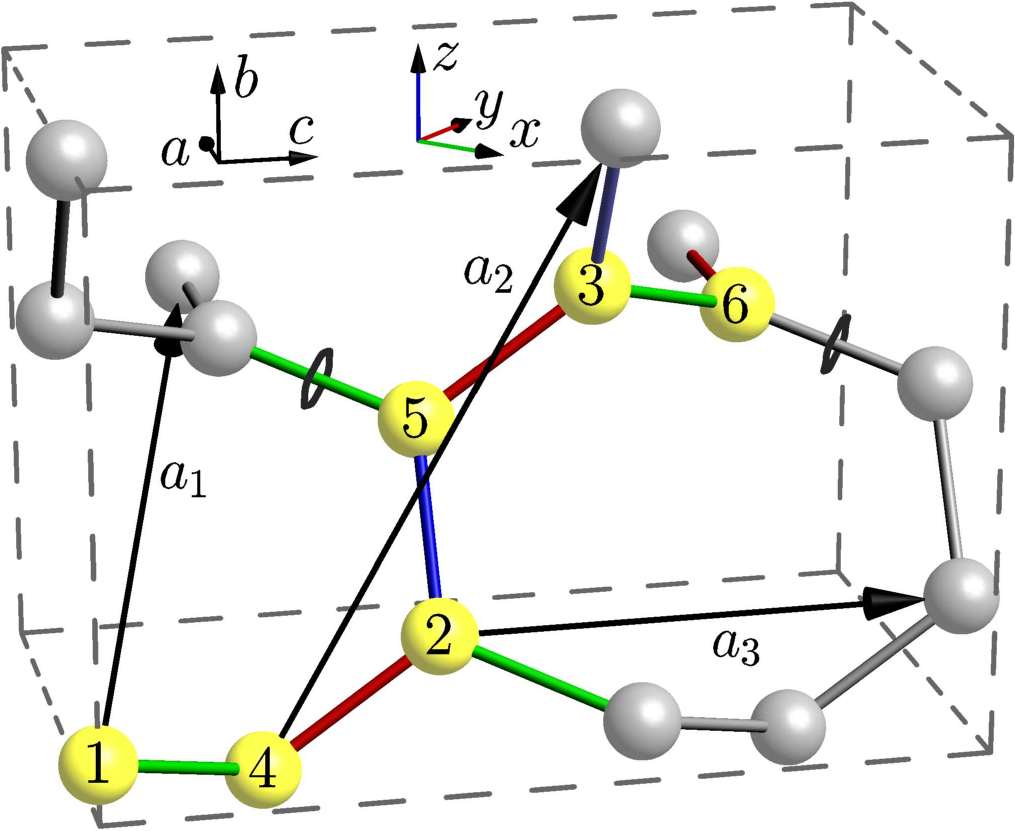}
	\caption{(Color Online) The primitive unit cell of the (8,3)b lattice. For $i<j$ the bonds without circles carry the gauge $u_{ij} = 1$ and those with circles have $u_{ij} = -1$. There are six distinct bonds: $d^{x}_{16} = (-1,0,0)$, $d^{x}_{24} = d^x_{35} = \frac{1}{\sqrt{3}}(0,-1,-\sqrt{2})$, $d^y_{14/12} = d^y_{36/56} = \frac{1}{2\sqrt{3}}( \sqrt{3},\mp 1,\pm 2\sqrt{2})$, and $d^z_{25/34} = \frac{1}{2}(\pm 1,\sqrt{3},0)$. The unit vectors are $a_1 = (\frac{1}{2},\frac{1}{2\sqrt{3}},\frac{2}{5\sqrt{6}})$, $a_2 = (0,\frac{2}{2\sqrt{3}},\frac{4}{5\sqrt{6}})$, and $a_3 = (0,0,\frac{6}{5\sqrt{6}})$.}
	\label{fig:ETB_primitive}
\end{figure}

We will study the Kitaev QSL model on three tri-coordinated lattices with 120 degree bond angles: the 2D honeycomb lattice, the 3D hyperhoneycomb\cite{Mandal} and the 
so-called\cite{Obrien} (8,3)b lattices.\cite{lattice_foot} Primitive unit cells for these lattices are portrayed in Figs. \ref{fig:HC_primitive}, \ref{fig:3DKitaev-basicunitcell-Hermanns}, and \ref{fig:ETB_primitive}, respectively. 

The Kitaev model on the honeycomb lattice and hyperhoneycomb lattices has zero-flux as its ground state, \cite{Kitaev,Mandal} though while for the former it follows directly from the application of the Lieb's theorem,\cite{Lieb} for the latter it has only been demonstrated numerically.\cite{Mandal,Kimchi,Kim}
For the honeycomb lattice all elementary plaquettes have six sites, and for the hyperhoneycomb lattice they are ten-sided. In both cases, the zero-flux ground state can be achieved by choosing a gauge with $u_{\left<ij\right>^\alpha} = 1$ for $i$ belonging to an odd-numbered sublattice so that $j$ is even.

The case of the (8,3)b lattice (Fig. \ref{fig:ETB_primitive}) is special. It has the required mirror planes to constrain its flux completely using Lieb's theorem, marking the first completely solved 3D Kitaev spin liquid.\cite{Obrien} The (8,3)b lattice has elementary plaquettes with eight and twelve bonds. According to Lieb's theorem, in the ground state all these plaquettes must carry $\pi$ flux, which can be achieved in the gauge with $u_{\left<ij\right>^\alpha} = 1$ for any $i < j$ except on the circled bonds in Fig. \ref{fig:ETB_primitive}. In this gauge choice, $u_{r,r'} = -1$ for $r = r_5$ and $r' = r_6 - a_3$. 	 

By fixing a gauge, performing a Fourier transformation into momentum space and using matrix notation, we get the following quadratic Majorana fermion Hamiltonian:
\begin{align}\label{Hc}
H &= \sum_{\textrm{all } {\mathbf k}} \mathbf{c}_{-{\mathbf k}}^T \mathrm{H}_k \mathbf{c}_{\mathbf k} .
\end{align} 
Due to different number of sublattices in the unit cell, the resulting band structure for the Majorana spinons consists of one, two, and three bands of fermionic quasiparticles for the honeycomb, hyperhoneycomb, and (8,3)b lattice, respectively.
For the honeycomb lattice, the band structure is essentially that of graphene, with a pair of Dirac nodes at the corners of the Brillouin zone. On the hyperhoneycomb the Majorana spinon band structure exhibits a 1D Fermi ring, while the (8,3)b lattice has bulk Weyl nodes. We will explore the topological nature of these band structures further in Sec. \ref{BandStructSec}. 

\subsection{Magnetic field perturbations and Majorana fermion band structures}\label{magnetic}

Next we introduce a parameter which, for the honeycomb and hyperhoneycomb Kitaev QSLs, breaks the symmetry protecting the gapless boundary modes. In both cases, this can be achieved by adding a weak magnetic field, which alters both the structure of the bulk Fermi surface, and the nature of the edge states. Specifically, Kitaev\cite{Kitaev} showed that the Majorana fermion band structure of the honeycomb lattice model, which has gapless Dirac points, can be gapped by the presence of a magnetic field perturbation ${\bf h}$ whose components $h^x, h^y$, and $h^z$ are all nonzero. On the hyperhoneycomb lattice Hermanns {\it et al}\cite{Hermanns,Obrien} showed that such a magnetic field gaps out most of the Fermi ring, leaving a pair of Weyl points.
Here we review the arguments of Kitaev's original paper, whose basis will be important when we consider the effects of the magnetic field in the Raman response. We discuss the ramifications of this symmetry-breaking for the protected surface states in Sec. \ref{BandStructSec}.

\begin{figure}
	\centering
	\includegraphics[width=0.8\linewidth,valign=t]{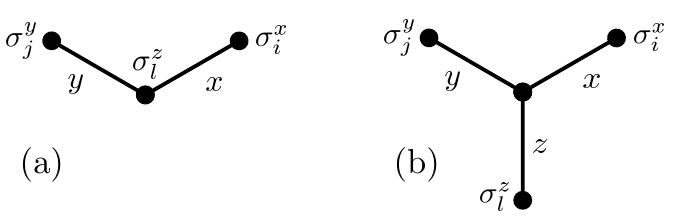} 
	\caption{The perturbation that leads to second-neighbor \ms hopping $c_i c_j$ for a pair of $x$ and $y$ bonds is the product of the three spin components $\sigma_i^x \sigma_l^z \sigma_j^y$ in the two configurations shown here.}
	\label{fig:3-spin_term}
\end{figure}

The magnetic field perturbation is given by
\begin{align}\label{pert}
V = \sum_j\sum_{\alpha=x,y,x} h^\alpha \sigma^\alpha_j.
\end{align}
Provided that the energy scale of the perturbation is below the local flux gap, one can deduce the approximate impact of $V$ on the Majorana spinons perturbatively within the zero-flux sector.\cite{Kitaev} In this approximation, the first non-vanishing time-reversal symmetry breaking terms appear at third order in $h$. One such term involves four sites pictured in Fig.~\ref{fig:3-spin_term}(b). This term leads to a four-spinon interaction that is irrelevant at low energies,\cite{Kitaev,Hermanns} and is ignored here. The other term, illustrated in Fig.~\ref{fig:3-spin_term}(a), leads to the effective interaction of the form $\sigma^x_i \sigma^z_l \sigma^y_j$ between three adjacent sites $i,l,$ and $j$, where $x,y,z$ labels are fixed by the character of the two bonds connecting these sites. 

More specifically, the interaction on the adjacent bond pairs $\ll ij \gg^\gamma \equiv \left<il\right>^\beta \left<ll\right>^\alpha$, where $\gamma$ is complementary to $\alpha$ and $\beta$ ($\epsilon^{\alpha \beta \gamma} \ne 0$), is $\sigma^\alpha_i \sigma_l^\gamma \sigma_j^\beta$. Within the Kitaev Majorana fermion description, this term is a next-nearest neighbor (NNN) hopping term for the \mssns
\begin{align} \label{2Hops}
H_h &= \sum_{\ll ij \gg^\gamma} \kappa_{ij} 
\sigma^{\alpha}_i \sigma^\gamma_{l} \sigma^{\beta}_j\nonumber\\
&= i \sum_{\ll ij \gg ^\gamma} \kappa^\gamma_{ij} c_i c_j, 
\end{align} 
where the NNN hopping amplitude is determined by $\kappa^\gamma_{ij}=\kappa_{ij}\tilde{u}_{\ll jl \gg ^\gamma}$ with
$\tilde{u}_{\ll jl \gg ^\gamma} = {u}_{\left<il \right>^\alpha} {u}_{\left<lj \right>^\beta}$. Therefore, once we have fixed the gauges $u_{\left<ij\right>^\alpha}$ in the parent spin liquid, the magnitudes and phases of these second-neighbor hopping terms are completely fixed by the magnetic field, whose role role is described by an effective interaction
$\kappa_{ij} \sim h^x h^y h^z / (\Delta^{il} \Delta^{lj})$, where $\Delta^{il}$ denotes the gap to the flux excitation created by changing the sign of the bond variable $u_{\langle il\rangle}$. 

Including the effective interaction (\ref{2Hops}) due to a magnetic field, the quadratic Hamiltonians describing the Majorana fermion band structure are as follows: (We use superscripts $\infty, \, 0 ,\, 8b$ to denote the honeycomb, the hyperhoneycomb and (8,3)b lattices, respectively)
\\
 ({\it i}) The Hamiltonian matrix of the honeycomb lattice reads
\begin{align}
\mathrm{H}^{\infty}_{\mathbf k} &= \frac{i}{2} \left( \begin{array}{cc}\label{HCH}
F_{\mathbf k} & \Gamma_{\mathbf k} \\
-\Gamma_{\mathbf k}^* & -F_{\mathbf k} \\
\end{array} \right) \hspace{1.0cm}
\Gamma_{\mathbf k} = J^z + J^x e^{ik_1} + J^y e^{ik_2} \nonumber \\
F_{\mathbf k} &= \kappa^z e^{i (k_2-k_1)} - \kappa^x e^{i k_2} + \kappa^y e^{i k_1} - \text{c.c.} 
\end{align} 
where $\mathbf{c}_{\mathbf k}^T = (c_{A,{\mathbf k}},c_{B,{\mathbf k} })$, $k_i = {\mathbf k}\cdot{\mathbf a}_i$.\\
 ({\it ii}) The Hamiltonian matrix of the hyperhoneycomb lattice reads
\begin{align}
H^{0}_{\mathbf k} &= \frac{i}{2} \left( \begin{array}{cc}
F_{\mathbf k} & \Gamma_{\mathbf k} \\
-\Gamma_{\mathbf k} ^\dagger & G_{\mathbf k} \\
\end{array} \right) \hspace{1.3cm} 
\Gamma_{\mathbf k} = \left( \begin{array}{cc}
J^z & A_{\mathbf k} e^{-ik_3} \\
B_{\mathbf k} & J^z \\
\end{array} \right) \nonumber \\
{F}_{\mathbf k} &= \left( \begin{array}{cc}
-\kappa^z_{14}(e^{ik_1}-e^{-ik_1}) & \delta_k \\
-\delta_k^* & -\kappa^z_{32}(e^{ik_2}-e^{-ik_2}) \\
\end{array} \right) \nonumber\\
{G}_{\mathbf k} &= \left( \begin{array}{cc}
\kappa^z_{32}(e^{ik_2}-e^{-ik_2}) & -\delta_{\mathbf k} \\
\delta_{\mathbf k} ^* & \kappa^z_{14}(e^{ik_1}-e^{-ik_1})\\
\end{array} \right) \nonumber \\
A_\mathbf{k} &= J^{x}_{14} + J^{14}_{14} e^{ik_1} \hspace{1 cm} B_\mathbf{k} = J^{x}_{32} + J^{y}_{32} e^{ik_2} \nonumber \\
\delta_k &= \kappa^y_{32} - \kappa^y_{14} e^{-ik_3} - \kappa^x_{32} e^{-ik_2} + \kappa^x_{14} e^{-i(k_3-k_1)}, 
\label{Hcs}
\end{align}
where $\mathbf{c}_{\mathbf k} ^T = (c_{1,{\mathbf k} },c_{3,{\mathbf k} },c_{2,{\mathbf k} },c_{4,{\mathbf k} })$.
Note that for both honeycomb and hyperhoneycomb lattices the diagonal blocks $F_{\mathbf k} , G_{\mathbf k} $ vanish for $\kappa =0$, leaving only the off-diagonal terms which couple sites on different sublattices. The lower indices of $\kappa$ -interaction specify a bond that is involved in the process leading to that term. This notation will be later useful for us for defining corresponding Raman operators. \\ 
({\it iii}) The zero-field Hamiltonian matrix of the (8,3)b lattice reads
\begin{align}\label{Hss}
\mathrm{H}^{8b}_k &= \frac{i}{2} \left( \begin{array}{cccccc}
0 & J^x_{12} e^{-ik_3} & 0 & J^z_{14} & 0 & J^y_{16} e^{-i(k_1+k_3)} \\
& 0 & 0 & J^y_{24} & J^z_{25} & 0 \\
& & 0& J^x_{34} e^{ik_2} & J^y_{35} & J^z_{36} \\
& & & 0 & 0 & 0 \\
& & & & 0 & -J^x_{56} e^{-ik_3} \\
& & & & & 0
\end{array} \right), 
\end{align}
where $\mathbf{c}_{\mathbf k}^T = (c_{1,{\mathbf k}},c_{2,{\mathbf k}k},c_{3,{\mathbf k}},c_{4,{\mathbf k}},c_{5,{\mathbf k}},c_{6,{\mathbf k}})$ and the lower triangle of the previous matrix has not been filled in for compactness, but is related to the upper triangle by the Hermiticity of $H^{8b}_k$. For the (8,3)b lattice we did not include the effects of the three-spin perturbation, because it does not constitute an important change in the symmetries within the effective Majorana description, in which time-reversal symmetry is already broken. 

\subsection{Fermi-surface topologies}\label{semimetal}\label{BandStructSec}

All three quadratic Hamiltonians in Eqs.~(\ref{HCH}-\ref{Hss}) describe band structures with protected gapless surface states. Among the three lattices we study, the honeycomb and hyperhoneycomb have {\it symmetry-protected} boundary flat bands, which become partially gapped upon introducing the symmetry-breaking magnetic field. The (8,3)b lattice is a Weyl semimetal, which has {\it topologically protected} Fermi arc surface states.\cite{Obrien,Hermanns}
 
Next we review the nature of the symmetry protecting the boundary flat bands, as its understanding is crucial to determining which Raman polarizations can couple to the gapless surface modes. 
 
To discuss the implications of time-reversal (TR) symmetry within the quadratic band structure we must first find a representation of the TR operator $\mathcal{T}$ in the Majorana spinon basis that recovers the known action of $\mathcal{T}$ on the original spin Hamiltonian. Since the action of $\mathcal{T}$ on the composite Majorana fermions is not directly observable there is indeed some choice for this representation. 
First, for the symmetry action to be entirely treated in the quadratic band structure $\mathcal{T}$ must act trivially on the $u_{ij}$. Second, due to the factor of $i$ in the quadratic Hamiltonian, $\mathcal{T}$ must act non-trivially on the Majorana spinons.
Then, since all of the lattices considered here are bipartite, TR symmetry can be represented by 
\begin{align} \label{TR}
c_{A,j} & \to c_{A,j}, c_{B,j} \to -c_{B,j} \nonumber \\
b^{\alpha}_{A,j} & \to - b^\alpha_{A,j}, b^\alpha_{B,j} \to b^\alpha_{B,j} 
\end{align} 
Provided the Hamiltonian is comprised only of NN Kitaev exchange (i.e. in the absence of a magnetic field) this gives $u_{ij} \to u_{ij}$ and $i c^A_j c^B_j \to i c^A_j c^B_j $ and recovers the correct action on the spins: $\sigma^j \rightarrow - \sigma^j$ at each site. 
However, the transformation (\ref{TR}) does not respect the translation invariance of every lattice. Indeed since the $a_1$ unit vector of the (8,3)b lattice relates sites on opposite sublattices translation invariance and $\mathcal{T}$ cannot be represented by commuting operators entirely within the Majorana spinons. Therefore, within this description the TR transformation breaks lattice translations and effectively doubles the unit cell in the $a_1$ direction. 

To pursue the consequences of this we consider the representation of $\mathcal{T}$ in the unit cell for the cases in which it is compatible with lattice translations. Time reversal takes $H_{\mathbf k} \to \mathcal{S} H_{-{\mathbf k}}^* \mathcal{S}$, where $\mathcal{S}$ is the sublattice-resolved gauge transformation 
\begin{align}\label{SS}
\mathcal{S} = \left(\begin{array}{cc}
\1 & 0 \\
0 & -\1
\end{array}\right),
\end{align}
where $\1$ is a $1 \times 1$ ($2 \times 2$) identity matrix on the honeycomb (hyperhoneycomb) lattice. For the (8,3)b lattice with a doubled unit cell the identity matrix $\1$ would have dimension $6 \times 6$. This symmetry becomes more useful if we multiply it by another symmetry that takes ${\mathbf k}\to -{\mathbf k}$, so that we can obtain a true symmetry of the matrix $H_{\mathbf k}$. Since our quadratic Hamiltonian of Majoranas essentially describes a spinless superconductor, one option is to use a particle-hole symmetry $f_{k} \leftrightarrow f^\dagger_{-k}$ for a given representation of Dirac fermions $f$ in terms of the Majoranas $c$. One such representation is 
\begin{align}
\begin{pmatrix} f_k \\ f^\dag_{-k} \end{pmatrix}= \mathcal{P}^\dagger \begin{pmatrix} c^{A}_k \\ c^{B}_{k} \end{pmatrix} \ , \ \ \ 
\mathcal{P}^\dagger = \left(\begin{array}{cc}
\1 & \1 i \\
\1 & -\1 i
\end{array}\right).
\end{align}
This gives the most trivial particle hole symmetry $\mathcal{P}$ acting as $H_{\mathbf k} \to \mathcal{P}^\dagger H_{-{\mathbf k} } \mathcal{P} = -H^*_{-{\mathbf k}}$. 
The product of $\mathcal{P}$ and $\mathcal{T}$ gives the sublattice-resolved gauge transformation Eq.~(\ref{SS}), $\mathcal{T} \mathcal{P} = \mathcal{S}$, given in Eq.~(\ref{SS}). Then $H_{\mathbf k} \to -H_{\mathbf k}$ under $\mathcal{S}$, making $\mathcal{S}$ a chiral symmetry.\cite{chiral_foot} We will refer to $\mathcal{S}$ as sublattice symmetry since it can be thought of as a gauge transformation that acts non-trivially on one of the two sublattices.

For both the honeycomb and hyperhoneycomb models, this sublattice symmetry guarantees the block-off-diagonal form of the matrices $\mathrm{H}_{\mathbf k}$. In the low-energy subspace, where we keep only the two bands that intersect at the Fermi surface, this also guarantees that the Hamiltonian takes the form $\mathrm{H}_{\mathbf k} = \vec{\sigma} \cdot \vec{d}_{\mathbf k}$ for some vectors $\vec{d}_{\mathbf k}$ such that $d_{\mathbf k}^z = 0$. The zero energy eigenvalues (which comprise the Fermi surface in our systems)
occur at the intersection of the surfaces defined by $d^x_{\mathbf k} = 0 $ and $d^y_{\mathbf k} = 0$. Hence, generically, the Fermi surfaces are lines in 3D and points in 2D. In this sense, the chiral sublattice symmetry $\mathcal{S}$ is responsible for the Fermi-ring that appears in the hyperhoneycomb lattice\cite{Schaffer} and the Dirac points in the honeycomb lattice, both of which are in the symmetry class BDI.\cite{Ryu10} 

For the (8,3)b lattice\cite{Obrien} the action of $\mathcal{T}$ can be represented as $\mathcal{S} H_{\mathbf k} \mathcal{S} = - H_{{\mathbf k}+k_1/2+k_3/2}$, where the Hamiltonian $ H_{\mathbf k}$ is given by Eq.~(\ref{Hss}). This model therefore lacks sublattice symmetry, putting it in the symmetry class $D$. This explains the different codimension of the Fermi-surface on this lattice. The perturbation $\kappa$, which breaks time-reversal symmetry $\mathcal{T}$ and hence $S$, also takes both of the other lattices into the symmetry class $D$.\cite{Hermanns} In each of these cases, $d_{\mathbf k}^z$ is generically not zero so that the Fermi ``surface" occurs at the intersections of three surfaces, which generically occurs at Weyl points. 

\subsection{Invariants and boundary modes}
While two quadratic Hamiltonians in the symmetry class BDI give examples of gapless band structures with symmetry-protected gapless boundary states,
the bulk Weyl nodes of the (8,3)b lattice lead to topologically protected surface Fermi arcs. 
Here we review the nature of these surface states, as well as the role played by symmetry to ensure their existence.

One way to see that these gapless boundary modes must exist is to identify a suitable topological invariant of the bulk band structure.\cite{Ryu10} These invariants cannot be changed unless TR symmetry, which prevents the Fermi-surfaces of these two systems from being gapped, is broken. 
The topological invariant protected by chiral sublattice symmetry can be computed for a test loop $\gamma$ in momentum space~ \cite{Schaffer} by
\begin{align}\label{nu}
\nu[\gamma] &= \frac{1}{4\pi i} \int_\gamma dk \tr\left[ H^{-1}_k \mathcal{S} \partial_k H_k \right] \nonumber \\
&= \frac{1}{2\pi} \int_\gamma dk \im\left[ \frac{\partial_k \Gamma_k}{\Gamma_k} \right] .
\end{align}
For the honeycomb lattice, $\Gamma_k$ is given in Eq.~(\ref{HCH});
evaluating the integral gives $|\nu| = 1$ for arcs $\gamma$ containing a single Dirac point and $\nu_\gamma = 0$ otherwise. The sign of $\nu$ is determined by the orientation of the loop, and which of the Dirac points is included.
Similarly on the hyperhoneycomb lattice, for which $\Gamma_k$ is given in Eq.~(8), we find $|\nu| = 1$ only for arcs that are linked with the Fermi-ring. 

\begin{figure}
	\centering
	\includegraphics[width=0.58\linewidth]{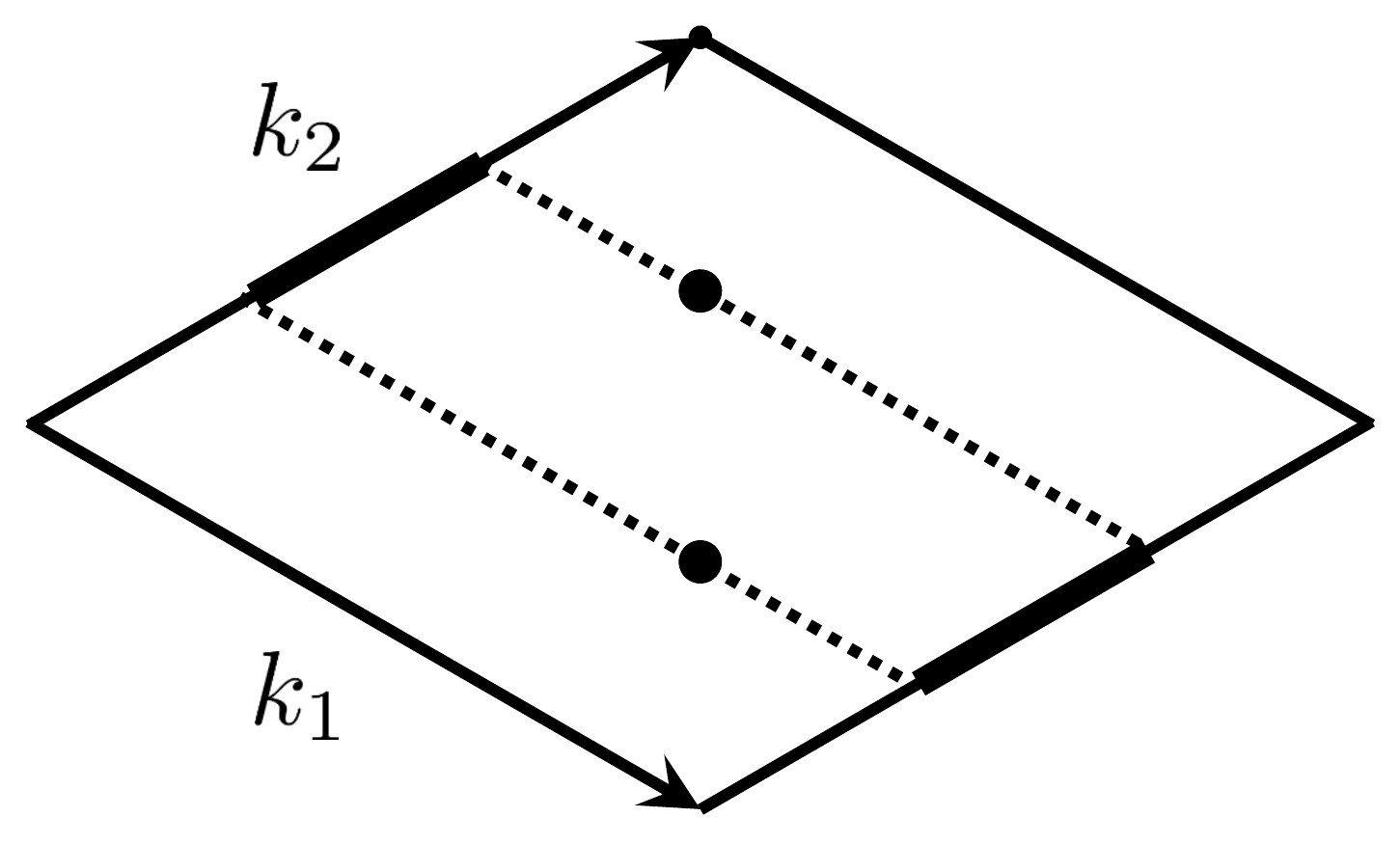}
	\caption{The 2D BZ with two Dirac points. On the $k_2$ boundaries we have highlighted the regions that have flat bands when the lattice is made finite in the $a_1$ direction.}
	\label{fig:2D_BZ}
\end{figure}

As an example, let us compute the topological invariant $\nu[\gamma]$ for the honeycomb lattice system. (The same calculation for the hyperhoneycomb lattice can be found in Ref. \onlinecite{Schaffer_supp}.) 
The Hamiltonian along any loop $k_\gamma$ reduces to an effective 1D Hamiltonian $H_{k_\gamma}$. Here we will be particularly interested in the loops obtained by fixing one of the momenta, e.g., $k_2$ and traversing the Brillouin zone in the $k_1$-direction, for which a partial Fourier transform gives a Hamiltonian in a mixed representation using real space along the $a_1$-direction and momentum space along the $k_2$-direction. With these variables one can rigorously consider making the $a_1$-direction finite, allowing a direct treatment of the boundary modes for a given $k_2$, which is still a good quantum number. 

Proceeding with the calculation, we find that for a given $k_2$ the integral (\ref{nu}) is given by
\begin{align}
\nu[\gamma_{k_2}] &= \frac{1}{2\pi} \im \int_0^{2\pi} d k \frac{ i J^x e^{ik}}{J^z + J^x e^{i k} + J^y e^{i k_2}} \nonumber \\
&= \frac{1}{2\pi} \im \int_{C} {dx} \frac{1}{x + (J^z+J^y e^{i k_2})/J^x } \nonumber \\
& = \left \{ \begin{array}{cc}
1 \quad & J^x > |J^z + J^y e^{ik_2}| \\
0 \quad & \text{otherwise},
\end{array} \right.
\end{align}
where the evaluation was done by changing variables to $x = e^{ik}$ and evaluating the contour integral around the unit circle $C$. In the case $J^x = J^y = J^z$, which we focus on here, this is nontrivial if $|1+e^{ik_2}| < 1$, which is satisfied with $k_2 \in [\frac{2 \pi}{3},\frac{4 \pi}{3}]$. This corresponds to the region between the projected Dirac points in the edge BZ, as depicted in bold in Fig. \ref{fig:2D_BZ}. For $k_2$ in this region there is a symmetry-protected Majorana mode on the edge of the system, represented by thick lines in the figure. In fact, the one-dimensional Hamiltonian obtained by taking $k_2$ as a parameter and considering $k_1$ as the 1D momentum is precisely a Kitaev Majorana chain for which the sublattice symmetry protects a Majorana end mode. \cite{Kitaev01,Fidkowski11} In Appendix \ref{app_end}, we derive the boundary modes for the finite Majorana chain with careful consideration of the effect of the perturbation $\kappa$, which later will be useful for characterizing the Raman response. 

\begin{figure}
	\centering
	\includegraphics[width=0.49\linewidth]{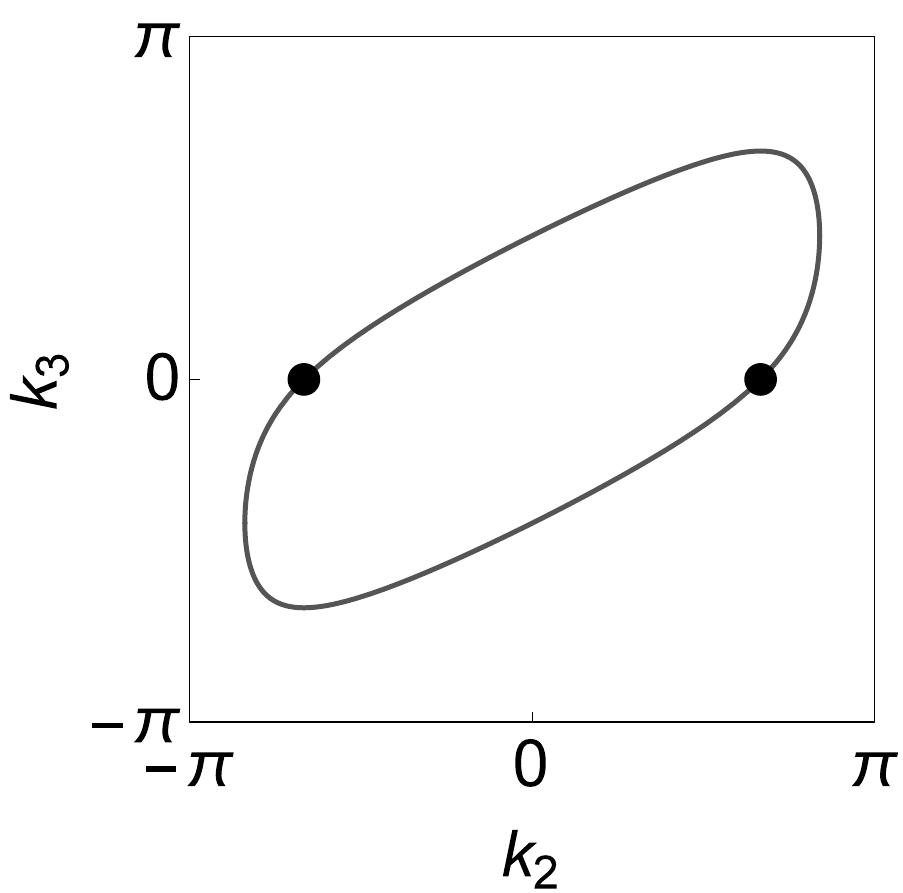}
	\includegraphics[width=0.49\linewidth]{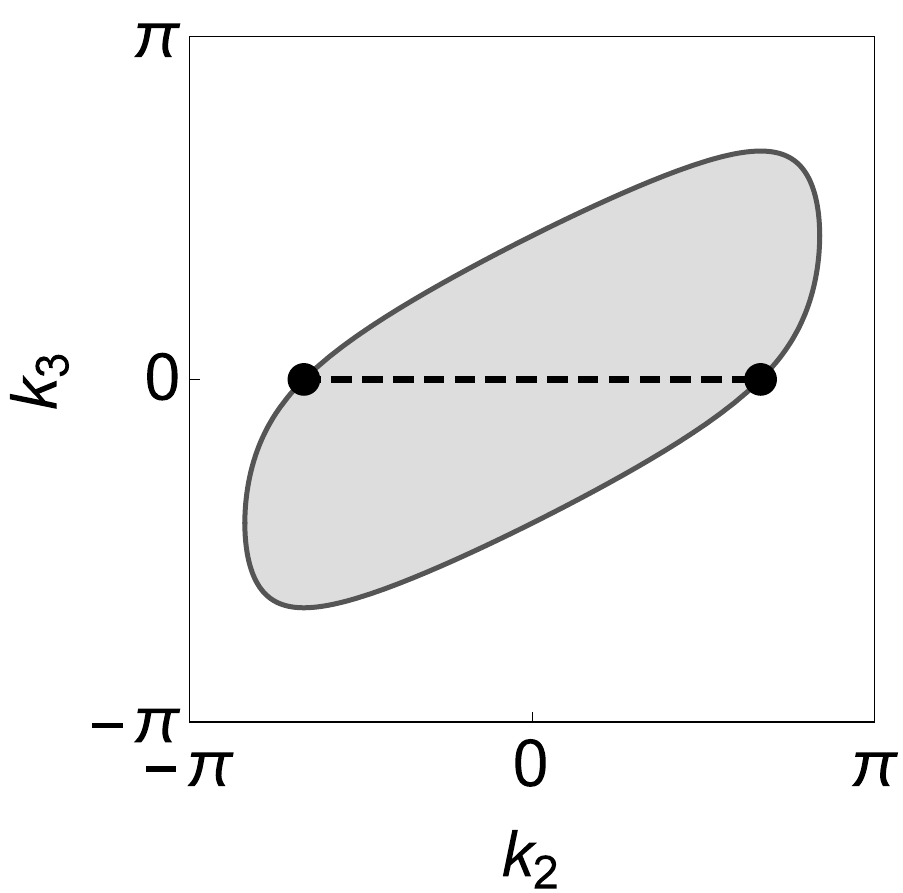}
	\caption{The Brillouin zone projected on the surface obtained by cutting normal to the $a_1$ direction. Illustrated are the projection of the Fermi-ring and the initial Weyl points (left), and their corresponding surface modes - the flat band filling the ring and the Fermi-arc connecting the points (right).}
	\label{fig:projectedBZ2}
\end{figure}

In the absence of TR symmetry gapless boundary modes may still occur, in this case protected by topology rather than by symmetry. Hamiltonians belonging to the symmetry class $D$ can have a non-trivial value for the following invariant which corresponds to the fundamental group element associated with the Berry curvature of the eigenfunctions of the filled bands:
\begin{align}
\nu_\Sigma = \frac{1}{4\pi} \int_\Sigma \vec{m}_k \cdot ( \partial_{k_1} \vec{m}_k \times \partial_{k_2} \vec{m}_k ) dk_1 dk_2 .
\end{align} 
Here $\vec{m}_k = \vec{d}_k/|\vec{d}_k|$ is a unit vector defined in terms of the decomposition of the Hamiltonian $H_k = \vec{\sigma} \cdot \vec{d}_k$ near the band crossing into Pauli matrices, and $\Sigma$ is an arbitrary surface in momentum space. In 2D, the only non-trivial choice for $\Sigma$ is to wrap the entire BZ, in which case this integral gives the Chern number. 
The Chern number for the Kitaev model on a honeycomb lattice perturbed by a weak magnetic field computed in the Kitaev's original paper was shown to be equal to $\sgn \kappa$.\cite{Kitaev} There it was also shown that breaking TR symmetry stabilizes a topologically ordered phase with a topologically protected chiral edge state. 
In 3D we obtain $\nu_\Sigma = \sum_j c_j$, where $c_j$ is the topological charge of Weyl node $j$. Nonzero projections of $\nu_\sigma$ to a surface imply the existence of topologically protected surface Fermi arcs.\cite{Wan}

In Fig. \ref{fig:projectedBZ2} we illustrate the relationship between the bulk Fermi surface and gapless surface modes for the hyperhoneycomb lattice, both with and without time reversal symmetry. When the lattice is finite in the $a_1$-direction (see Fig. \ref{fig:3DKitaev-basicunitcell-Hermanns}), a surface flat band appears inside the projection of the bulk Fermi ring onto the boundary surface if TR symmetry is unbroken. Applying a weak magnetic field gaps the Fermi ring everywhere except at a pair of Weyl nodes, whose positions
 in the limit $\kappa \to 0$, along with the projection of the Fermi ring, are shown on the left of Fig. \ref{fig:projectedBZ2}. On the right, we show the corresponding surface modes: the flat band filling the projected ring for $\kappa=0$, and the Fermi-arc connecting the projected Weyl points for small $\kappa$.


				
	\section{Theory of Raman scattering in Kitaev-like Mott insulators} \label{RamanSec}

	Here we develop the framework for computing the resonant Raman response. As stated in the introduction, though we focus particularly on the terms relevant to the exact Kitaev QSL, we derive the Raman vertices for generic JKK-type systems, in which nearest-neighbor exchange is mediated predominantly by intermediate oxygen sites, leading to large Kitaev-type interactions.\cite{Jackeli} The results of this section are therefore also applicable to other settings where resonant exchange processes need to be considered in these systems. 

	\subsection{The JKK systems}
	
We begin with a detailed review of the exchange processes in JKK systems. In iridates and ruthenates -- the two families of JKK systems currently realized experimentally -- the magnetic degree of freedom arises from electrons occupying $t_{2g}$-orbitals. For these materials, it is convenient to describe the low-energy spin state of the $d^5$-configuration of Ir$^{4+}$ and Ru$^{3+}$ ions by using a hole description. In the local axes bound to the oxygen octahedron the $t_{2g}$ orbitals are
$\left\vert X\right\rangle =\left\vert yz\right\rangle $,
$\left\vert Y\right\rangle =\left\vert zx\right\rangle $, and
$\left\vert Z\right\rangle=\left\vert xy\right\rangle $.
	
The bond symmetry of a pair of non-distorted edge-sharing octahedra restricts the independent hopping integrals to three terms, which we call $t_1, t_2,$ and $t_3$. \cite{Rau14,Rau16} $t_2$ is special because it leads to the Kitaev interaction, and because it is the only term that comes from oxygen-mediated exchange paths, which are dominant in JKK systems. $t_1$ and $t_3$ come from direct hopping processes between iridium sites. Following Rau. {\it et al},\cite{Rau14} we include one additional hopping integral $t_4$ to account for the other terms that exist when the local structure deviates from perfect octahedra, such as by trigonal or monoclinic distortions.\cite{trigfoot} Then the hopping integrals between the $t_{2g}$ orbital states in the basis $(\ket{X},\ket{Y},\ket{Z})^T$ for a NN $z$-bond take the form
\begin{align}\label{hop}
{\hat t}_{\rm \small NN}^z = 
\left(
\begin{array}{ccc}
t_1 & t_2 & t_4 \\
t_2 & t_1 & t_4 \\
t_4 & t_4 & t_3
\end{array}\right).
\end{align}	
The forms of ${\hat t}_{\rm \small NN}^x$ and ${\hat t}_{\rm \small NN}^y$ are obtained by permuting the basis of orbitals. On the honeycomb lattice rotation symmetry ensures that the $t_i$ have the same magnitude and structure on all three bonds. 
On the 3D tri-coordinated lattices, the hopping integrals of $Z$ and $X$ or $Y$ bonds may be significantly different, however in this work, for simplicity, we will treat them as being of the same order of magnitude. 

The second-neighbor hopping ${\hat{t}}_{\rm NNN}^{z}$ has the same symmetry as the product $\hat{t}_{\rm NN}^{x}\hat{t}_{\rm NN}^{y}$ and is therefore not a symmetric matrix. We choose a simplified form given by
\begin{align}\label{hop2}
\hat{t}^{z}_{\rm \small NNN} = 
\left(
\begin{array}{ccc}$ $
0 & t_s & 0 \\
t_s + t_a & 0 & 0 \\
0 & 0 & 0
\end{array}\right).
\end{align}
The origin of these processes is a bit more complicated, and is different for RuCl$_3$ than for the iridate A$_2$IrO$_3$ compounds. In the former case, 
there exist only the Ruthenium octahedra in the lattice, and the hoppings between second neighbors are primarily due to direct overlap of Ru orbitals; consequently these terms are expected to be small. In A$_2$IrO$_3$, an extended s-orbital of the element A$=$Na,Li in the center of the octahedron makes a sizable contribution to second-neighbor hoppings along the path Ir-O-A-O-Ir. The form (\ref{hop2}) for this hopping is justified by numerical work on the iridates.\cite{Yamaji14,Foyevtsova13,kim15,Winter16} 
In addition, with $t_a < t_s$ this model is sufficient to explain the dominant Kitaev interactions together with non-vanishing Heisenberg exchange between second neighbors believed to apply to the iridates.\cite{Sizyuk} 

To proceed, it is useful to express the above hopping matrix elements in the basis of the angular momentum eigenstates diagonalizing the spin-orbit coupling, $|{\mathbf j}_{eff}={\mathbf J},J_{z}\rangle$. 
These states are energetically split into a low-energy Kramers doublet $|\frac{1}{2},J_{z}\rangle$ and a higher-energy quartet $|\frac{3}{2},J_{z}\rangle$. (In the presence of lattice distortions these are not exact eigenstates, but are adiabatically related to them; therefore we use the same notation in both cases). 

In order to obtain a hopping matrix in the hole picture, we substitute electronic creation/annihilation operator with those of holes.
In the single-hole eigenbasis this results in a hopping matrix $T^{a,a'}$, where $a,a'$ run over these six angular momentum eigenstates; $T^{a,a'}$ can be obtained from Eq. (\ref{hop}) using the appropriate change of basis between the $|X \rangle, |Y \rangle, |Z \rangle$ orbitals and the $|{\mathbf J},J_{z}\rangle$ eigenstates. 
In this basis, the hopping Hamiltonian takes the form
\begin{align}
H_t= \sum_{n,n',a,a'} T^{a,a'}_{n,n'} \psi^\dagger_{n,a} \psi_{n',a'},
\end{align}
where $n,\, n'$ are site indices, $\psi_{n,a}^{\dag }$ and $\psi_{n,a}$ are the hole creation and annihilation operators in the angular momentum eigenstate indexed by $a$. 
		
To describe the exchange couplings, the on-site interactions are also essential. We treat the on-site interaction physics with the Kanamori Hamiltonian, for which Hund's coupling $J_H$ plays an essential role: \cite{Rau14} 
\begin{align}\label{Kanamori}
H_{\text{int}} = \sum_n \left[\frac{U-3J_H}{2}(N_n-5)^2 - 2J_H S_n^2 - \frac{J_H}{2} L_n^2\right],
\end{align}
where $L_n=1$ is the effective orbital angular momentum on the site $n$ and $N_n$ is the number of electrons in the $t_{2g}$ orbitals. For the case of no holes, which is a filled-doublet, this Hamiltonian is magnetically trivial. However, for the case of a two-hole state the interactions are essential. There are $15$ two-hole states, and we indicate a basis of product states of single-particle eigenstates by $|\mu \rangle = |\psi^{(1)}\rangle |\psi^{(2)}\rangle=| {\bf J}^{(1)}, J_z^{(1)};{\bf J}^{(2)}, J_z^{(2)} \rangle$, and the two-hole angular momentum eigenstates of (\ref{Kanamori}) by $\xi$. 
Ignoring the lattice distortions, the eigenstates $\ket{\xi}$ can be obtained from the single-particle eigenstates simply by using the Clebsch-Gordon coefficients. 
	
	\subsection{Raman scattering in Mott insulators}
	
The basic processes leading to the Raman response are similar to those leading to exchange interactions, except that the electron hopping is assisted by photons. Consequently, the operator describing Raman processes is proportional to the spin-exchange couplings, weighted by polarization-dependent factors that determine the ability of the photons to control the magnitude of an electron hopping along certain bonds.\cite{Loudon,fleury68,Shastry,Perkins08,Ko,Perkins13,Knolle14-2,Perreault15,Perreault16,Nasu2016}
	
In our derivation, we will follow the $\mathbf{T}$-matrix formulation of time-dependent perturbation theory for Raman scattering. \cite{Shastry,Ko,Bruus04}
At zero temperature the Raman intensity
can be written as a correlation function of Raman operators $R$:
\begin{align}\label{I}
I(\omega) = 2\pi \int d\omega e^{i\omega t}\braket{R(t) R(0)},
\end{align}
where $\omega=\omega_{\rm in}-\omega_{\rm out}$ is the total energy transferred to the system, and in the following we assume that $\omega \ll \omega_{\rm in (out)}$. 
For a Mott-insulator, the Raman operator is 
\begin{align}\label{R1}
R= -P H_t^{	{\boldsymbol \epsilon}_{\rm out} } (H-i\eta)^{-1} 
H_t^{ {\boldsymbol \epsilon}_{\rm in}} P,
\end{align}
where $P$ is the projector onto states with a fixed electron occupancy per site, and ${\boldsymbol \epsilon}_{\rm in}$ and $ {\boldsymbol \epsilon}_{\rm out}$ are the incoming and outgoing photon polarization vectors, respectively. 
$H_t^{\epsilon}$ is the electron/photon vertex for the polarization $\epsilon$ given by
\begin{align}
H_t^{\boldsymbol{\epsilon}} = \left(\frac{i e}{\hbar c}\right)\sum_{n,n'}\sum_{{\mbox{$ \begin{smallmatrix}
a= 1,2 \\ a'=1,...,6 
\end{smallmatrix}$}}} (\mathbf{d}_{n,n'}\cdot \boldsymbol{\epsilon}) T^{a,a'}_{n,n'} \psi^\dagger_{n,a} \psi_{n',a'}, 
\end{align}
where $a =1,2$ runs through the low-energy doublet (i.e. the states that can be occupied before scattering) and $a'=1,..,6$ runs through all of the single hole angular momentum eigenstates. We use $\mathbf{d}_{n,n'}$ to denote the spatial vector from the lattice site $n$ to site $n'$. 

\noindent The full Hamiltonian in the resolvent $(-H+i\eta)^{-1}$ can be written as $H=H_t+H_{U}$, where for convenience, we define the interaction term $H_U$ relative to the initial photon energy, $H_U=H_{\text{int}} - \omega_{\rm in}$, with $H_{\text{int}}$ given in Eq.~(\ref{Kanamori}). The resolvent $(-H+i\eta)^{-1}$ can be formally expanded to give
\begin{align}\label{RT}
R= P H_t^{\boldsymbol{\epsilon}_{\text{out}}} \left[H_U^{-1} + H_U^{-1}H_t H_U^{-1} + ... \right] H_t^{\boldsymbol{\epsilon}_{\text{in}}} P,
\end{align}
where we have dropped the finite (negative) imaginary part $-i\eta$ in the inverse operators for simpler expression.

In the presence of a magnetic field, the resolvent in Eq.~(\ref{RT}) has an additional small parameter proportional to $h/U$:
\begin{align}
\left[H_U + H_t + H_h\right]^{-1} = H_U^{-1}\left[\1 + H_t H_U^{-1} + H_h H_U^{-1} + ... \right].
\nonumber
\end{align} 
Hence, in the regime $h \ll t$ we can neglect the magnetic field during the Raman process.\cite{Perreault16}

If $t/(U-\omega_{\rm in})\equiv t/U_\omega$ is small, electron hopping is strongly suppressed, and the derivation of the Raman operator proceeds as it does for a spin-exchange Hamiltonian. The lowest-order terms contributing to $R$ are linear in $t/U_\omega$ and have the well-known Loudon-Fleury (LF) form~\cite{fleury68}
\begin{align}\label{RF}
R& = \sum_{n,n';\alpha,\beta} (\mathbf{d}_{n,n'}\cdot \boldsymbol{\epsilon}_{\text{in}}) (\mathbf{d}_{n,n'}\cdot \boldsymbol{\epsilon}_{\text{out}}) H_{n,n'}^{\alpha,\beta}\sigma^\alpha_n \sigma^\beta_{n'} ,
\end{align}
where $H_{n,n'}^{\alpha,\beta}$ defines the generic spin-exchange Hamiltonian on the bonds $\langle n,n'\rangle$. 

It is useful to review the algebra required to compute the spin-exchange processes contributing to $H_{n,n'}^{\alpha,\beta}$, which we will generalize to include further hoppings when we examine resonant Raman scattering in the next subsection. In the basis of single-particle eigenstates $a,a'$ on site $1$ and $b,b'$ on site $2$, 
we have
\begin{align}
H_{n,n'}^{\alpha,\beta} &= \sum_{\mbox{$\begin{smallmatrix}
a,a' = 1,2 \\ b,b' = 1,...,6 
\end{smallmatrix}$}} \sigma_{a'a}^\alpha [T_{n,n'}]_{a b} \Sigma_{b b'}^\beta [T_{n',n}]_{b' a'} \\
&= \tr\left[ \sigma^\alpha_n {T}_{n,n'} \Sigma^\beta_{n'} {T}_{n,n'} \right]. \label{exch}
\end{align} 
Here $\sigma^\alpha$ is the $\alpha$ Pauli matrix acting on the low-energy doublet states, and the interaction between the $6$ two-hole intermediate states and the low-energy ${\bf J}= 1/2$ doublet on site $n'$ is described by the matrix element $\Sigma^\alpha_{bb'}$. 

Computing the $\Sigma$ matrix is a non-trivial but a straightforward procedure.
We do it in three steps: {\it i}) diagonalize the two-hole Hamiltonian, {\it ii}) rewrite each eigenstate $|\xi\rangle$ of the two-hole Hamiltonian in the basis of the product states of the single-particle angular momentum eigenstates $|\mu\rangle$ using Clebsch-Gordan coefficients described by a matrix $C_{\mu,\xi}$ as in
\begin{align}
[H_U]^{-1}_{\mu\mu'} &= C_{\mu,\xi} \frac{1}{\epsilon_{\xi}} C_{\xi,\mu'}^\dagger \\
[H_U]^{-1}_{ab;b'a'} &= [H_U]^{-1}_{\nu_{a,b},\nu_{b',a'}} \sgn(a-b)\sgn(b'-a').
\end{align}
Here $\nu_{a,b} = 1,...,15$ is the index of a unique product state given the indices for two single-hole eigenstates $a$ and $b$. Note that $\nu_{a,b} = \nu_{b,a}$ and the fermion statistics are taken care of by the explicit $\sgn$ factors.
Finally, {\it iii}) compute the matrix element between an electron coming into single-hole eigenstate $a'$ and going out from $a$ with the local low-energy doublet's $\alpha$ component. 
\begin{align}
\Sigma^\alpha_{aa'} &= - \sigma_{b'b}^\alpha [H_U]^{-1}_{a,b;b'a'}. 
\end{align}

\subsection{Raman matrix elements beyond nearest neighbors}\label{res}

\begin{figure}
	\centering
	\includegraphics[width=\linewidth]{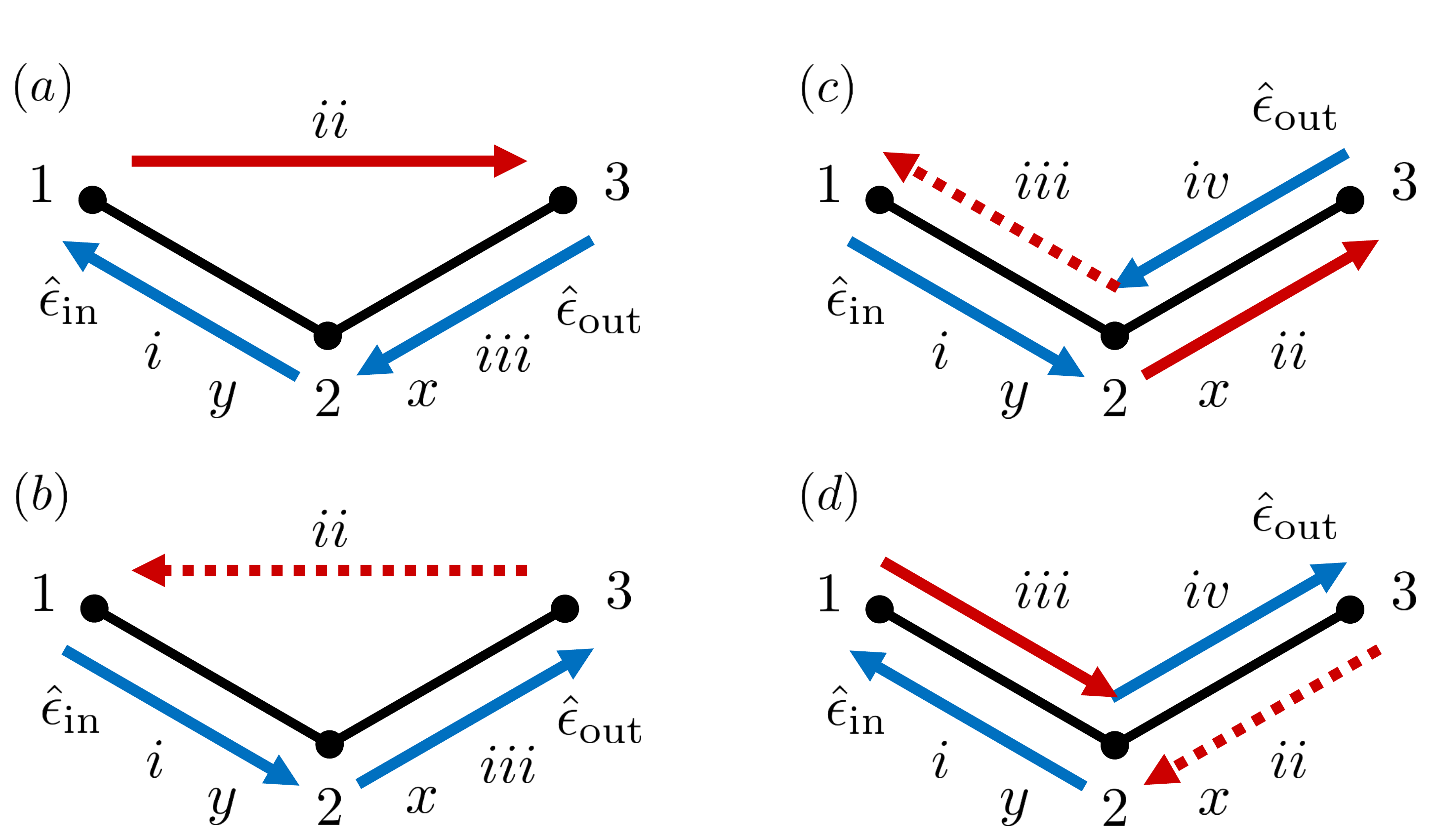}
	\caption{(Color Online) The three- and four-hop pathways on a tri-coordinated lattice with second-neighbor hopping. The Roman numerals indicate the order in which the hops occur. The blue arrows represent hops mediated by photons, which connect the half-filled sector to the one with one holon and one doublon. The solid red represents a doublon hop and the dashed red arrows are for holon hops. For the three hop processes there are similar processes starting at the other sites, indicated by the subscript, and each path has a reverse, indicated by a prime.}
	\label{fig:pathways}
\end{figure}

Having reviewed the steps by which the standard Loudon-Fleury Raman vertex is obtained in JKK systems, we now derive the analogous operator for resonant Raman processes. Resonant Raman scattering involves driving the system at photon frequencies where $t/U_\omega$, though less than $1$, is not overwhelmingly small. In this case, processes involving multiple electron hops can contribute significantly to the Raman response. To compute such contributions, we have to consider terms that are of subleading order in $t/U_\omega$. Here we will include terms generated by both three and four hop processes, which contribute to an effective 3-spin effective Raman vertex. This 3-spin vertex is of particular interest as, unlike the Loudon-Fleury vertex, it can couple to the symmetry-protected gapless boundary modes of the honeycomb and hyperhoneycomb Kitaev QSLs.

The general three-spin terms of interest have the form \cite{Shastry,Shastry2,Ko}
\begin{align}\label{Rres1}
R_{\text{res}} &= i \sum_{\ll ilj \gg} M_{ilj}^{\alpha \gamma \beta} 
\sigma^{\alpha}_i \sigma^\gamma_{l} \sigma^{\beta}_j \times A_{ilj} \\ 
A_{ilj} & = \left[\left(\boldsymbol{\epsilon}_{\text{in}} \cdot \mathbf{d}_{ji} \right) \left(\boldsymbol{\epsilon}_{\text{out}} \cdot \mathbf{d}_{il} \right) - \left(\boldsymbol{\epsilon}_{\text{out}} \cdot \mathbf{d}_{ji} \right) \left(\boldsymbol{\epsilon}_{\text{in}} \cdot \mathbf{d}_{il} \right) \right], \nonumber
\end{align} 
where the polarization-dependent factor $A_{ilj}$ is only non-zero in polarization channels that are anti-symmetric in the exchange of {\it in} and {\it out} polarizations. These anti-symmetric channels vanish in the non-resonant Loudon-Fleury Raman operator. In the following, we denote the symmetrized channel as $R_{\alpha \beta} = (R_{\alpha \beta} + R_{\beta \alpha})/2$ and the anti-symmetrized one as $R_{[\alpha \beta]} = (R_{\alpha \beta} - R_{\beta \alpha})/2$. 

The remainder of this section is devoted to computing the matrix element $M^{yzx}_{123}$ for the leading-order exchange processes involving 3 and 4 hops. 
We will focus on the subset of the resulting chiral three-spin terms that project into the zero-flux sector of the Kitaev Hamiltonian discussed in the previous section, as these are the only ones that can contribute at energies below the flux gap. In practice this means including terms that follow the `Kitaev' symmetry such that the spin components in Eq.~(\ref{Rres1}) for an $\alpha$ and $\beta$ bond sharing a site $2$ has the spin component of the outer sites determined by the connecting bond.

As shown in Fig. \ref{fig:pathways}, as an example, we choose sites $1,2,$ and $3$ such that $1$ and $2$ are connected by a $y$ bond and $2$ and $3$ are connected by an $x$ bond and compute the matrix element $M^{yzx}_{123}$. The form of $M^{\alpha \gamma \beta}_{ilj}$ for other NN bond pairs is identical. We proceed primarily under the assumption that the conditions for realizing the Kitaev model are near perfect, so that the dominant hopping terms are the ones mediated by oxygens and the direct hopping terms are perturbatively small.

We first consider terms involving three hops, which necessarily involve one hop across a NNN bond. There are twelve such 3-hop processes. To describe them, it is convenient to use the language of doublons and holons \cite{Ko} by calling the two-hole state the doublon (as it involves two excitations) and the completely filled state the holon (representing a lack of excitations). We label the processes that have a doublon hopping clockwise by $(a_1), (a_2),$ and $(a_3)$ when the initial hop is from site $1,2,$ or $3$ respectively; the process $(a_2)$ is illustrated in Fig. \ref{fig:pathways}(a). The corresponding counter-clockwise processes are $(a_1'), (a_2'),$ and $(a_3')$. The remaining six processes (see Fig. \ref{fig:pathways}(b)) are analogous, except with an intermediate holon hop; we label the clockwise (counter-clockwise) processes $(b_j)$ ($(b_j')$) respectively. For each such process, the contribution to $M_{ilj}^{\alpha \gamma \beta}$ is given by a trace of the relevant hopping and interaction matrices. For the process $(a_2)$, for example, we obtain $\tr\left[{T}_{\rm NN}^x \Sigma^\alpha {T}_{\rm NNN}^{z} \Sigma^\beta {T}_{\rm NN}^y \sigma^\gamma \right]$. (More technical details can be found in the Mathematica notebook included in the ancillary files.)

All of the processes considered here turn out to yield contributions are equal in magnitude, and of opposite sign, to their time-reversed partners, and consequently cannot contribute to Raman channels that are even under the exchange of in and out polarizations. However, the odd combination of these events $[(a_1)-(a_1') + (a_2)-(a_2')+(a_3)-(a_3')]$ plus an analogous one for the (b$_i$)'s yields a non-vanishing Raman matrix element. Adding together the contributions of all possible three-hop processes gives a contribution to the three-spin Raman matrix element:
\begin{align}\label{R3}
M_{123}^{yzx} &= -i \frac{4 J^2 t_2^2 (3{t}_a+5{t}_s)}{9(J-U_\omega)^2(3J-U_\omega)^2}\left[1 + \mathcal{O}\left(\frac{t_4 t_1}{t_2^2}\right)\right] 
\end{align} 
where we have used that $t_2 \gg t_1,t_3,t_4$. Notice that the term (\ref{R3}) appears only at $\mathcal{O}(J^2/U_\omega^4)$. This occurs because the $t_2$ hopping term does not allow hopping directly between the low energy ${\bf J} = 1/2$ doublets, and requires Hund's coupling to mediate the interaction with this low-energy spin. There are thus no three spin terms coming from this process in the absence of Hund's coupling.

In addition to Eq. (\ref{R3}), there are other 3-hop terms at the same order $\mathcal{O}(J^2/U_\omega^4)$ that do \textit{not} project into the zero-flux sector. As these processes are suppressed at low energies, we do not present them here.

The 4-hop processes are of two types. The first type of processes are those in which an electron hops traverse the simple path $1\to 2 \to 3 \to 2 \to 1$ (and the analogous process starting at the site $3$). For these paths, the photon is absorbed and re-emitted on the same bond, making them their own time-reversal partners. Consequently, these processes can only contribute to the symmetric Raman channels. 

The second type of processes involve one intermediate holon hop and one intermediate doublon hop, instead of two doublon hops. The holon hop must happen when the doublon is on site 1 or 3, yielding the two distinct types of paths in Fig.~\ref{fig:pathways} (c) and (d), as well as their time-reversed partners, which we label $(c')$ and $(d')$ (not shown). Again the sum of time-reversal pairs vanishes, but their difference yields a Raman term that is odd under time reversal. Each term gives a contribution to the matrix element $M^{\alpha \beta \gamma}_{i j k}$ of the form $\tr\left[T_{\rm \small NN}^y \sigma^\alpha T_{\rm\small NN}^x (\Sigma')^{\beta'} T_{\rm \small NN}^x \Sigma^\beta T_{\rm\small NN}^y \sigma^\gamma \right]$, where $\Sigma'$ represents the interactions for staying at the same site during two hops, and the bond labels $x,y$ correspond to the pathway shown in Fig. \ref{fig:pathways}(c). Note that $\beta$ and $\beta'$ correspond to the same site 2. Adding up $[(c)-(c') + (d)-(d')]$ gives
\begin{align}\label{R4}
M_{123}^{yzx} &= i \left[3t_2^2 - t_2(t_3+11t_1) - t_4(3t_4+(t_3+2t_1)) \right]  \nonumber \\
& \times {t}_4 (2t_1+t_3) \frac{32 J^2 (2J-U_\omega)}{81(J-U_\omega)^3(3J-U_\omega)^3} \ .
\end{align} 
Note that the $t_4$ term is non-zero only in the presence of the trigonal distortion. Both Eqs. (\ref{R3}) and (\ref{R4}) multiply the same polarization factors and 3-spin term $S_1^z S_2^y S_3^x$ in Eq.~(\ref{Rres1}). 

There are also other three-spin terms that do not project into the zero flux sector appearing at the order $J^2/U_\omega^5$, which do not require the symmetry-breaking hopping $t_4$. Importantly, there are no three-spin terms at lower orders in $J/U_\omega$. All three-spin terms due to the 4-hop processes vanish if there is only oxygen-mediated hopping. This is easy to understand since the holon has just one electron hopping to an empty site so that Hund's coupling cannot be involved. Then since the $t_2$ hop is not allowed between low energy ${\bf J} = 1/2$ states, these states can only be connected through direct hopping terms. 

In summary, the 3-spin term in Eq. (\ref{Rres1}) can appear even in materials with no direct electron hopping processes, provided that second-neighbor oxygen-mediated hopping is present. Alternatively, if we include direct hopping processes, all of the processes shown in 
Fig.~\ref{fig:pathways} can contribute to the 3-spin resonant Raman response. We emphasize that the 3-spin term needs not to be of the same order as the Loudon-Fleury term in the Raman vertex, since it will be the dominant contribution to the anti-symmetric polarization channel, to which the Loudon-Fleury term does not contribute.

			
\section{Raman scattering results}	\label{ResultsSec}
				
In Section \ref{BandStructSec}, we reviewed the topological nature of the band structures of Kitaev QSLs on the honeycomb, hyperhoneycomb, and (8,3)b lattices. In the former two cases, both with and without a magnetic field perturbation, we argued that 
resonant Raman scattering is, in principle, well-suited to detect the corresponding protected gapless boundary modes. We now present detailed results for the resulting Raman spectra, both for the thin film systems where we expect the gapless boundary modes to be visible in the low-energy spectrum, and for the bulk systems. 

\subsection{Raman spectra for strips and slabs}\label{ResRaman2}

\begin{figure*}
	\centering
	\includegraphics[width=\linewidth,valign=t]{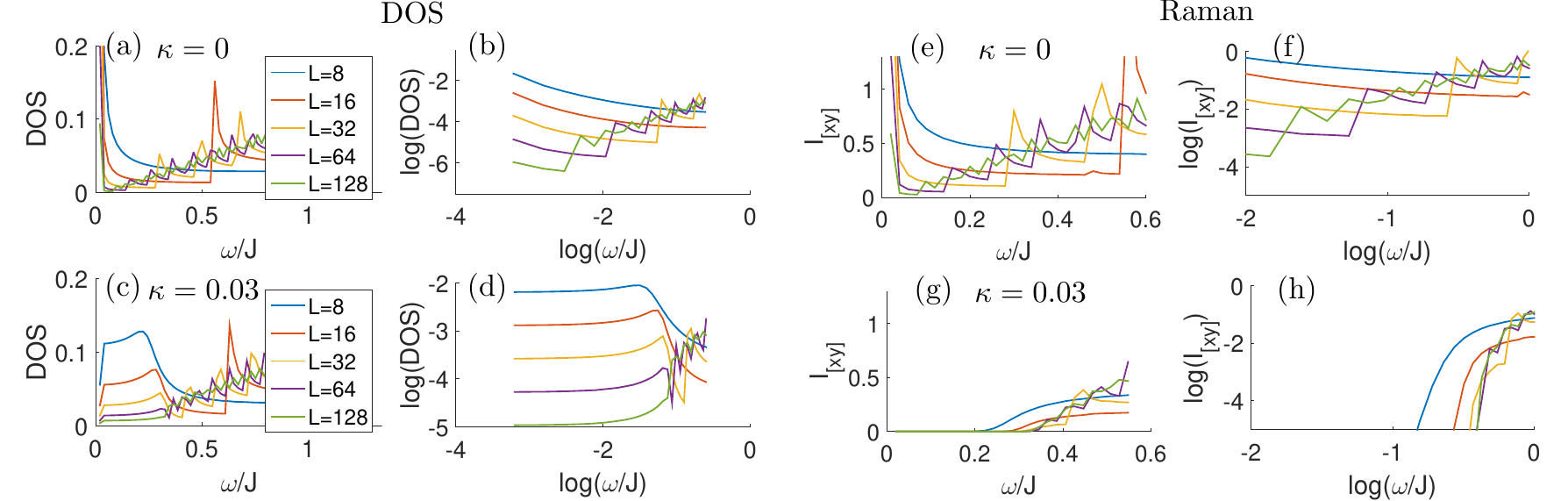}
	\caption{The low-energy DOS for the honeycomb lattice plotted for (a) $\kappa=0$ and (c) $\kappa=0.03$ for different slab widths $L$, measured in unit cells in the $a_1$ direction. The low-energy peaks and plateaus in the DOS are due to (a) the edge flat bands and (c) topological edge modes. (b) and (d) show the corresponding log-log plots. The crossover between power laws describing the surface contribution to ones describing the bulk is clearly seen. (e), (f), (g), and (h) are the same plots for	the resonant Raman intensity $I_{[ab]}$ in the antisymmetric $[ab]$ channel. The suppression of low-frequency modes in (g) compared to (c) is due to the matrix element effects discussed in section \ref{ResRaman2}. All spectra are obtained using the methods outlined in Ref. \onlinecite{Perreault15}.
	}
	\label{fig:slab_HC}
\end{figure*}

\begin{figure*}
	\centering
	\includegraphics[width=\linewidth,valign=t]{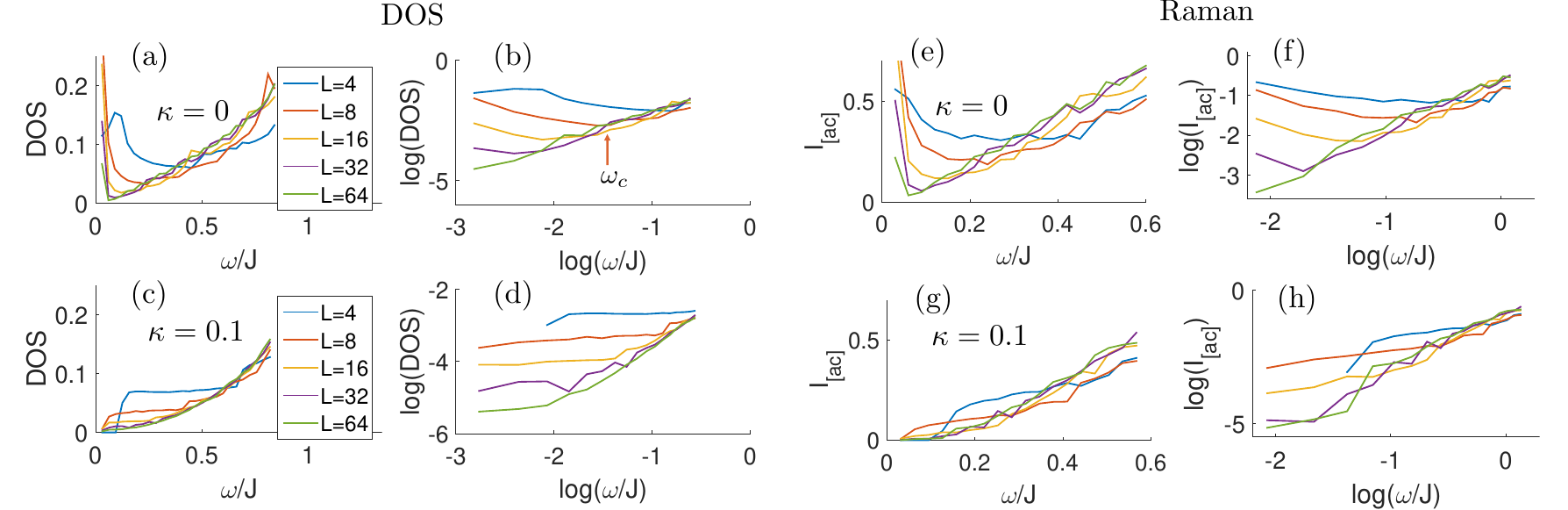}
	\caption{The low-energy DOS for the hyperhoneycomb lattice plotted for (a) $\kappa=0$ and (c) $\kappa=0.1$ for different slab widths $L$, measured in unit cells in the $a_1$ direction. Note that the unit cell size is $4$ compared to $2$ on the honeycomb lattice so that the same lengths are taken when counted in numbers of sites. Similar to the honeycomb lattice case the low-energy peaks and plateaus in the DOS are due to (a) the surface flat bands and (c) surface Fermi-arcs.
	}
	\label{Fig4}
\end{figure*}

\begin{figure*}
	\centering
	\includegraphics[width=\linewidth,valign=t]{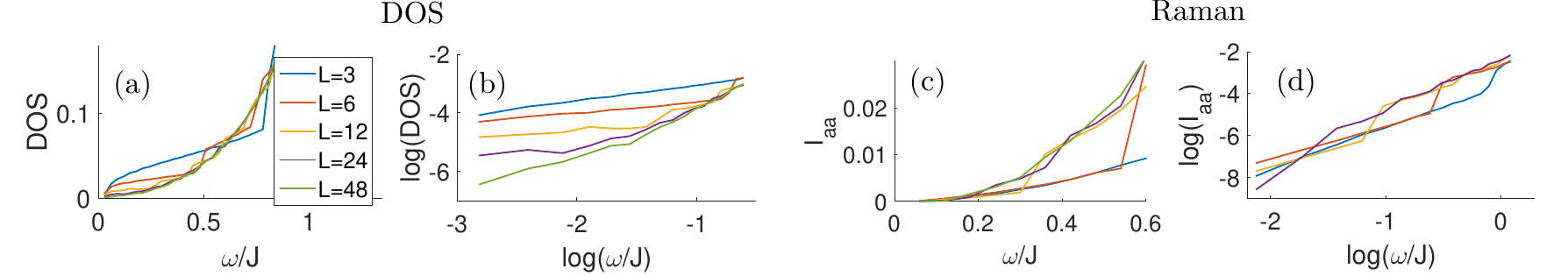}
	\caption{The low-energy DOS and the Raman intensity $I_{aa}$ for the (8,3)b lattice plotted for different slab widths $L$, measured in unit cells in the $a_1$ direction (unit cells have six sites on this lattice). The low-energy plateaus in the DOS are due to surface Fermi-arcs in this case similar to the $\kappa > 0$ case on the hyperhoneycomb lattice.
	}
	\label{fig:slab(8,3)b}
\end{figure*}
 
To study the Raman response of the topological surface modes, we consider systems that are infinite in two directions, but have a finite length $L$ in the stacking direction $a_1$. In the following, we measure $L$ in units of $a_1$. For all of the cases considered here, given a fixed $L$ (on the order of a few tens of $a_1$), there is an energy below which the DOS is dominated by the two surfaces, leading to a possibility to detect signatures of the surface modes in the Raman spectrum. 

Specifically, in the honeycomb and hyperhoneycomb cases with $\kappa = 0$, the flat surface bands lead to a $\omega \approx0$ peak in the DOS, defined as $DOS(\omega) \equiv \sum_{{\rm eigenstates} \, \xi} \delta(\omega - \epsilon_\xi)$ for finite $L$ (Figs.~\ref{fig:slab_HC} and \ref{Fig4} (a,b)). Because the finite thickness allows weak back-scattering between the top and bottom surfaces, at finite $L$ the surface modes do not form a true flat band, and the observed peak is neither infinitely sharp nor strictly at $\omega=0$, though it becomes increasingly sharply focused there in the limit of large $L$. The height of this peak relative to the rest of the spectrum also decreases with $L$, however, due to the decreasing surface-to-bulk ratio. 

As discussed in Sec.~\ref{RamanSec}, for both of these systems the sublattice symmetry ensures that the boundary flat bands can be seen only in the antisymmetric Raman channels. At low frequencies the resulting Raman spectrum in the $[ac]$ channel (Figs.~\ref{fig:slab_HC} and \ref{Fig4} (e,f)) closely tracks the DOS, as anticipated, giving a qualitative signature of the topological surface flat bands. 

When $\kappa \ne 0$, or for (8,3)b lattice, the DOS is expected to tend to a constant at zero energy, since most of the states in the flat surface band are gapped, leaving only a surface Fermi arc. This low-energy plateau is clearly visible in the DOS, as shown in Figs.~\ref{fig:slab_HC} and \ref{Fig4}~(c,d) and Fig.~\ref{fig:slab(8,3)b}(a,b). However, in the Raman response the effect is strongly suppressed, as shown in Figs.~\ref{fig:slab_HC} and \ref{Fig4} (g,h), and Fig.~\ref{fig:slab(8,3)b}(c,d). This suppression is present in both non-resonant Raman channels, such as $I_{(a,a)}$, {\it and} resonant Raman channels such as $I_{[a,c]}$. It is most striking for the honeycomb and (8,3)b lattices, though it is also present for the hyperhoneycomb lattice.


These unexpected results indicate that for surfaces with broken TR-symmetry the Raman spectrum does not simply reflect the DOS. Instead, the contribution of the surface modes to Raman scattering is also suppressed by matrix element effects, which occur when two boundary mode excitations cannot be created on the same surface without momentum transfer. As we now discuss, these processes are suppressed because the Raman scattering is essentially a $\Delta q = 0$ process,\cite{Devereaux07} which in the Kitaev spin liquids excites a pair of spinons on neighboring lattice sites. 

Let us investigate how this affects each of the lattices. 
With $\kappa > 0$, the gapped 2D honeycomb lattice has chiral Majorana edge modes -- meaning that all of these Majoranas on the top (bottom) edge of our strip will be right (left) movers. It follows that creating a pair of such excitations on (say) the top edge requires a net momentum transfer, which cannot be accomplished with Raman processes. Although a Raman process could, in principle, create one surface mode with momentum $\vec{p}$ and one bulk mode with momentum $-\vec{p}$, the bulk modes are gapped, so that such a spinon pair cannot be created at arbitrarily low frequencies. Thus, in this case the low-frequency behavior seen in the DOS is not observed in the Raman spectrum. 

On the other hand, if we leave time-reversal symmetry intact on the honeycomb lattice the edge modes have a very different character: as explained in Sec. \ref{ModelSec}, they now consist of a flat band over the range $\frac{2\pi}{3}<k_2 <\frac{4\pi}{3}$, where $k_2$ represents the momentum along the edge. States in this symmetry-protected flat band necessarily consist of both right-movers and left-movers, since the symmetry protection requires a degeneracy at each $k_2$ value in the flat-band region. Thus in this case a Raman process can create a pair of boundary low-energy spinons on the same edge, and the resonant Raman response tracks the DOS. 

It is worth emphasizing that for the honeycomb and hyperhoneycomb lattices even with time-reversal symmetry not all Raman channels can couple to the gapless boundary modes. \cite{Perreault16} This is because the boundary modes are sublattice polarized, which poses a problem for a two-spinon operator that respects (i.e. is odd/chiral under) sublattice symmetry. The Loudon-Fleury, or non-resonant Raman operator is exactly such an operator, taking its two-spinon form from the Hamiltonian itself, and cannot see the boundary modes of TR-symmetric systems. However, the low-energy terms that appear in resonant Raman processes are able to probe these modes.

A useful perspective on the difference between the TR- invariant and TR-breaking cases can be obtained by the mapping of the full 2D system onto a series of 1D Majorana chains. 
For a strip of the honeycomb lattice with $N$ unit cells along the $a_1$ direction, and $k_2$ the conserved crystal momentum along the strip, we view the Hamiltonian matrix $H_{k_2}$ as a one-parameter family of 1D Majorana chains. For $\frac{2\pi}{3}<k_2 <\frac{4\pi}{3}$ the fact that there are symmetry-protected zero-energy edge states of the full 2D system follows from the existence of a zero-energy boundary mode in each 1D Hamiltonian. 
These boundary modes are sublattice polarized -- which is unsurprising since the zero-energy flat bands are protected by sublattice symmetry. In Appendix \ref{app_end}, we demonstrate the origin of this sublattice polarization by explicitly solving for the boundary modes of the finite chain. This explicit solution shows that for a given $k_2$, finite size effects dictate that the true eigenstates of the Hamiltonian are superpositions of states that have zero-energy Majorana boundary modes at each of the chain's endpoints. In other words, for a given $k_2$, there is an equal probability for the corresponding boundary spinon to live on either edge of the system. For the TR-broken case, however, the boundary eigenstates (which are no longer required to have exactly zero energy) are localized purely on one end of the system for $k_2>0$, and on the opposite end for $k_2<0$, consistent with our expectations for chiral edge states. 

\begin{figure*}
	\centering
	\includegraphics[width=\linewidth]{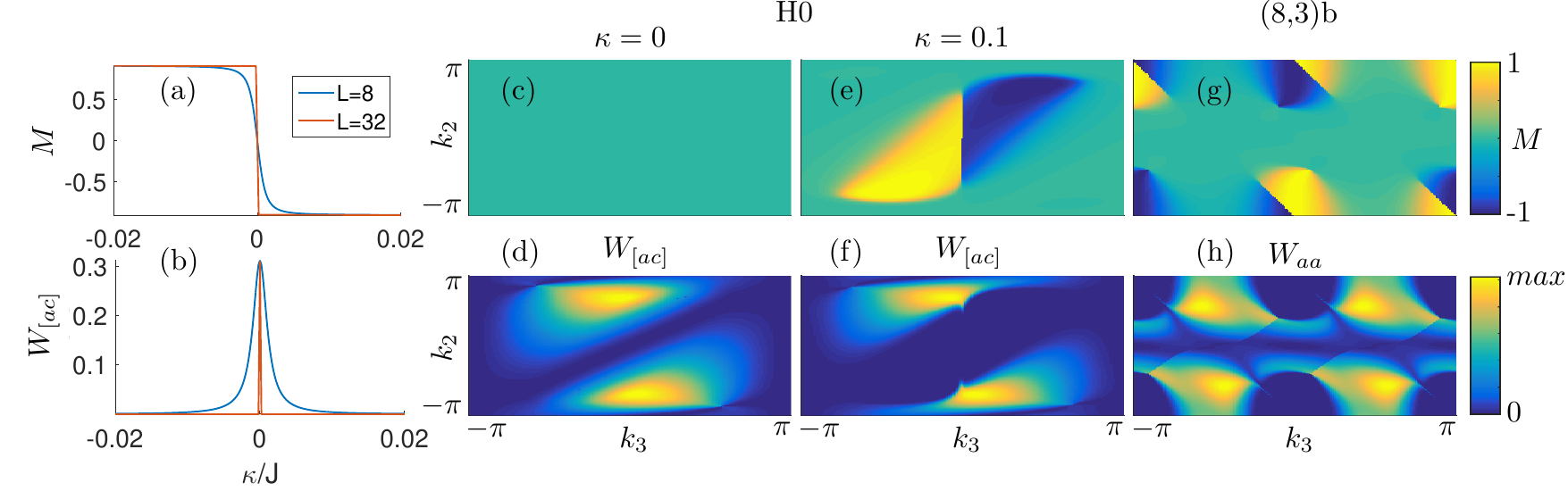}
	\caption{(a) The surface polarization and (b) the distribution of the Raman weight $W_{[ac]}$ in the lowest-energy band as a function of $\kappa/J$ at an arbitrary point on the flat band of the hyperhoneycomb lattice. (c) and (e) are colormaps of the surface polarization, and (d) and (f) surface-band Raman weights for the hyperhoneycomb lattice at $\kappa = 0$ and $\kappa = 0.1$, respectively. (g) The surface polarization and (h) Raman weight $W_{aa}$ of the surface band for the unperturbed (8,3)b system. All of the other Raman channels (not shown) similarly vanish in the surface-polarized regions away from the Fermi-arcs.}
	\label{fig:surf_pol_plots2}
\end{figure*}

Using this explicit solution we can also investigate how, at fixed $k_2$, the edge state becomes polarized to one boundary or the other as we turn on the TR-breaking perturbation. 
In Appendix \ref{app_end}, we show that the nature of the boundary eigenstates is determined by a competition between the energy scale of the TR-breaking perturbation $\epsilon_k = |F_k| \sim \kappa$, and the finite size splitting $E_{FS}$. If $E_{FS} > \epsilon_k$, the eigenstate at a given $k_2$ has an amplitude on both boundaries, and the Raman operator still reflects the boundary density of states at low energies. However if $E_{FS} < \epsilon_k$, the boundary mode at a given $k_2$ lives entirely on one of the two boundaries, and the local Raman operator cannot create a pair of boundary excitations. 

The situation on the hyperhoneycomb lattice is very similar to the honeycomb case just described. Again the symmetry-protected surface flat bands are necessarily sublattice-polarized, and hence can couple to (resonant) Raman processes. However, breaking time-reversal symmetry polarizes the boundary mode for each surface momentum such that it lives almost entirely either on the upper or the lower surface, leading to a strong suppression of the resulting Raman weights. This is illustrated in Fig.~\ref{fig:surf_pol_plots2}, which shows the extent of this surface polarization, together with the corresponding Raman weights, on the hyperhoneycomb lattice both with and without TR symmetry, as well as on the (8,3)b lattice. 
In all cases, the figure shows slab configurations with $N=16$ unit cells in the $a_1$ direction and open boundary conditions. 
The signed surface polarization is determined by computing
$
M = \langle \psi_{\epsilon} | {\bf M} | \psi_{\epsilon} \rangle
$,
where $\psi_{\epsilon}$ are the surface states at energy $\epsilon\rightarrow 0$ and ${\bf M}$ is a diagonal operator in the Majorana spinon basis ${\bf M} = \text{diag}(1,...,1,0,...,0,-1,...,-1)$, where the number of nonzero elements corresponds to two unit cells on each end. (The number of sites per unit cell is 2, 4, and 6 for the honeycomb, hyperhoneycomb, and (8,3)b lattices respectively) Then $M = 1 (-1)$ means that the boundary mode is polarized to the upper (lower) edge or surface; eigenstates with equal amplitudes on both surfaces have $M=0$. 

The associated Raman weight is the matrix element 
\begin{align}
W_{[ac]}({\bf k}) = \tensor*[_0]{\braket{R_{[ac]} | a^\dagger_{{\bf k}} a^\dagger_{-{\bf k}}} }{_0}  \tensor*[_0]{\braket{a^\dagger_{{\bf k}} a^\dagger_{-{\bf k}} | R_{[ac]}^\dagger } }{_0} 
\end{align}
where $R_{[ac]} = (R_{ac} - R_{ca})/2$, and $a^\dagger_{{\bf k}}$ creates a surface- polarized excitation with crystal momentum $k$.
Importantly, because the Raman operator acts locally in space, it creates or destroys a pair of excitations localized to the same surface.

Figs.~\ref{fig:surf_pol_plots2} (a) and (b) show the appearance of surface polarization and the corresponding vanishing of the Raman weight $W_{[ac]}$ as the perturbation $\kappa$ is turned on for a particular $k$-point on the surface BZ of the hyperhoneycomb lattice. As expected, for larger systems the vanishing is more immediate due to the exponentially smaller finite size splitting (see Appendix \ref{app_end}). Fig.~\ref{fig:surf_pol_plots2}(c)-(f) illustrate how this occurs in the Brillouin zone. For $\kappa=0$, eigenstates in the flat band on the hyperhoneycomb lattice have equal amplitude on each boundary, so that the signed surface polarization vanishes (see \ref{fig:surf_pol_plots2} (c)). As $\kappa$ is increased, the states originally in the flat band quickly become polarized to a single surface for each surface ${\bf k}$ value, as seen in (e). Correspondingly Raman weight is pushed out of the flat band region, remaining only near the surface projection of the Weyl points. The surface polarization is opposite on either side of the Fermi-arc, and also odd under $k\to -k$ as required by inversion symmetry. For the value of $\kappa$ shown, the Raman weight is significant only very close to the Fermi arc (where the energy scale due to the perturbation $\kappa$ vanishes, and finite-size effects dominate) and on the portion of the Fermi-arc near the surface projection of the Weyl points. This is because the finite size effects are strongest near the projected Weyl nodes, falling off as a power law rather than exponentially in the slab thickness. On the honeycomb lattice, where the bulk is fully gapped, the total Raman weight of the boundary modes vanishes much more quickly with $\kappa$, as observed above. 

On the (8,3)b lattice the sublattice symmetry is always broken, and there is never a flat surface band. Instead the modes near the Fermi-arcs are always surface-polarized, except for extremely thin slabs. In Figs.~\ref{fig:surf_pol_plots2} (f) and (g) we see small pockets in the surface BZ around the Fermi-arcs, at which the surface bands are polarized to a single surface of the system in a way that switches when we cross the Fermi-arc and that respects inversion symmetry. All of the Raman channels, of which one representative $W_{aa}$ is shown, vanish in these pockets except possibly in a small region very close to the Fermi-arc, whose size depends on the size of the system. 

In summary, our analysis reveals that Raman scattering is an effective probe of {\it non-chiral} topological boundary modes, for which zero-momentum transfer processes can excite a pair of Majorana spinons at the same edge or surface. For chiral boundary modes, however, the coupling between the Raman operator and these surface pairs is very strongly suppressed, since a spinon on the top boundary with momentum $k_2$ generally has a partner spinon of momentum $-k_2$ that is localized to the bottom surface, and vice versa. This suppression is controlled by the ratio of the TR-breaking energy scale to the scale of finite-sized splitting. 

\subsection{Bulk Raman spectra at zero magnetic field} \label{BulkSec}

In addition to its potential to detect surface flat band states, the Raman response is a useful probe of the bulk spinon density of states, which can also be suggestive of spin liquid physics. In our previous works,\cite{Perreault15,Knolle14-2} we have analyzed the zero-field case of sublattice-symmetric Kitaev models, identifying spin-liquid signatures in the polarization dependence of the bulk Raman spectra. Here we will review these results and describe how the anti-symmetric Raman channels add new measurable quantities to the Raman response in these previously-studied lattices. We also extend our analysis to the (8,3)b lattice, which does not exhibit sublattice symmetry.

\begin{figure*} 
	\centering
	\includegraphics[width=\linewidth]{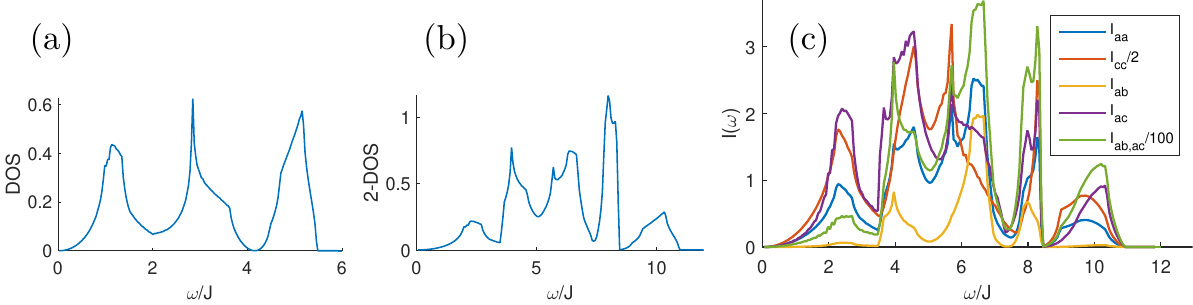}
	\caption{(Color Online) (a) DOS and (b) 2-DOS (see Eq.~(\ref{2DOS})) for the (8,3)b lattice with no external applied magnetic field. (c) All five of the independent non-vanishing Raman spectra reflect the qualitative features of the 2-DOS shown in (b).}
	\label{fig:ETb}
\end{figure*}

One key way in which the bulk Raman signal can provide information about the spin liquid state is through the number of independent polarization channels, which reflect both the lattice symmetries preserved by the spin-liquid state, and certain features of the Kitaev exchange interaction. Refs. \onlinecite{Knolle14-2,Perreault15} discuss this in detail for non-resonant Raman scattering on the honeycomb and hyperhoneycomb lattice, respectively; the key results derived there are reviewed in Appendix \ref{SIS}. Notably, Ref. \onlinecite{Knolle14-2} showed that on the honeycomb lattice with $J^x=J^y=J^z$ the non-resonant Raman response is independent of polarization. This is due to a combination of symmetry (see Appendix \ref{SIS}) and 
the fact that $R_{xx} + R_{xy} \sim H$, implying that $R_{xx} \sim -R_{xy}$ since the Hamiltonian $H$ does not create spinon excitations. This relationship has its origin in the strong similarity of the Hamiltonian and the Loudon-Fleury vertex which dominates the symmetric off-resonant Raman response. We therefore termed this a Loudon-Fleury (LF) relationship.\cite{Perreault15} Resonant Raman scattering allows for one additional independent Raman operator $R_{[x,y]}$, for a total of two independent Raman spectra $I_{xx}, I_{[x,y]}$.

For the hyperhoneycomb lattice, Ref. \onlinecite{Perreault15} showed that a combination of symmetries and Loudon-Fleury relationships leaves four independent non-resonant symmetric Raman spectra: $\{I_{aa},I_{cc},I_{ab},I_{ac} \}$. In this case an additional accidental equality at the operator level, which further gives $I_{cc} = 9I_{aa}$ leaving the three independent non-vanishing spectra $\{ I_{aa},I_{ab},I_{ac} \}$. (This is true for any Kitaev couplings $J^x= J^y \neq J^z$ which respect the underlying symmetry of the lattice). The antisymmetric resonant Raman channels add the three independent symmetry-allowed Raman spectra $\{I_{[ab]},I_{[ac]},I_{[ac],ac}\}$. At the operator level it turns out, again accidentally, that $R_{[ab]} = 0$ so that $I_{[ab]} = 0$, leaving an additional two independent non-vanishing Raman spectra. The details of the corresponding symmetry analysis are given in Appendix \ref{SIS}.

The (8,3)b lattice has point group symmetry $D_{3d}$.\cite{Obrien} For non-resonant Raman processes, this point group allows for the six independent non-vanishing Raman spectra $\{I_{aa},I_{cc},I_{aa,cc},I_{ab},I_{ac},I_{ab,ac} \}$. There is also one Loudon-Fleury relationship, giving $-2I_{aa,ac} = I_{cc}$. The resulting five independent non-zero polarization-symmetric spectra are plotted in Fig.~\ref{fig:ETb}. Similar to Ref. \onlinecite{Perreault15} we find the Raman intensity follows a momentum-locked two-particle DOS (2-DOS) defined as
\begin{align} \label{2DOS}
\rho_2(\omega) = \sum_{n,m;{\bf k}} \delta(\omega - \varepsilon_{m,{\bf k}} - \varepsilon_{n,{\bf k}}),
\end{align} 
where $\varepsilon_{m,{\bf k}}$ is the energy of the excitation at ${\bf k}$ (or $-{\bf k}$) in the $m$th band (Fig. \ref{fig:ETb}(b)). The three bands on this lattice lead to six two-particle peaks, which are clearly visible both in the 2-DOS and the Raman spectra. Interestingly, unlike the other lattices we have considered this one does not have states all the way to the energy $6J$, which is the maximum that follows from the tri-coordination. Resonant Raman scattering also introduces three anti-symmetric spectra $I_{[ab]}$, $I_{[ac]}$, and $I_{[ac],ac}$, (not shown here) whose features are qualitatively similar. 

We present the full frequency response of the honeycomb and hyperhoneycomb models in Appendix \ref{finitefrecSec}, where we also discuss the validity of the perturbed model at finite frequency.
				

\section{Discussion} \label{DiscussionSec}

In this paper, we studied the Raman scattering response in general Jackeli-Khaliullin-Kitaev systems, in which the Kitaev QSL phases could potentially arise. %
A systematic calculation of the Raman intensity in different systems, both for bulk and a slab geometry, showed that the Raman scattering response provides clear signatures of exotic 2D and 3D Kitaev QSL phases. 

One important result is the difference in power laws governing the low-frequency Raman response for bulk systems with Fermi lines and Fermi points in the Majorana spinon description. A second achievement is the characterization of the ability of Raman processes to couple to the boundary modes in these systems. In particular, we found that though the edge or surface modes in the symmetry class BDI do not couple to the usual non-resonant Loudon-Fleury vertex, 
they do couple to resonant Raman processes, and are therefore observable for sufficiently thin films. For the symmetry class D, where the gapless boundary modes are chiral, with modes at ${\bf k}$ and $-{\bf k}$ localized on opposite surfaces, momentum unresolved probes such as Raman scattering do not couple effectively to boundary states except in extremely thin films where the finite-size splitting dominates. The third accomplishment is a symmetry-group-based analysis of Raman scattering's polarization dependence allowing us to understand how the symmetry of the QSL state is reflected in the polarization dependence of the Raman response. 



Our most important results focus on the low-energy response of QSLs. However, since our analysis neglects terms that create excitations above the flux gap, the calculated resonant Raman spectra are valid only at energy scales that are low relative to the local flux gap. The local flux gap in the Kitaev QSL phase is small -- approximately 0.26 J,\cite{Kitaev} 0.13 J,\cite{Obrien,Kimchi14,Smith} and 0.16 J\cite{Obrien} for the honeycomb, hyperhoneycomb, and (8,3)b lattices, respectively. For the known JKK systems, this makes the low-energy response most appropriate for a Brillouin scattering setup.\cite{Perreault16} 

Targeting the low-energy response with resonant Raman scattering has several advantages over non-resonant techniques. First, the resonant processes can couple to the boundary flat-band states (sublattice polarized modes), while the non-resonant processes do not couple to them. Second, extracting the low-frequency Raman response of the spin liquid typically requires that one can accurately separate the contribution of acoustic phonons, which are expected to arise at similar energies to the two-spinon bands plotted here,\cite{Sandilands} from that of the \mssns. 
However, phonons are not expected to couple to the anti-symmetric channels 
as easily as the electronic excitations do. Specifically, such a coupling can occur only if the phonons have access to a resonant process involving another type of excitation,\cite{Rousseau,Klein} which in the Mott-resonant regime is an electron hop. Therefore as long as the electron-phonon coupling is not large, these processes will be suppressed with respect to the direct interaction between the photon and the electron in these channels, and the \mss should dominate the low-frequency Raman response, in particular in the anti-symmetric channels. 


Ultimately, 
the primary experimental challenge is to identify qualitative signatures of the Kitaev QSL phase. Encouragingly, some promising preliminary steps in this direction have been taken. For example, Raman~\cite{Sandilands} and inelastic neutron scattering experiments~\cite{Banerjee16} in $\alpha$-RuCl$_3$ have been interpreted in terms of weakly-confined fractionalized excitations by close comparison with controlled calculations of the corresponding response functions in Kitaev models.~\cite{Knolle14-1,Knolle14-2} For example, it was recently shown for the 2D honeycomb model that the temperature dependence of Raman scattering encodes the fermionic statistics of the fractionalized Majorana fermions, evidence of which is already visible in experiments on $\alpha$-RuCl$_3$ at temperatures much above the residual long-range magnetic order.~\cite{Nasu2016} In line with this, our work provides clear signatures of more exotic 2D and 3D Kitaev QSL phases which will be hopefully relevant for their experimental detection in the future. Moreover, we have established resonant Raman scattering on thin films as a probe of fractionalized boundary modes in general.

\section*{Acknowledgements}
We acknowledge helpful discussions with I. Rousochatzakis, D.L. Kovrizhin, R. Moessner, J. Rau, K. O'Brien, A. Smith, A. Edelman and Y. Sizyuk. BP acknowledges the support of the Torske Klubben Fellowship. The work of J.K. is supported by a Fellowship within the Postdoc-Program of the German Academic Exchange Service (DAAD). NP acknowledges the support from NSF DMR-1511768. FJB is supported by NSF DMR-1352271 and Sloan FG-2015-65927.

\appendix


\section{End modes of finite Majorana chains}\label{app_end}

As we found in section IV, the surface polarization (top/bottom) of the strips and slabs determines whether a local Raman operator can couple to it. In particular, in systems with inversion symmetry the surface modes at $k$ and $-k$ must be on opposite boundaries of the system, as only in this case can the two surface excitations be simultaneously probed by a Raman operator. 

In this Appendix, we study the restriction of the Kitaev Hamiltonian to 1D, as occurs when we fix a point in the boundary BZ. The resulting Hamiltonian describes a gapped 1D Majorana chain \cite{Kitaev01} in the same symmetry class as its parent lattice. We focus on the case applicable to the honeycomb and hyperhoneycomb lattices where the system is in class BDI but a perturbation $\kappa$ takes it into class D. 

	
\begin{figure}
	\centering
	\includegraphics[width=0.9\linewidth]{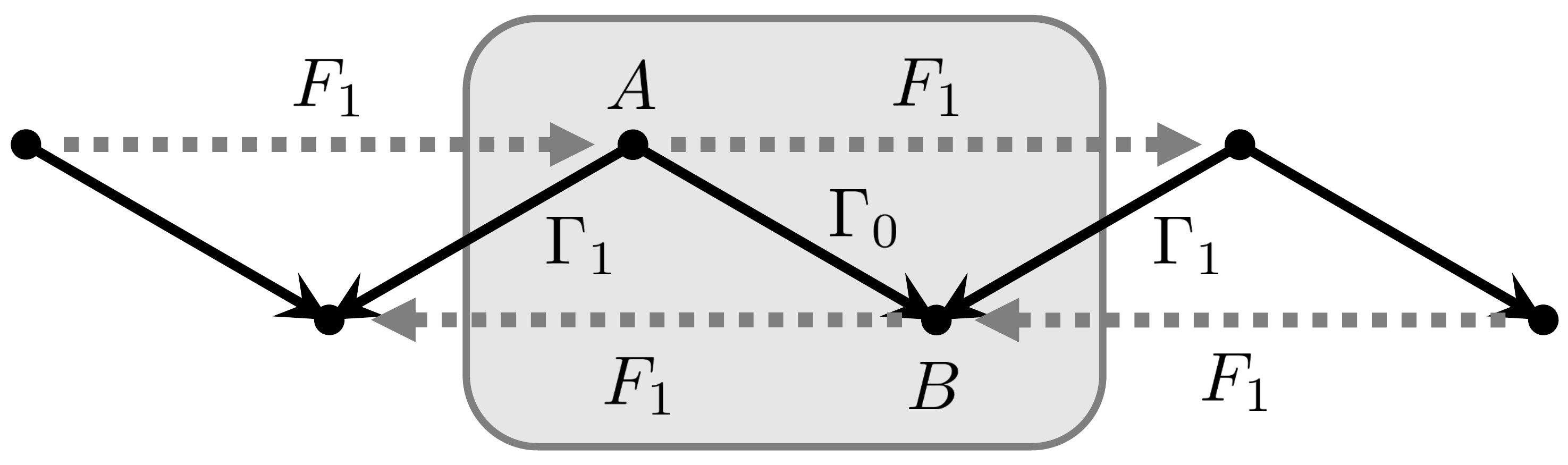}
	\caption{One unit cell of the 1D effective Hamiltonian on $k_1$ in real space.}
	\label{fig:1D_fig}
\end{figure}

Here we follow Ref. \onlinecite{Niu12} to consider recursion relations for zero-energy modes in finite chains. For concreteness, we consider the Hamiltonian obtained by fixing the $k_2$ wave vector in the Hamiltonian for the honeycomb lattice, Eq. (\ref{HCH}), and treating the result as a 1D Hamiltonian on crystal momentum $k_1$. In real space, this Hamiltonian can be visualized as shown in Fig. \ref{fig:1D_fig}, where $A$ and $B$ are the sublattice indices within a single unit cell. 

In momentum space, the Hamiltonian can be written in matrix form as
\begin{align}\label{1D}
\mathcal{H}_{1D} &= i\left( \begin{array}{cc}
F_1 e^{ik_1} - c.c. & \Gamma_0 + \Gamma_1 e^{ik_1} \\
-(\Gamma_0^* + \Gamma_1^* e^{-ik_1}) & -(F_1 e^{ik_1} - c.c.)  \end{array} \right),
\end{align}
where $\Gamma_0 = J^z + J^y e^{ik_2}$ and $\Gamma_1 = J^x$ are, respectively, the effective NN hopping and the hopping between 1D effective unit cells. The diagonal terms, describing the hopping between the sites of the same sublattice, come from $F_1 = -\kappa^z e^{-ik_2} + \kappa^y$. This is precisely the Hamiltonian of the bulk system. However, for our purposes the dependence on $k_2$ is only relevant when we want to characterize the Hamiltonian in the boundary BZ of a higher-dimensional system and we therefore drop $k_2$ for simpler notation, and take $\Gamma_0$, $\Gamma_1$, and $F_1$ as \emph{complex} hopping parameters for the \mssns. 

	
	
We first consider the effective finite chain pictured in Fig. \ref{fig:1D_fig} when TR-symmetry is unbroken ($F=0$). The Hamiltonian takes the form
\begin{align}
H_{1D} = i \Gamma_1 \left(
\begin{array}{cccccc}
0 &     \lambda &  &  &  & \\
-\lambda^* & 0 & 1 &  &   & \\
& -1 & 0 & \lambda &  & \\
& & \ddots & \ddots & \ddots & \\
& & &-\lambda^* & 0 & 1 \\
& & & & -1 & 0 \\
& & & & & -\lambda^*
\end{array}	 \right) ,
\end{align}
where $\lambda = \Gamma_0 / \Gamma_1$ and we have used that $\Gamma_1 \in \mathbb{R}$. For an eigenvector to be at zero energy, its action at every row must be zero. For an ansatz $A$, the equation $H_{1D} A= 0$ will give a recursion relation for its components.\cite{Niu12} Due to the sublattice symmetry, the recursion relation only relates terms in the eigenvectors associated with the same sublattice. We therefore take the ansatz $A_L^T = (A_1,0,A_2,0,\cdots,A_N,0)$ giving the recursion relation \cite{Niu12}
\begin{align}
-\lambda^* A_j + A_{j+1} = 0.
\end{align}
This is solved in the bulk by
\begin{align}
A_{j} &= (\lambda^*)^{j-1} A_1 = \frac{A_1}{\lambda^*} \exp\left[-j \log(1/\lambda^*)\right].
\end{align}
This solution is exponentially localized to one end or the other for $|\lambda| < 1 $ and $|\lambda| > 1$ respectively. Of course, there is another eigenvector $A_R$ that exists on the other sublattice. This one is related to the first one by inversion symmetry and is therefore localized on the opposite end.
\begin{align}
A_R = \mathcal{I} A_L = \left(\begin{array}{cccc}
&  & 1  \\ 
& 1 & \\
\iddots & &
\end{array}\right)A_L^*.
\end{align}
		
%

However, we have to ignored terms in the Hamiltonian at the boundary. More precisely, the Hamiltonian $H_{\text{bulk}}$, for which these are eigenvectors at zero energy, is the one without the first and last rows. 	
We, therefore, treat these additional terms as a perturbation, $V = H - H_{\text{bulk}}$, which is responsible for the finite-size effects:
\begin{align}\label{pertV}
V = i \left(
\begin{array}{cccc}
0 &      \Gamma_0 & \\
\ddots & \ddots & \ddots \\
& \ddots & \ddots & \\
& & -\Gamma_0^*  
\end{array}	 \right) ,
\end{align}
where all but the two elements are zero. For $A_R$ and $A_L$ to become approximate eigenvectors in the infinite limit, the boundary terms must act on an exponentially suppressed part of the eigenvectors. This occurs only if $|\lambda|<1$, corresponding to the non-trivial topological phase. In this case the eigenvectors are normalized if 
\begin{align}
	A_1 = \lambda^* \frac{(\lambda^*)^{N+1} - 1}{\lambda^* - 1}.
\end{align} 
Since the perturbation Eq.~(\ref{pertV}) is a part of the Hamiltonian, it is no surprise that its action exchanges the sublattices. 
So, although $\braket{A_L|V|A_L} = \braket{A_R|V|A_R}=0$, the off-diagonal matrix element 
\begin{align}
E_{\rm end}=\braket{A_L|V|A_R} = \left( i \Gamma_0 A_1 A_N + h.c. \right) \sim |\lambda|^N.
\end{align}
The finite-size splitting exists only to the extent that the two Majorana modes can interact and is thus exponentially suppressed. Moreover, since $V$ exchanges the two states, within this low-energy subspace the eigenstates are the even and odd combinations of $A_R$ and $A_L$. 

To understand how the lowering of the symmetry affects the endmodes, we now introduce the second-neighbor hopping perturbation, which can be written as 
\begin{align}
F = i F_1 \left(
\begin{array}{ccccccc}
0 & 0 & 1 &   &   &  \\
0 & 0 & 0 & -1&   & \\
1 & 0 & 0 & 0 & 1 & \\
0& -1 & 0 & 0 & 0 & -1 &  \\
& & \ddots & \ddots & \ddots & \ddots\\
& & & -1 & 0 & 0 
\end{array}	 \right) ,
\end{align}
where we have used that $F_1$ is pure imaginary, so that $-F^* = F$. Unlike the boundary terms, this perturbation splits the degeneracy between the states $A_L$ and $A_R$ but does not mix them within the low-energy eigenspace. Thus, $\braket{A_L|F|A_R} = 0$ and
\begin{align}
E_F &= \braket{A_L|F|A_L} = - \braket{A_R|F|A_R} \nonumber \\
&=\sum_{j=2}^{N-2} A_j^* (A_{j-2} + A_{j+2}) \sim |F_1| \sim \kappa.
\end{align}
Therefore, we generally expect the choice of low-energy basis to depend on a competition between these energy scales $E_F \sim |F_1| \sim \kappa$ and $E_{\rm end} \sim \exp\left(-N / |\lambda|\right)$.
As noted in the main text, this competition explains the vanishing of the Raman operator in the presence of the perturbation $\kappa$ in the parts of the Brillouin zone 
that otherwise hosted zero-energy modes. In addition, we see that near the transition, $\lambda \to 1$, the finite-size effects become important on much longer scales as the localization length grows. At the projection of gapless points in the surface BZ of the hyperhoneycomb lattice $\lambda$ goes to one, creating a delocalized state that is needed for the change of the topological index and hence the number of polarized surface modes. The increase in localization length near these gapless points explains why, for small perturbations and a finite system size, these points near the gapless states stay depolarized while in the rest of the former flat band the weight is disappearing.
		

\section{DOS power laws}\label{DOSPL}

As discussed in the main text, the low-energy density of states reflects the dimension of the Fermi-surface through its low-energy power law. Here we review how to obtain low-energy power laws for the DOS and heat capacity from the general arguments of Fermi liquid theory.\cite{ShankarRG}

We consider a system with $d$ space dimensions at zero temperature. Then $\epsilon_{\bf k} = 0$ is satisfied on some set of points whose dimension we call $d_f$, the Fermi-surface (FS) dimension. The co-dimension of the FS is $d_c = d - d_f$. We assume that as we go away from the FS in ${\bf k}$-space the dispersion of excitations obeys some power-law $\epsilon_{\bf k} \sim |\mathbf{q}|^p$, where $\mathbf{q}$ parametrizes the components of ${\bf k}$ that are perpendicular to the FS, $\mathbf{q}={\bf k}_{\perp {\rm FS}}$. 

First consider the DOS:
\begin{align}
\rho(\omega) &= \int_{BZ} \delta(\omega-\epsilon_k) d^dk \nonumber\\
&= \int_{\{k:\epsilon_k = 0\}} \frac{1}{|\nabla_k \epsilon_k|} d^dk \nonumber\\
&\sim \frac{1}{|\nabla_k \epsilon_k|} A_{FS} \hspace{.1 cm} |\mathbf{q}|^{d_c-1} \label{third_line} \\
&\approx \frac{1}{|\mathbf{q}|^{p-1}} A_{FS} \hspace{.1 cm} |\mathbf{q}|^{d_c-1} \nonumber\\
&= A_{FS} |\mathbf{q}|^{d_c - p} \nonumber\\
& \propto \omega^{(d_c - p)/p},
\end{align}
where $A_{FS}$ is the area of the FS (in the appropriate dimension $d_f$). The approximation made in (\ref{third_line}) is equivalent to saying that the number of $k$-points satisfying $\epsilon_{\bf k} = \epsilon_{{\bf k}'}$ for a fixed $k'$ scales with $|\mathbf{q}|^{d_c - 1}$, where $\mathbf{q}=({\mathbf k}-{\mathbf k}')_{\perp {\rm FS}}$. This can easily be checked in 2D and 3D for Fermi-points and Fermi-lines which realize $d_c = 1,2$, or $3$. 

The specific heat is $C_V=\frac{dU}{dT}$, where the total energy $U = \int_0^\infty d\omega \, \omega\, \rho(\omega)\, n_F(\omega)$ with $n_F(\omega)$ the Fermi-Dirac distribution function. Substituting the expression $\rho(\omega) \sim \omega^{(d_c - p)/p}$, ones finds
\begin{align}
C_V \propto T^{d_c/p},
\end{align}
up to a dimensionless integral. The unperturbed Kitaev models in class BDI then have the following scaling (since $p=1$)
\begin{align}
\rho(\omega) &\sim \left\{\begin{array}{lr}\alpha L^3 \omega + \beta L^2 /\omega & \hspace{7mm}\kappa = 0 \\
\alpha' L^3 \omega^2 + \beta' L^2 & \hspace{7mm}\kappa \ne 0
\end{array}\right. \\
C_V &\sim \left\{\begin{array}{lr}
a L^3 T^2 + b L^2 & \hspace{6mm} \kappa = 0 \\
a' L^3 T^3 + b' L^2 T & \hspace{6mm} \kappa \ne 0
\end{array}\right. ,
\end{align} 
where the second term represents the surface contribution. 
The case for the (8,3)b lattice (symmetry class D) is again the same as the perturbed case for these models. These DOS power laws are consistent with the observations in Sec. \ref{ResRaman2} of the main text.


\section{Symmetries and independent Raman spectra}\label{SIS}
Here we review the implications of symmetry for Raman scattering in the honeycomb, hyperhoneycomb, and (8,3)b lattices, giving the relevant details of the point-groups and elucidating the consequent relationships between different Raman channels. This provides both a review of the relevant results of Refs. \onlinecite{Knolle14-2,Perreault15}, and new results pertaining to the (8,3)b lattice. We also extend both analyses to include the anti-symmetric symmetry channels accessible through resonant Raman scattering.

\subsection{Honeycomb lattice}

The honeycomb lattice has point group $D_{3d}$ including three-fold rotations, two-fold reflections, and inversion. Raman scattering only couples to inversion-symmetric channels. $D_{3d}$ has three of these: $A_{1g}$, $A_{2g},$ and $E_g$. 
Within the 2D restriction of this group there are only two distinct quadratic operators. We use the $\sim$ symbol to indicate that two operators lead to the same spectra due to symmetry. Then the non-zero Raman operators are $R_{xx} \sim R_{yy}$ in the $A_{1g}$ channel and $R_{xy}$ in the $E_g$ channel, while the $A_{2g}$ channel does not support any symmetric quadratic operators. However, when we allow for anti-symmetric operators we find that the $A_{2g}$ channel has a contribution from the operator $R_{[xy]}$.
		
\begin{figure}
	\centering
	\includegraphics[width=.4\linewidth]{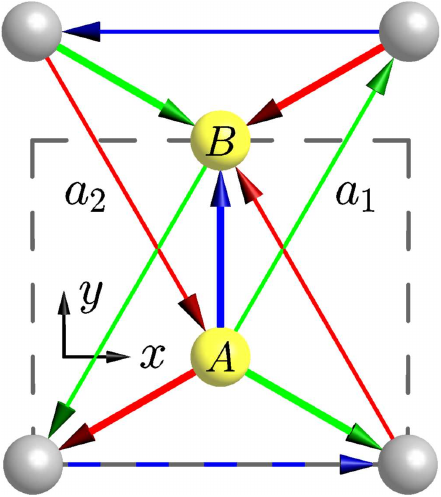}
	\caption{(Color Online) The primitive unit cell of the Honeycomb lattice with arrows to represent the sign of the hopping terms.}
	\label{fig:2Hops_HC}
\end{figure}
	
Next we consider the effect of second-neighbor spinon hopping, which comes from the magnetic field perturbation, and study affects the symmetry-group analysis. We illustrate the interaction with directed bonds in Fig. \ref{fig:2Hops_HC}, where the arrows indicate the directions for positive hopping.
One can then check that the three-spin perturbation breaks the reflection symmetries of the Hamiltonian that pass through sites (and hence pass through the $C_3$ rotation center, which is left intact). Therefore, for small magnetic fields the symmetry group is broken down to $S_6$. 
This removes the distinction between $A_{1g}$ and $A_{2g}$ allowing correlations between the $[xy]$ channel and the $xx \sim yy$ channels within the new $A_g$ channel ($I_{[xy],xx} \ne 0$). However, numerically we find $I_{[xy],xx} = 0$ for this particular model. 
	
	\subsection{Hyperhoneycomb lattice}	
	\begin{figure}
	\centering
	\includegraphics[width=0.68\linewidth]{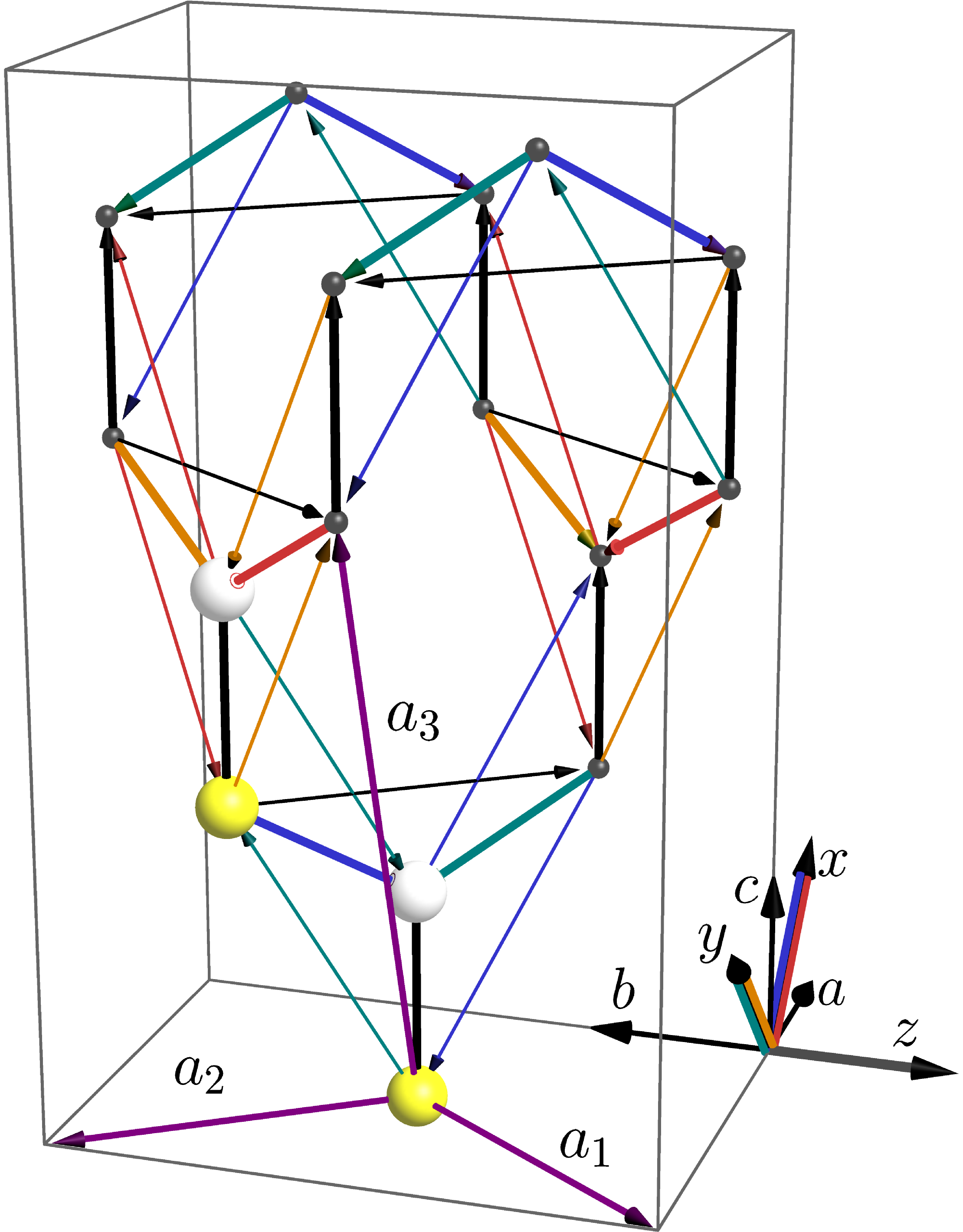}
	\caption{(Color Online) The hyperhoneycomb lattice with arrows to represent the sign of the hopping terms.}
	\label{fig:symm}
\end{figure}

\begin{table}\centering
	$\begin{array}{|r|c|c|c|}
	\hline
	& J & \kappa_z & \kappa_x \text{ or } \kappa_y \\ \hline
	C^a_2 & - & + & - \\ \hline
	C^b_2 & - & - & + \\ \hline
	C^c_2 & + & - & - \\ \hline
	T_{\frac{a_3 - a_1}{2}}\sigma^a & + & + & - \\ \hline
	T_{\frac{a_3}{2}}\sigma^b & + & - & + \\ \hline
	\left( T_{\frac{a_1}{2}} \text{ or } T_{\frac{a_2}{2}} \right) \sigma^c & - & - & - \\ \hline
	i & - & - & - \\ \hline
	T_{\frac{a_3}{2}}C^c_4 & + & - & +\\ \hline
	T_{\frac{a_3 - a_1}{2}}C^{-c}_{4} & + & - & + \\
	\hline
	\end{array}$
	\label{table}
	\caption{The action of the lattice point-group symmetries, and the effective screw axes, on the second-neighbor terms shown in Fig. \ref{fig:symm}. An even (+) (odd (-) ) entry denotes an unbroken (broken) symmetry. The relevant symmetries are \cite{Perreault15} (1) $C_2^\alpha$ symmetries in the orthogonal $\hat{\mathbf{a}}, \hat{\mathbf{b}},$ and $\hat{\mathbf{c}}$ directions about the center-points of $z$-bonds; (2) Inversion about centers of $x$ or $y$ bonds; (3) Effective screw axes composed of $C_4$ rotation, $(a_3-a_2)/2$ translation, {\it e.g.}, and the dilatation $a \to \sqrt{2}a$, $b \to b/\sqrt{2}$; and (4) Glide planes with their normals in the directions $\mathbf{a},\mathbf{b},$ and $\mathbf{c}$ with their reflections passing through the inversion centers.\cite{Perreault15}
	}
\end{table}
		
The hyperhoneycomb lattice has the point group $D_{2h}$. This group admits four inversion-symmetric Raman channels: $A_g, B_{1g}, B_{2g},$ and $B_{3g}$, which leads to nine independent non-zero spectra: $\{ I_{aa},I_{bb},I_{cc},I_{aa,bb},I_{aa,cc},I_{bb,cc},I_{ab},I_{ac},I_{bc} \}$. However, an effective $C_4$ screw axis along the $c$ direction \cite{Perreault15} effectively enlarges the point group symmetry to $D_{4h}$. 
We find then that $I_{bb} = 4I_{aa}$, $2I_{ac} = I_{bc}$, and $2I_{aa,cc} = I_{bb,cc}$. Thus a representative set of spectra is given by the following six representations: $\{ I_{aa},I_{cc},I_{aa,bb},I_{aa,cc},I_{ab},I_{ac} \}$. 
	
LF relationships further reduce the number of independent spectra. $R_{aa} + R_{bb} + R_{cc} = H$, which is guaranteed by the form of the lattice and the form of the symmetric Raman operator, leads to $-R_{cc} \sim R_{aa} + R_{bb}$. This identifies $I_{cc} = -3I_{aa,cc} = 5I_{aa} + 2I_{aa,bb}$, leaving only four independent spectra: $\{ I_{aa},I_{cc},I_{ab},I_{ac} \}$. It turns out $R_{aa} = 2R_{bb}$ as operators, which further gives $I_{cc} = 9I_{aa}$. Finally, there are three independent non-zero spectra: $\{ I_{aa},I_{ab},I_{ac} \}$. 
	
Anti-symmetric resonant Raman operators add three operators to $D_{2h}$, $\{R_{[ab]},R_{[ac]},R_{[bc]}\}$, which correspond, respectively, to $B_{1g},B_{2g},B_{3g}$ irreducible representations. However, the higher effective symmetry of $D_{4h}$ implies that $[ab]$ is relegated to its own channel ($A_{2g}$) and therefore cannot mix with any other channels. It also leads to $R_{[bc]} = 2 R_{[ac]}$. At the operator level it turns out that $R_{[ab]} = 0$ so that $I_{[ab]} = 0$. This leads to the three additional symmetry-allowed independent spectra $\{I_{[ab]},I_{[ac]},I_{[ac],ac}\}$. 	
	
The second-neighbor hopping terms for the hyperhoneycomb lattice are illustrated in Fig. \ref{fig:symm}. We find that the low-energy perturbation preserves both inversion and the glide planes along the $c$-axis, while breaking the other symmetries. To verify this, it suffices to compare the sign changes obtained by each term in the full Majorana spinon Hamiltonian collected in Table \ref{table}. This leaves the point group $C_{2h}$ at small fields. For this group, all nine possible quadratic operators are distinct. The only symmetry constraints are to organize them into two channels $A_g$ and $B_g$ that do not mix with $4$ and $2$ symmetric spectra respectively and $1$ and $2$ antisymmetric spectra respectively. This gives $13$ or $25$ independent spectra depending on whether we include symmetric-only or also anti-symmetric operators. 
	
\begin{figure*}[htb] 
	{\includegraphics[width=\linewidth, trim = 0mm 0mm 0mm 0mm, clip]{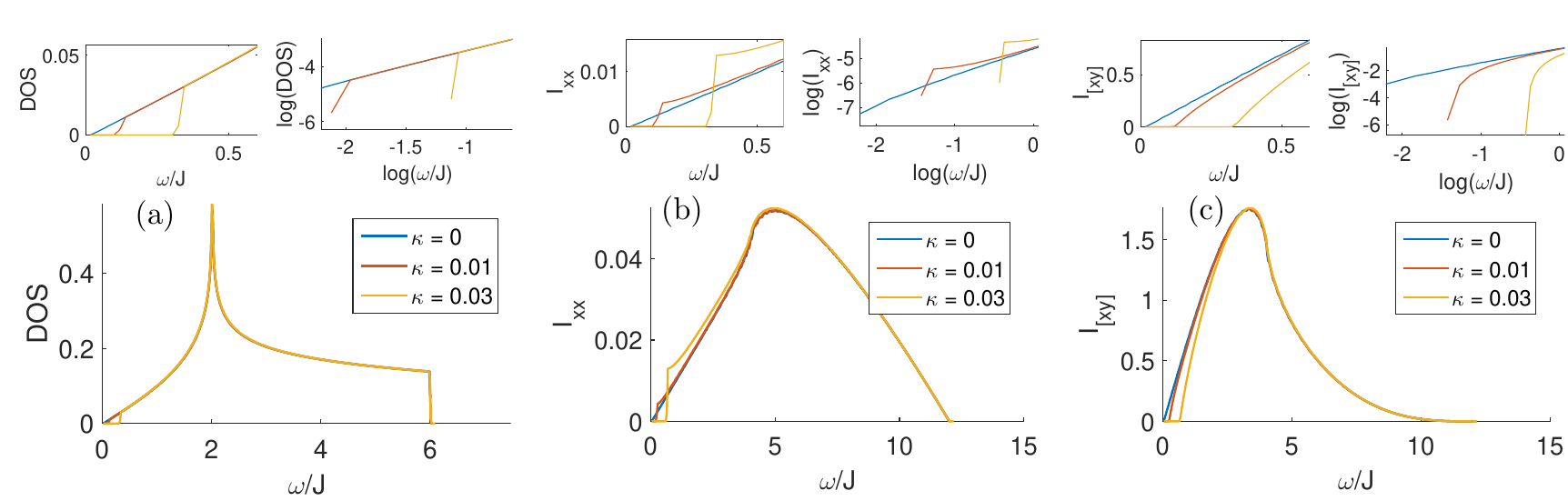}}	
	\caption{(Color Online) (a) The DOS (b) the Raman intensity in the $xx$ channel and (c) the antisymmetric Raman spectrum $[xy]$ each for three values of $\kappa$. The lower panels show Raman spectra over their entire frequency range; the upper panels show a close-up of the low-frequency behavior on both a linear and log scale, demonstrating the sharp gap.
	}
	\label{bulk_HC}
\end{figure*}
	
\subsection{ (8,3)b lattice}
	
The (8,3)b lattice has space group R3m.\cite{Obrien} The corresponding point group is $D_{3d}$. This is the same as the group obtained for honeycomb layers stacked by a unit vector normal to the plane, but in that case there is no contribution from the third direction. Realizations of the corresponding $E_g$ representation now come in two types: $R_{ab}$ and $R_{ac} \sim R_{bc}$. The $A_{1g}$ channel now has two independent representations, $R_{aa} \sim R_{bb}$ and $R_{cc}$. Keeping only the spectra that are allowed to be non-zero by symmetry, the six distinct spectra are represented by the set $\{I_{aa},I_{cc},I_{aa,cc},I_{ab},I_{ac},I_{ab,ac} \}$. \cite{Perreault15} As for the hyperhoneycomb lattice, there is also one Loudon-Fleury (LF) relationship between these spectra, giving $-2I_{aa,ac} = I_{cc}$. Anti-symmetric Raman processes appear in both the $A_{2g}$ and $E_g$ channels and are represented by $R_{[ab]}$ and $R_{[ac]} \sim R_{[bc]}$. This leads to the three distinct anti-symmetric Raman spectra $I_{[ab]}$, $I_{[ac]}$, and $I_{[ac],ac}$. 
		
\section{Finite frequency response in a magnetic field}\label{finitefrecSec}

\begin{figure}
	{\includegraphics[width=\linewidth, trim = 0mm 0mm 0mm 0mm, clip]{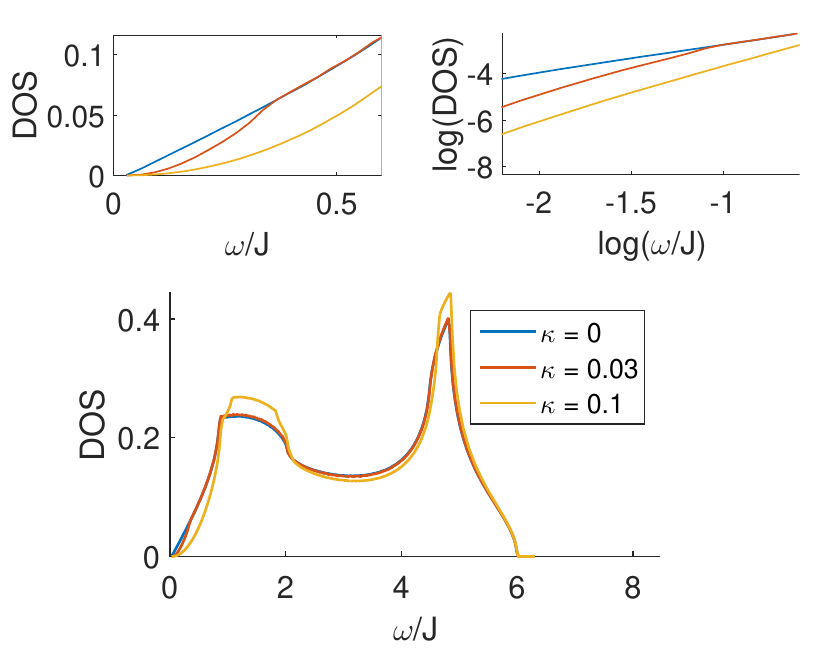}}
	\caption{(Color Online) The DOS of the hyperhoneycomb lattice for the \mss as a function of $\kappa$. Inset are a low-energy zoom (left) and a log-log plot to illustrate the power laws (right), which are $1.2,2.1,2.4\pm0.1$ at low energies.}
	\label{DOS_HC}
\end{figure}

\begin{figure}
	\centering 
	\includegraphics[width=.85\linewidth]{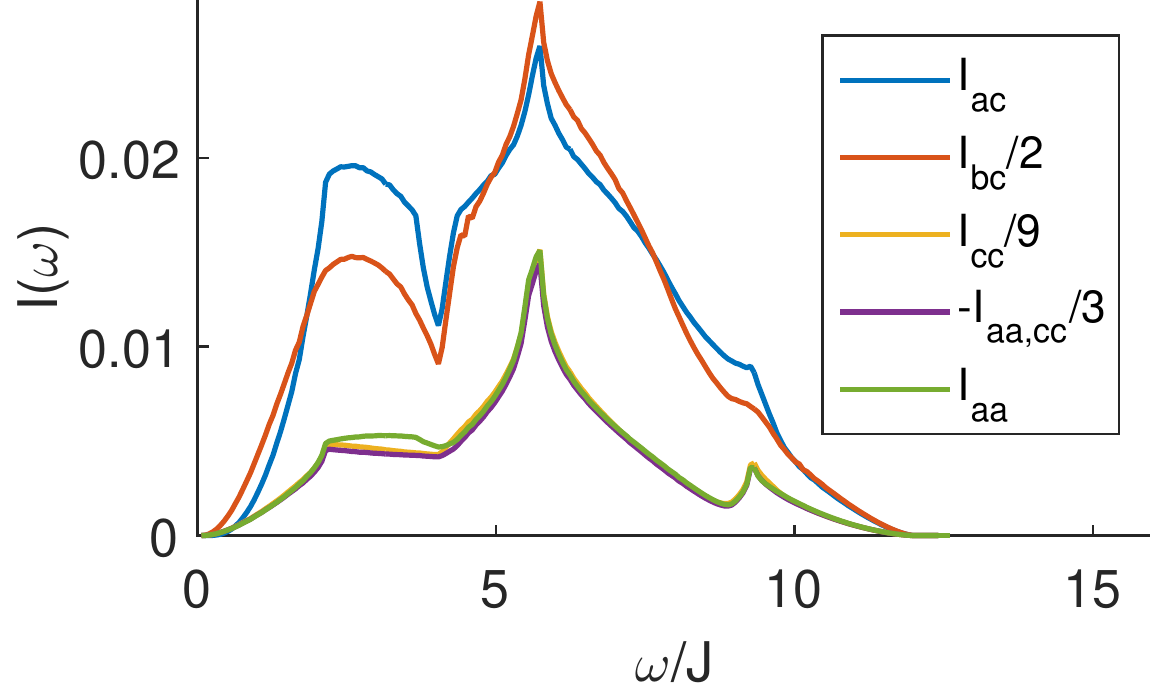}
	\caption{(Color Online) Here we have plotted an example of the breaking of the symmetry relationship $I_{aa} = -I_{aa,cc}/3 = I_{cc}/9$ and the LF-relationship $I_{ac} = I_{bc}/2$ in the presence of the magnetic field perturbation of $\kappa = 0.15$ (chosen for effect). The symmetry requirement $I_{ac,cc}=0$ holds because it is odd under the glide-plane symmetry that remains.}
	\label{fig:exampl}
\end{figure}

\begin{figure*}[htb] 
	\centering 
	\includegraphics[width=\linewidth]{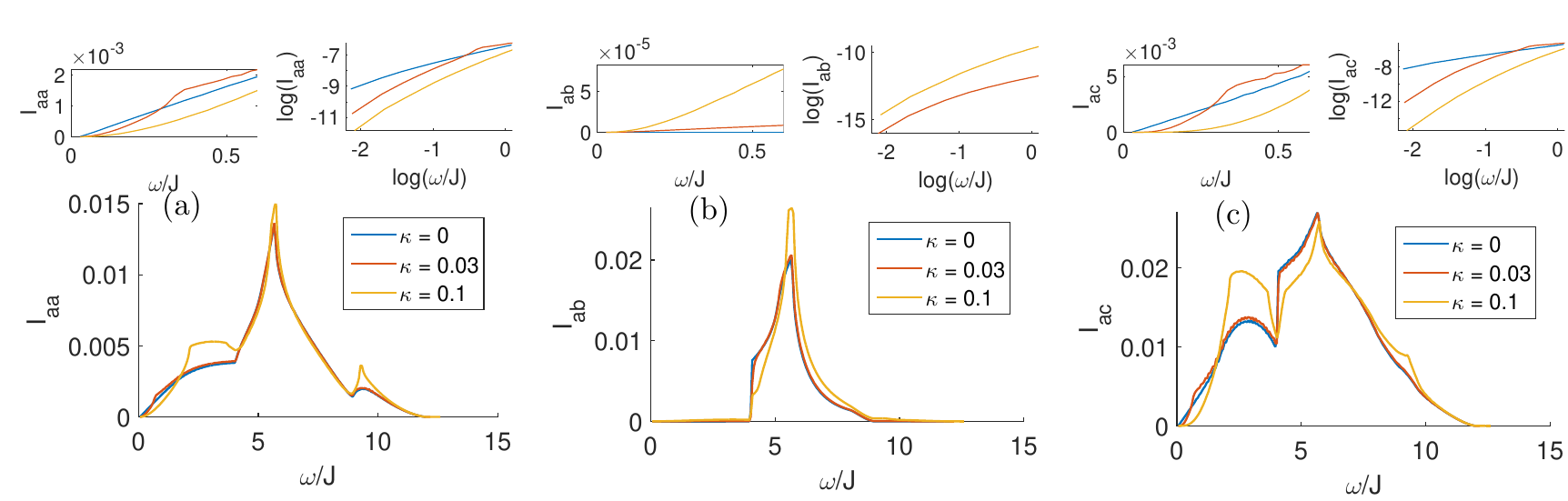}
	\caption{(Color Online) Raman spectra for the hyperhoneycomb lattice, shown for three channels ($aa, ab,$ and $ac$) representative of the linearly independent spectra in the limit $\kappa\rightarrow 0$. 
	}
	\label{bulk}
\end{figure*}

\begin{figure}
	\centering 
	\includegraphics[width=.85\linewidth]{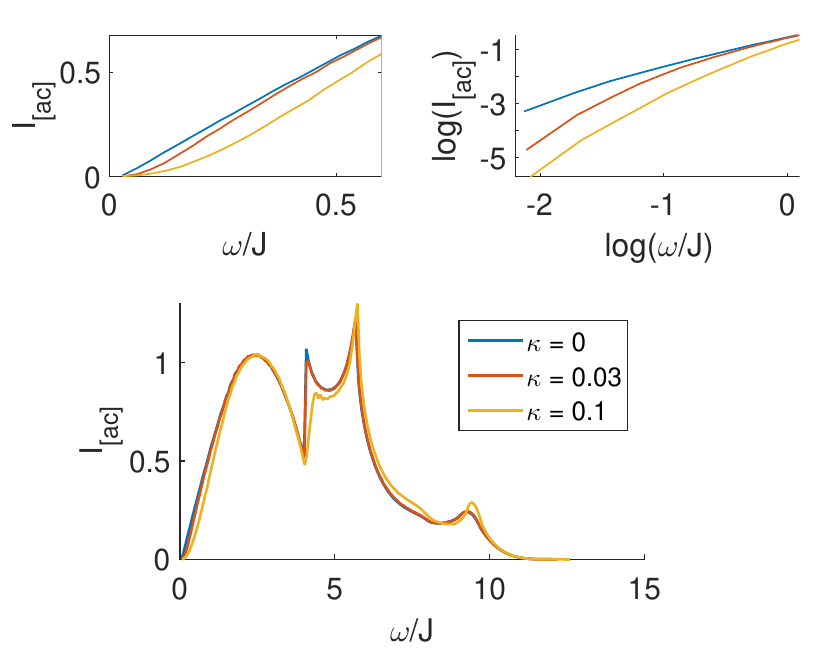}
	\caption{(Color Online) The Raman intensity in the anti-symmetric or rotational symmetry channel $[ac]$ of the hyperhoneycomb lattice.}
	\label{fig:[ac]}
\end{figure}

\subsection{Reliability of the projected model at finite frequency} \label{CaveatSec}

For the honeycomb and hyperhoneycomb lattices, it is interesting to consider how the magnetic field, which fundamentally alters the nature of the low-energy band structure, changes the bulk Raman response. 
Below we discuss how the second-neighbor hopping terms, generated at weak magnetic field in the effective spinon Hamiltonian, affect the Raman spectra in these two cases. 
Before presenting the results, however, we review several caveats in comparing Raman spectra for the Hamiltonian derived perturbatively in Sec. (\ref{magnetic}) to actual Raman spectra at finite magnetic field. Specifically, though we expect the spectra to match well at sufficiently low energy, there are several potential differences for Raman processes involving excitations above the flux gap $\Delta$, which is a fraction of $J$ in these systems. In practice, this means that caution must be exercised when comparing our results below to the exact finite-field result at all but the lowest energy scales.

The first caveat is that the three-spin terms discussed in Section \ref{magnetic} comprise the leading-order correction at finite magnetic field only at energy scales that are small compared to $\Delta$. For a Raman processes that generates a pair of spinons with total energy $\omega > 2\Delta$, the action of the original magnetic field perturbation (which creates one flux and one spinon), may be able to relax the system due to the interaction of the initial spinons with these new particles. In particular, if the spinon generated by the magnetic field perturbation is able to annihilate one of the two present in the Raman process at hand, the energy of intermediate state in perturbation theory may be lower than initial state energy $\omega$, which would lead to a breakdown of the perturbation theory used in Section \ref{magnetic} when applied to Raman excited states.
	
The second concern at higher energies is that, of the 3-spin terms described in Section \ref{magnetic}, our second-neighbor hopping model ignores those that generate four-spinon interactions (shown in Fig. \ref{fig:3-spin_term}(b)). While these terms are irrelevant at low energies in the renormalization-group sense, they do become important at energies where the DOS is not small, which is roughly near or above $J$. 

The third potential complication is that the resonant Raman operators have contributions from terms that do not project to zero flux, which we have ignored in our calculations. This approximation is certainly valid when describing Raman spectra at energy scales below the flux gap, as is relevant to our analysis of the topological surface states, but is questionable at higher energies for the anti-symmetric channels. 

Though we will not address the problems listed above here, the effect of flux perturbations to the LF Raman operator was considered in Ref.~\onlinecite{Knolle14-2}. There we argued that their qualitative effect is to produce a peak at twice the local flux gap. 
		
\subsection{Bulk response in a magnetic field} \label{BfieldResp}

We now study the Raman spectra of the effective model in which the magnetic field generates second-neighbor hopping terms for the \mssns, but does not otherwise alter the system. Despite the caveats outlined above, we will plot the results across the entire spinon bandwidth to illustrate the symmetry effects of the perturbation and to get an idea of its qualitative effects. In both the honeycomb and hyperhoneycomb lattices, a weak magnetic field significantly alters the polarization dependence, since the second-neighbor hopping terms in Eq. (\ref{2Hops}) break much of the lattice symmetry. It also changes the low-energy power law due to the change in Fermi-surface dimension. In particular, since the dispersion is generically linear about the Fermi-surface in the cases considered here, the limiting DOS is $\rho(\omega) \sim \omega^{d_c - 1}$, where $d_c = d- d_f$ is the co-dimension of the Fermi-surface, whose dimension is $d_f$ ($d$ is space dimension). A similar power law rule exists for the surface BZ as well. 
	
First, we consider the honeycomb lattice in the presence of a weak magnetic field $\kappa>0$. As detailed in Appendix \ref{SIS}, the second-neighbor hopping terms reduce the symmetry of the honeycomb lattice. As anticipated in Ref. \onlinecite{Perreault15}, they also break the LF-relationship, since they add a term to the Hamiltonian without affecting the Raman operator at the same order. However, rather surprisingly (see the Appendix), in this case the number of independent Raman spectra does not increase. The resulting two independent non-vanishing spectra $I_{xx}$ and $I_{[xy]}$, together with the DOS, are plotted in Fig. \ref{bulk_HC}. In this case the spectra are qualitatively unaltered at high energies, and the only qualitative effect of the second-neighbor hopping terms is to gap both low-energy spectra. 

These results illustrate an important point: the point group symmetries broken by the bare magnetic field are not necessarily broken by the second-neighbor hopping terms. For instance, a generic magnetic field breaks all of the lattice symmetries except inversion in all of the lattices considered here. However, on the honeycomb lattice the rotation symmetry remains in the low-energy theory where the magnetic field is replaced by second-neighbor hopping, while the reflection symmetries of this lattice are broken even by small magnetic fields. 

As shown for the hyperhoneycomb lattice in Fig. \ref{DOS_HC}, the magnetic field has very little impact on the DOS at high energies. The low-energy power law changes from linear at $\kappa=0$ to quadratic for $\kappa>0$. However, the dramatic reduction in symmetry has a significant impact on the high-energy Raman spectra of the hyperhoneycomb lattices, as illustrated in Fig. \ref{fig:exampl}. The effect is particularly pronounced for the $I_{ac}$ and $I_{bc}$ spectra which take on a low-energy power law that is quartic and quadratic respectively. To compare the effects on the different Raman channels that exist without the perturbation we have collected those spectra for different values of the perturbation in Fig. \ref{bulk}. A similar plot for the anti-symmetric channel $I_{[ac]}$ is included in Fig. \ref{fig:[ac]}.

Since time-reversal symmetry cannot be represented in terms of the \mss within the primitive unit cell of the (8,3)b lattice, there is no change in the internal symmetries if we perturb it with a magnetic field. For this reason we do not consider the effect of a finite magnetic field on this lattice. We note, however, that in the absence of a magnetic field its low-energy bulk Raman spectrum falls off with a larger power than that of they hyperhoneycomb, since this system has bulk Weyl nodes rather than a bulk Fermi ring.

\bibliography{References2}

\begin{thebibliography}{80}%
\makeatletter
\providecommand \@ifxundefined [1]{%
 \@ifx{#1\undefined}
}%
\providecommand \@ifnum [1]{%
 \ifnum #1\expandafter \@firstoftwo
 \else \expandafter \@secondoftwo
 \fi
}%
\providecommand \@ifx [1]{%
 \ifx #1\expandafter \@firstoftwo
 \else \expandafter \@secondoftwo
 \fi
}%
\providecommand \natexlab [1]{#1}%
\providecommand \enquote  [1]{``#1''}%
\providecommand \bibnamefont  [1]{#1}%
\providecommand \bibfnamefont [1]{#1}%
\providecommand \citenamefont [1]{#1}%
\providecommand \href@noop [0]{\@secondoftwo}%
\providecommand \href [0]{\begingroup \@sanitize@url \@href}%
\providecommand \@href[1]{\@@startlink{#1}\@@href}%
\providecommand \@@href[1]{\endgroup#1\@@endlink}%
\providecommand \@sanitize@url [0]{\catcode `\\12\catcode `\$12\catcode
  `\&12\catcode `\#12\catcode `\^12\catcode `\_12\catcode `\%12\relax}%
\providecommand \@@startlink[1]{}%
\providecommand \@@endlink[0]{}%
\providecommand \url  [0]{\begingroup\@sanitize@url \@url }%
\providecommand \@url [1]{\endgroup\@href {#1}{\urlprefix }}%
\providecommand \urlprefix  [0]{URL }%
\providecommand \Eprint [0]{\href }%
\providecommand \doibase [0]{http://dx.doi.org/}%
\providecommand \selectlanguage [0]{\@gobble}%
\providecommand \bibinfo  [0]{\@secondoftwo}%
\providecommand \bibfield  [0]{\@secondoftwo}%
\providecommand \translation [1]{[#1]}%
\providecommand \BibitemOpen [0]{}%
\providecommand \bibitemStop [0]{}%
\providecommand \bibitemNoStop [0]{.\EOS\space}%
\providecommand \EOS [0]{\spacefactor3000\relax}%
\providecommand \BibitemShut  [1]{\csname bibitem#1\endcsname}%
\let\auto@bib@innerbib\@empty
\bibitem [{\citenamefont {Klitzing}\ \emph {et~al.}(1980)\citenamefont
  {Klitzing}, \citenamefont {Dorda},\ and\ \citenamefont
  {Pepper}}]{Klitzing1980}%
  \BibitemOpen
  \bibfield  {author} {\bibinfo {author} {\bibfnamefont {K.~v.}\ \bibnamefont
  {Klitzing}}, \bibinfo {author} {\bibfnamefont {G.}~\bibnamefont {Dorda}}, \
  and\ \bibinfo {author} {\bibfnamefont {M.}~\bibnamefont {Pepper}},\ }\href
  {\doibase 10.1103/PhysRevLett.45.494} {\bibfield  {journal} {\bibinfo
  {journal} {Phys. Rev. Lett.}\ }\textbf {\bibinfo {volume} {45}},\ \bibinfo
  {pages} {494} (\bibinfo {year} {1980})}\BibitemShut {NoStop}%
\bibitem [{\citenamefont {Thouless}\ \emph {et~al.}(1982)\citenamefont
  {Thouless}, \citenamefont {Kohmoto}, \citenamefont {Nightingale},\ and\
  \citenamefont {den Nijs}}]{Thouless1982}%
  \BibitemOpen
  \bibfield  {author} {\bibinfo {author} {\bibfnamefont {D.~J.}\ \bibnamefont
  {Thouless}}, \bibinfo {author} {\bibfnamefont {M.}~\bibnamefont {Kohmoto}},
  \bibinfo {author} {\bibfnamefont {M.~P.}\ \bibnamefont {Nightingale}}, \ and\
  \bibinfo {author} {\bibfnamefont {M.}~\bibnamefont {den Nijs}},\ }\href
  {\doibase 10.1103/PhysRevLett.49.405} {\bibfield  {journal} {\bibinfo
  {journal} {Phys. Rev. Lett.}\ }\textbf {\bibinfo {volume} {49}},\ \bibinfo
  {pages} {405} (\bibinfo {year} {1982})}\BibitemShut {NoStop}%
\bibitem [{\citenamefont {Hasan}\ and\ \citenamefont {Kane}(2010)}]{Hasan2010}%
  \BibitemOpen
  \bibfield  {author} {\bibinfo {author} {\bibfnamefont {M.~Z.}\ \bibnamefont
  {Hasan}}\ and\ \bibinfo {author} {\bibfnamefont {C.~L.}\ \bibnamefont
  {Kane}},\ }\href {\doibase 10.1103/RevModPhys.82.3045} {\bibfield  {journal}
  {\bibinfo  {journal} {Rev. Mod. Phys.}\ }\textbf {\bibinfo {volume} {82}},\
  \bibinfo {pages} {3045} (\bibinfo {year} {2010})}\BibitemShut {NoStop}%
\bibitem [{\citenamefont {Qi}\ and\ \citenamefont {Zhang}(2011)}]{Qi2011}%
  \BibitemOpen
  \bibfield  {author} {\bibinfo {author} {\bibfnamefont {X.-L.}\ \bibnamefont
  {Qi}}\ and\ \bibinfo {author} {\bibfnamefont {S.-C.}\ \bibnamefont {Zhang}},\
  }\href {\doibase 10.1103/RevModPhys.83.1057} {\bibfield  {journal} {\bibinfo
  {journal} {Rev. Mod. Phys.}\ }\textbf {\bibinfo {volume} {83}},\ \bibinfo
  {pages} {1057} (\bibinfo {year} {2011})}\BibitemShut {NoStop}%
\bibitem [{\citenamefont {Wan}\ \emph {et~al.}(2011{\natexlab{a}})\citenamefont
  {Wan}, \citenamefont {Turner}, \citenamefont {Vishwanath},\ and\
  \citenamefont {Savrasov}}]{wan11}%
  \BibitemOpen
  \bibfield  {author} {\bibinfo {author} {\bibfnamefont {X.}~\bibnamefont
  {Wan}}, \bibinfo {author} {\bibfnamefont {A.~M.}\ \bibnamefont {Turner}},
  \bibinfo {author} {\bibfnamefont {A.}~\bibnamefont {Vishwanath}}, \ and\
  \bibinfo {author} {\bibfnamefont {S.~Y.}\ \bibnamefont {Savrasov}},\ }\href
  {\doibase 10.1103/PhysRevB.83.205101} {\bibfield  {journal} {\bibinfo
  {journal} {Phys. Rev. B}\ }\textbf {\bibinfo {volume} {83}},\ \bibinfo
  {pages} {205101} (\bibinfo {year} {2011}{\natexlab{a}})}\BibitemShut
  {NoStop}%
\bibitem [{\citenamefont {Vafek}\ and\ \citenamefont
  {Vishwanath}(2014)}]{Vafek2014}%
  \BibitemOpen
  \bibfield  {author} {\bibinfo {author} {\bibfnamefont {O.}~\bibnamefont
  {Vafek}}\ and\ \bibinfo {author} {\bibfnamefont {A.}~\bibnamefont
  {Vishwanath}},\ }\href {\doibase 10.1146/annurev-conmatphys-031113-133841}
  {\bibfield  {journal} {\bibinfo  {journal} {Annual Review of Condensed Matter
  Physics}\ }\textbf {\bibinfo {volume} {5}},\ \bibinfo {pages} {83} (\bibinfo
  {year} {2014})}\BibitemShut {NoStop}%
\bibitem [{\citenamefont {Bernevig}\ \emph {et~al.}(2006)\citenamefont
  {Bernevig}, \citenamefont {Hughes},\ and\ \citenamefont
  {Zhang}}]{BernevigHughesZhang}%
  \BibitemOpen
  \bibfield  {author} {\bibinfo {author} {\bibfnamefont {B.~A.}\ \bibnamefont
  {Bernevig}}, \bibinfo {author} {\bibfnamefont {T.~L.}\ \bibnamefont
  {Hughes}}, \ and\ \bibinfo {author} {\bibfnamefont {S.-C.}\ \bibnamefont
  {Zhang}},\ }\href {\doibase 10.1126/science.1133734} {\bibfield  {journal}
  {\bibinfo  {journal} {Science}\ }\textbf {\bibinfo {volume} {314}},\ \bibinfo
  {pages} {1757} (\bibinfo {year} {2006})}\BibitemShut {NoStop}%
\bibitem [{\citenamefont {Teo}\ \emph {et~al.}(2008)\citenamefont {Teo},
  \citenamefont {Fu},\ and\ \citenamefont {Kane}}]{Teo08}%
  \BibitemOpen
  \bibfield  {author} {\bibinfo {author} {\bibfnamefont {J.~C.~Y.}\
  \bibnamefont {Teo}}, \bibinfo {author} {\bibfnamefont {L.}~\bibnamefont
  {Fu}}, \ and\ \bibinfo {author} {\bibfnamefont {C.~L.}\ \bibnamefont
  {Kane}},\ }\href {\doibase 10.1103/PhysRevB.78.045426} {\bibfield  {journal}
  {\bibinfo  {journal} {Phys. Rev. B}\ }\textbf {\bibinfo {volume} {78}},\
  \bibinfo {pages} {045426} (\bibinfo {year} {2008})}\BibitemShut {NoStop}%
\bibitem [{\citenamefont {Weng}\ \emph {et~al.}(2015)\citenamefont {Weng},
  \citenamefont {Fang}, \citenamefont {Fang}, \citenamefont {Bernevig},\ and\
  \citenamefont {Dai}}]{DaiBernevig15}%
  \BibitemOpen
  \bibfield  {author} {\bibinfo {author} {\bibfnamefont {H.}~\bibnamefont
  {Weng}}, \bibinfo {author} {\bibfnamefont {C.}~\bibnamefont {Fang}}, \bibinfo
  {author} {\bibfnamefont {Z.}~\bibnamefont {Fang}}, \bibinfo {author}
  {\bibfnamefont {B.~A.}\ \bibnamefont {Bernevig}}, \ and\ \bibinfo {author}
  {\bibfnamefont {X.}~\bibnamefont {Dai}},\ }\href {\doibase
  10.1103/PhysRevX.5.011029} {\bibfield  {journal} {\bibinfo  {journal} {Phys.
  Rev. X}\ }\textbf {\bibinfo {volume} {5}},\ \bibinfo {pages} {011029}
  (\bibinfo {year} {2015})}\BibitemShut {NoStop}%
\bibitem [{\citenamefont {Hsieh}\ \emph {et~al.}(2008)\citenamefont {Hsieh},
  \citenamefont {Qian}, \citenamefont {Wray}, \citenamefont {Xia},
  \citenamefont {Hor}, \citenamefont {Cava},\ and\ \citenamefont
  {Hasan}}]{Hsieh08}%
  \BibitemOpen
  \bibfield  {author} {\bibinfo {author} {\bibfnamefont {D.}~\bibnamefont
  {Hsieh}}, \bibinfo {author} {\bibfnamefont {D.}~\bibnamefont {Qian}},
  \bibinfo {author} {\bibfnamefont {L.}~\bibnamefont {Wray}}, \bibinfo {author}
  {\bibfnamefont {Y.}~\bibnamefont {Xia}}, \bibinfo {author} {\bibfnamefont
  {Y.~S.}\ \bibnamefont {Hor}}, \bibinfo {author} {\bibfnamefont {R.~J.}\
  \bibnamefont {Cava}}, \ and\ \bibinfo {author} {\bibfnamefont {M.~Z.}\
  \bibnamefont {Hasan}},\ }\href {\doibase 10.1038/nature06843} {\bibfield
  {journal} {\bibinfo  {journal} {Nature}\ }\textbf {\bibinfo {volume} {452}},\
  \bibinfo {pages} {970} (\bibinfo {year} {2008})}\BibitemShut {NoStop}%
\bibitem [{\citenamefont {Xu}\ \emph {et~al.}(2015)\citenamefont {Xu},
  \citenamefont {Belopolski}, \citenamefont {Alidoust}, \citenamefont
  {Neupane}, \citenamefont {Bian}, \citenamefont {Zhang}, \citenamefont
  {Sankar}, \citenamefont {Chang}, \citenamefont {Yuan}, \citenamefont {Lee},
  \citenamefont {Huang}, \citenamefont {Zheng}, \citenamefont {Ma},
  \citenamefont {Sanchez}, \citenamefont {Wang}, \citenamefont {Bansil},
  \citenamefont {Chou}, \citenamefont {Shibayev}, \citenamefont {Lin},
  \citenamefont {Jia},\ and\ \citenamefont {Hasan}}]{Xu15}%
  \BibitemOpen
  \bibfield  {author} {\bibinfo {author} {\bibfnamefont {S.-Y.}\ \bibnamefont
  {Xu}}, \bibinfo {author} {\bibfnamefont {I.}~\bibnamefont {Belopolski}},
  \bibinfo {author} {\bibfnamefont {N.}~\bibnamefont {Alidoust}}, \bibinfo
  {author} {\bibfnamefont {M.}~\bibnamefont {Neupane}}, \bibinfo {author}
  {\bibfnamefont {G.}~\bibnamefont {Bian}}, \bibinfo {author} {\bibfnamefont
  {C.}~\bibnamefont {Zhang}}, \bibinfo {author} {\bibfnamefont
  {R.}~\bibnamefont {Sankar}}, \bibinfo {author} {\bibfnamefont
  {G.}~\bibnamefont {Chang}}, \bibinfo {author} {\bibfnamefont
  {Z.}~\bibnamefont {Yuan}}, \bibinfo {author} {\bibfnamefont {C.-C.}\
  \bibnamefont {Lee}}, \bibinfo {author} {\bibfnamefont {S.-M.}\ \bibnamefont
  {Huang}}, \bibinfo {author} {\bibfnamefont {H.}~\bibnamefont {Zheng}},
  \bibinfo {author} {\bibfnamefont {J.}~\bibnamefont {Ma}}, \bibinfo {author}
  {\bibfnamefont {D.~S.}\ \bibnamefont {Sanchez}}, \bibinfo {author}
  {\bibfnamefont {B.}~\bibnamefont {Wang}}, \bibinfo {author} {\bibfnamefont
  {A.}~\bibnamefont {Bansil}}, \bibinfo {author} {\bibfnamefont
  {F.}~\bibnamefont {Chou}}, \bibinfo {author} {\bibfnamefont {P.~P.}\
  \bibnamefont {Shibayev}}, \bibinfo {author} {\bibfnamefont {H.}~\bibnamefont
  {Lin}}, \bibinfo {author} {\bibfnamefont {S.}~\bibnamefont {Jia}}, \ and\
  \bibinfo {author} {\bibfnamefont {M.~Z.}\ \bibnamefont {Hasan}},\ }\href
  {\doibase 10.1126/science.aaa9297} {\bibfield  {journal} {\bibinfo  {journal}
  {Science}\ }\textbf {\bibinfo {volume} {349}},\ \bibinfo {pages} {613}
  (\bibinfo {year} {2015})}\BibitemShut {NoStop}%
\bibitem [{\citenamefont {Lv}\ \emph {et~al.}(2015)\citenamefont {Lv},
  \citenamefont {Weng}, \citenamefont {Fu}, \citenamefont {Wang}, \citenamefont
  {Miao}, \citenamefont {Ma}, \citenamefont {Richard}, \citenamefont {Huang},
  \citenamefont {Zhao}, \citenamefont {Chen}, \citenamefont {Fang},
  \citenamefont {Dai}, \citenamefont {Qian},\ and\ \citenamefont {Ding}}]{Lv}%
  \BibitemOpen
  \bibfield  {author} {\bibinfo {author} {\bibfnamefont {B.~Q.}\ \bibnamefont
  {Lv}}, \bibinfo {author} {\bibfnamefont {H.~M.}\ \bibnamefont {Weng}},
  \bibinfo {author} {\bibfnamefont {B.~B.}\ \bibnamefont {Fu}}, \bibinfo
  {author} {\bibfnamefont {X.~P.}\ \bibnamefont {Wang}}, \bibinfo {author}
  {\bibfnamefont {H.}~\bibnamefont {Miao}}, \bibinfo {author} {\bibfnamefont
  {J.}~\bibnamefont {Ma}}, \bibinfo {author} {\bibfnamefont {P.}~\bibnamefont
  {Richard}}, \bibinfo {author} {\bibfnamefont {X.~C.}\ \bibnamefont {Huang}},
  \bibinfo {author} {\bibfnamefont {L.~X.}\ \bibnamefont {Zhao}}, \bibinfo
  {author} {\bibfnamefont {G.~F.}\ \bibnamefont {Chen}}, \bibinfo {author}
  {\bibfnamefont {Z.}~\bibnamefont {Fang}}, \bibinfo {author} {\bibfnamefont
  {X.}~\bibnamefont {Dai}}, \bibinfo {author} {\bibfnamefont {T.}~\bibnamefont
  {Qian}}, \ and\ \bibinfo {author} {\bibfnamefont {H.}~\bibnamefont {Ding}},\
  }\href {\doibase 10.1103/PhysRevX.5.031013} {\bibfield  {journal} {\bibinfo
  {journal} {Phys. Rev. X}\ }\textbf {\bibinfo {volume} {5}},\ \bibinfo {pages}
  {031013} (\bibinfo {year} {2015})}\BibitemShut {NoStop}%
\bibitem [{\citenamefont {K\"onig}\ \emph {et~al.}(2007)\citenamefont
  {K\"onig}, \citenamefont {Wiedmann}, \citenamefont {Br\"une}, \citenamefont
  {Roth}, \citenamefont {Buhmann}, \citenamefont {Molenkamp}, \citenamefont
  {Qi},\ and\ \citenamefont {Zhang}}]{koenig07}%
  \BibitemOpen
  \bibfield  {author} {\bibinfo {author} {\bibfnamefont {M.}~\bibnamefont
  {K\"onig}}, \bibinfo {author} {\bibfnamefont {S.}~\bibnamefont {Wiedmann}},
  \bibinfo {author} {\bibfnamefont {C.}~\bibnamefont {Br\"une}}, \bibinfo
  {author} {\bibfnamefont {A.}~\bibnamefont {Roth}}, \bibinfo {author}
  {\bibfnamefont {H.}~\bibnamefont {Buhmann}}, \bibinfo {author} {\bibfnamefont
  {L.~W.}\ \bibnamefont {Molenkamp}}, \bibinfo {author} {\bibfnamefont {X.~L.}\
  \bibnamefont {Qi}}, \ and\ \bibinfo {author} {\bibfnamefont {S.~C.}\
  \bibnamefont {Zhang}},\ }\href {\doibase 10.1126/science.1148047} {\bibfield
  {journal} {\bibinfo  {journal} {Science}\ }\textbf {\bibinfo {volume}
  {318}},\ \bibinfo {pages} {766} (\bibinfo {year} {2007})}\BibitemShut
  {NoStop}%
\bibitem [{\citenamefont {Wen}\ and\ \citenamefont {Niu}(1990)}]{Wen1990}%
  \BibitemOpen
  \bibfield  {author} {\bibinfo {author} {\bibfnamefont {X.~G.}\ \bibnamefont
  {Wen}}\ and\ \bibinfo {author} {\bibfnamefont {Q.}~\bibnamefont {Niu}},\
  }\href {\doibase 10.1103/PhysRevB.41.9377} {\bibfield  {journal} {\bibinfo
  {journal} {Phys. Rev. B}\ }\textbf {\bibinfo {volume} {41}},\ \bibinfo
  {pages} {9377} (\bibinfo {year} {1990})}\BibitemShut {NoStop}%
\bibitem [{\citenamefont {Levin}\ and\ \citenamefont
  {Stern}(2012)}]{Levin2012}%
  \BibitemOpen
  \bibfield  {author} {\bibinfo {author} {\bibfnamefont {M.}~\bibnamefont
  {Levin}}\ and\ \bibinfo {author} {\bibfnamefont {A.}~\bibnamefont {Stern}},\
  }\href {\doibase 10.1103/PhysRevB.86.115131} {\bibfield  {journal} {\bibinfo
  {journal} {Phys. Rev. B}\ }\textbf {\bibinfo {volume} {86}},\ \bibinfo
  {pages} {115131} (\bibinfo {year} {2012})}\BibitemShut {NoStop}%
\bibitem [{\citenamefont {Swingle}\ \emph {et~al.}(2011)\citenamefont
  {Swingle}, \citenamefont {Barkeshli}, \citenamefont {McGreevy},\ and\
  \citenamefont {Senthil}}]{Swingle11}%
  \BibitemOpen
  \bibfield  {author} {\bibinfo {author} {\bibfnamefont {B.}~\bibnamefont
  {Swingle}}, \bibinfo {author} {\bibfnamefont {M.}~\bibnamefont {Barkeshli}},
  \bibinfo {author} {\bibfnamefont {J.}~\bibnamefont {McGreevy}}, \ and\
  \bibinfo {author} {\bibfnamefont {T.}~\bibnamefont {Senthil}},\ }\href
  {\doibase 10.1103/PhysRevB.83.195139} {\bibfield  {journal} {\bibinfo
  {journal} {Phys. Rev. B}\ }\textbf {\bibinfo {volume} {83}},\ \bibinfo
  {pages} {195139} (\bibinfo {year} {2011})}\BibitemShut {NoStop}%
\bibitem [{\citenamefont {Levin}\ \emph {et~al.}(2011)\citenamefont {Levin},
  \citenamefont {Burnell}, \citenamefont {Koch-Janusz},\ and\ \citenamefont
  {Stern}}]{Levin11}%
  \BibitemOpen
  \bibfield  {author} {\bibinfo {author} {\bibfnamefont {M.}~\bibnamefont
  {Levin}}, \bibinfo {author} {\bibfnamefont {F.~J.}\ \bibnamefont {Burnell}},
  \bibinfo {author} {\bibfnamefont {M.}~\bibnamefont {Koch-Janusz}}, \ and\
  \bibinfo {author} {\bibfnamefont {A.}~\bibnamefont {Stern}},\ }\href
  {\doibase 10.1103/PhysRevB.84.235145} {\bibfield  {journal} {\bibinfo
  {journal} {Phys. Rev. B}\ }\textbf {\bibinfo {volume} {84}},\ \bibinfo
  {pages} {235145} (\bibinfo {year} {2011})}\BibitemShut {NoStop}%
\bibitem [{\citenamefont {Maciejko}\ \emph {et~al.}(2012)\citenamefont
  {Maciejko}, \citenamefont {Qi}, \citenamefont {Karch},\ and\ \citenamefont
  {Zhang}}]{Maciejko12}%
  \BibitemOpen
  \bibfield  {author} {\bibinfo {author} {\bibfnamefont {J.}~\bibnamefont
  {Maciejko}}, \bibinfo {author} {\bibfnamefont {X.-L.}\ \bibnamefont {Qi}},
  \bibinfo {author} {\bibfnamefont {A.}~\bibnamefont {Karch}}, \ and\ \bibinfo
  {author} {\bibfnamefont {S.-C.}\ \bibnamefont {Zhang}},\ }\href {\doibase
  10.1103/PhysRevB.86.235128} {\bibfield  {journal} {\bibinfo  {journal} {Phys.
  Rev. B}\ }\textbf {\bibinfo {volume} {86}},\ \bibinfo {pages} {235128}
  (\bibinfo {year} {2012})}\BibitemShut {NoStop}%
\bibitem [{\citenamefont {Lu}\ and\ \citenamefont
  {Vishwanath}(2016)}]{LuVishwanath}%
  \BibitemOpen
  \bibfield  {author} {\bibinfo {author} {\bibfnamefont {Y.-M.}\ \bibnamefont
  {Lu}}\ and\ \bibinfo {author} {\bibfnamefont {A.}~\bibnamefont
  {Vishwanath}},\ }\href {\doibase 10.1103/PhysRevB.93.155121} {\bibfield
  {journal} {\bibinfo  {journal} {Phys. Rev. B}\ }\textbf {\bibinfo {volume}
  {93}},\ \bibinfo {pages} {155121} (\bibinfo {year} {2016})}\BibitemShut
  {NoStop}%
\bibitem [{\citenamefont {Mesaros}\ and\ \citenamefont
  {Ran}(2013)}]{MesarosRan}%
  \BibitemOpen
  \bibfield  {author} {\bibinfo {author} {\bibfnamefont {A.}~\bibnamefont
  {Mesaros}}\ and\ \bibinfo {author} {\bibfnamefont {Y.}~\bibnamefont {Ran}},\
  }\href {\doibase 10.1103/PhysRevB.87.155115} {\bibfield  {journal} {\bibinfo
  {journal} {Phys. Rev. B}\ }\textbf {\bibinfo {volume} {87}},\ \bibinfo
  {pages} {155115} (\bibinfo {year} {2013})}\BibitemShut {NoStop}%
\bibitem [{\citenamefont {{Barkeshli}}\ \emph {et~al.}(2014)\citenamefont
  {{Barkeshli}}, \citenamefont {{Bonderson}}, \citenamefont {{Cheng}},\ and\
  \citenamefont {{Wang}}}]{LongQPaper}%
  \BibitemOpen
  \bibfield  {author} {\bibinfo {author} {\bibfnamefont {M.}~\bibnamefont
  {{Barkeshli}}}, \bibinfo {author} {\bibfnamefont {P.}~\bibnamefont
  {{Bonderson}}}, \bibinfo {author} {\bibfnamefont {M.}~\bibnamefont
  {{Cheng}}}, \ and\ \bibinfo {author} {\bibfnamefont {Z.}~\bibnamefont
  {{Wang}}},\ }\href@noop {} {\  (\bibinfo {year} {2014})},\ \Eprint
  {http://arxiv.org/abs/1410.4540} {arXiv:1410.4540} \BibitemShut {NoStop}%
\bibitem [{\citenamefont {Anderson}(1973)}]{Anderson1973}%
  \BibitemOpen
  \bibfield  {author} {\bibinfo {author} {\bibfnamefont {P.}~\bibnamefont
  {Anderson}},\ }\href {\doibase 10.1016/0025-5408(73)90167-0} {\bibfield
  {journal} {\bibinfo  {journal} {Materials Research Bulletin}\ }\textbf
  {\bibinfo {volume} {8}},\ \bibinfo {pages} {153 } (\bibinfo {year}
  {1973})}\BibitemShut {NoStop}%
\bibitem [{\citenamefont {Lacroix}\ \emph {et~al.}(2011)\citenamefont
  {Lacroix}, \citenamefont {Mendels},\ and\ \citenamefont
  {Mila}}]{Lacroix2011}%
  \BibitemOpen
  \bibinfo {editor} {\bibfnamefont {C.}~\bibnamefont {Lacroix}}, \bibinfo
  {editor} {\bibfnamefont {P.}~\bibnamefont {Mendels}}, \ and\ \bibinfo
  {editor} {\bibfnamefont {F.}~\bibnamefont {Mila}},\ eds.,\ \href {\doibase
  10.1007/978-3-642-10589-0} {\emph {\bibinfo {title} {{Introduction to
  Frustrated Magnetism: Materials, Experiments, Theory (Springer Series in
  Solid-State Sciences)}}}},\ \bibinfo {edition} {2011th}\ ed.\ (\bibinfo
  {publisher} {Springer},\ \bibinfo {year} {2011})\BibitemShut {NoStop}%
\bibitem [{\citenamefont {Balents}(2010)}]{Balents}%
  \BibitemOpen
  \bibfield  {author} {\bibinfo {author} {\bibfnamefont {L.}~\bibnamefont
  {Balents}},\ }\href {\doibase 10.1038/nature08917} {\bibfield  {journal}
  {\bibinfo  {journal} {Nature}\ }\textbf {\bibinfo {volume} {464}},\ \bibinfo
  {pages} {199} (\bibinfo {year} {2010})}\BibitemShut {NoStop}%
\bibitem [{\citenamefont {Savary}\ and\ \citenamefont {Balents}()}]{savary16}%
  \BibitemOpen
  \bibfield  {author} {\bibinfo {author} {\bibfnamefont {L.}~\bibnamefont
  {Savary}}\ and\ \bibinfo {author} {\bibfnamefont {L.}~\bibnamefont
  {Balents}},\ }\href {http://arxiv.org/abs/1601.03742} {\bibinfo  {journal}
  {arXiv:1601.03742}\ }\BibitemShut {NoStop}%
\bibitem [{\citenamefont {Pesin}\ and\ \citenamefont {Balents}(2010)}]{Pesin}%
  \BibitemOpen
\bibfield  {journal} {  }\bibfield  {author} {\bibinfo {author} {\bibfnamefont
  {D.}~\bibnamefont {Pesin}}\ and\ \bibinfo {author} {\bibfnamefont
  {L.}~\bibnamefont {Balents}},\ }\href {\doibase 10.1038/nphys1606} {\bibfield
   {journal} {\bibinfo  {journal} {Nature Physics}\ }\textbf {\bibinfo {volume}
  {6}},\ \bibinfo {pages} {376} (\bibinfo {year} {2010})}\BibitemShut {NoStop}%
\bibitem [{\citenamefont {Knolle}\ \emph
  {et~al.}(2014{\natexlab{a}})\citenamefont {Knolle}, \citenamefont {Chern},
  \citenamefont {Kovrizhin}, \citenamefont {Moessner},\ and\ \citenamefont
  {Perkins}}]{Knolle14-2}%
  \BibitemOpen
  \bibfield  {author} {\bibinfo {author} {\bibfnamefont {J.}~\bibnamefont
  {Knolle}}, \bibinfo {author} {\bibfnamefont {G.-W.}\ \bibnamefont {Chern}},
  \bibinfo {author} {\bibfnamefont {D.~L.}\ \bibnamefont {Kovrizhin}}, \bibinfo
  {author} {\bibfnamefont {R.}~\bibnamefont {Moessner}}, \ and\ \bibinfo
  {author} {\bibfnamefont {N.~B.}\ \bibnamefont {Perkins}},\ }\href {\doibase
  10.1103/PhysRevLett.113.187201} {\bibfield  {journal} {\bibinfo  {journal}
  {Phys. Rev. Lett.}\ }\textbf {\bibinfo {volume} {113}},\ \bibinfo {pages}
  {187201} (\bibinfo {year} {2014}{\natexlab{a}})}\BibitemShut {NoStop}%
\bibitem [{\citenamefont {Perreault}\ \emph {et~al.}(2015)\citenamefont
  {Perreault}, \citenamefont {Knolle}, \citenamefont {Perkins},\ and\
  \citenamefont {Burnell}}]{Perreault15}%
  \BibitemOpen
  \bibfield  {author} {\bibinfo {author} {\bibfnamefont {B.}~\bibnamefont
  {Perreault}}, \bibinfo {author} {\bibfnamefont {J.}~\bibnamefont {Knolle}},
  \bibinfo {author} {\bibfnamefont {N.~B.}\ \bibnamefont {Perkins}}, \ and\
  \bibinfo {author} {\bibfnamefont {F.~J.}\ \bibnamefont {Burnell}},\ }\href
  {\doibase 10.1103/PhysRevB.92.094439} {\bibfield  {journal} {\bibinfo
  {journal} {Phys. Rev. B}\ }\textbf {\bibinfo {volume} {92}},\ \bibinfo
  {pages} {094439} (\bibinfo {year} {2015})}\BibitemShut {NoStop}%
\bibitem [{\citenamefont {Perreault}\ \emph {et~al.}(2016)\citenamefont
  {Perreault}, \citenamefont {Knolle}, \citenamefont {Perkins},\ and\
  \citenamefont {Burnell}}]{Perreault16}%
  \BibitemOpen
  \bibfield  {author} {\bibinfo {author} {\bibfnamefont {B.}~\bibnamefont
  {Perreault}}, \bibinfo {author} {\bibfnamefont {J.}~\bibnamefont {Knolle}},
  \bibinfo {author} {\bibfnamefont {N.~B.}\ \bibnamefont {Perkins}}, \ and\
  \bibinfo {author} {\bibfnamefont {F.}~\bibnamefont {Burnell}},\ }\href
  {http://arxiv.org/abs/1601.02623} {\bibfield  {journal} {\bibinfo  {journal}
  {arXiv:1601.02623}\ } (\bibinfo {year} {2016})}\BibitemShut {NoStop}%
\bibitem [{\citenamefont {Hal{\'a}sz}\ \emph {et~al.}(2016)\citenamefont
  {Hal{\'a}sz}, \citenamefont {Perkins},\ and\ \citenamefont
  {Brink}}]{Halasz16}%
  \BibitemOpen
  \bibfield  {author} {\bibinfo {author} {\bibfnamefont {G.~B.}\ \bibnamefont
  {Hal{\'a}sz}}, \bibinfo {author} {\bibfnamefont {N.~B.}\ \bibnamefont
  {Perkins}}, \ and\ \bibinfo {author} {\bibfnamefont {J.~v.~d.}\ \bibnamefont
  {Brink}},\ }\href {http://arxiv.org/abs/1605.03272} {\bibfield  {journal}
  {\bibinfo  {journal} {arXiv:1605.03272}\ } (\bibinfo {year}
  {2016})}\BibitemShut {NoStop}%
\bibitem [{\citenamefont {Kitaev}(2006)}]{Kitaev}%
  \BibitemOpen
  \bibfield  {author} {\bibinfo {author} {\bibfnamefont {A.}~\bibnamefont
  {Kitaev}},\ }\href@noop {} {\bibfield  {journal} {\bibinfo  {journal} {Annals
  of Physics}\ }\textbf {\bibinfo {volume} {321}},\ \bibinfo {pages} {2}
  (\bibinfo {year} {2006})}\BibitemShut {NoStop}%
\bibitem [{\citenamefont {Mandal}\ and\ \citenamefont
  {Surendran}(2009)}]{Mandal}%
  \BibitemOpen
  \bibfield  {author} {\bibinfo {author} {\bibfnamefont {S.}~\bibnamefont
  {Mandal}}\ and\ \bibinfo {author} {\bibfnamefont {N.}~\bibnamefont
  {Surendran}},\ }\href {\doibase 10.1103/PhysRevB.79.024426} {\bibfield
  {journal} {\bibinfo  {journal} {Phys. Rev. B}\ }\textbf {\bibinfo {volume}
  {79}},\ \bibinfo {pages} {024426} (\bibinfo {year} {2009})}\BibitemShut
  {NoStop}%
\bibitem [{\citenamefont {O'Brien}\ \emph {et~al.}(2016)\citenamefont
  {O'Brien}, \citenamefont {Hermanns},\ and\ \citenamefont {Trebst}}]{Obrien}%
  \BibitemOpen
  \bibfield  {author} {\bibinfo {author} {\bibfnamefont {K.}~\bibnamefont
  {O'Brien}}, \bibinfo {author} {\bibfnamefont {M.}~\bibnamefont {Hermanns}}, \
  and\ \bibinfo {author} {\bibfnamefont {S.}~\bibnamefont {Trebst}},\ }\href
  {\doibase 10.1103/PhysRevB.93.085101} {\bibfield  {journal} {\bibinfo
  {journal} {Phys. Rev. B}\ }\textbf {\bibinfo {volume} {93}},\ \bibinfo
  {pages} {085101} (\bibinfo {year} {2016})}\BibitemShut {NoStop}%
\bibitem [{\citenamefont {Yao}\ and\ \citenamefont
  {Kivelson}(2007)}]{YaoKivelson}%
  \BibitemOpen
  \bibfield  {author} {\bibinfo {author} {\bibfnamefont {H.}~\bibnamefont
  {Yao}}\ and\ \bibinfo {author} {\bibfnamefont {S.~A.}\ \bibnamefont
  {Kivelson}},\ }\href {\doibase 10.1103/PhysRevLett.99.247203} {\bibfield
  {journal} {\bibinfo  {journal} {Phys. Rev. Lett.}\ }\textbf {\bibinfo
  {volume} {99}},\ \bibinfo {pages} {247203} (\bibinfo {year}
  {2007})}\BibitemShut {NoStop}%
\bibitem [{\citenamefont {Hermanns}\ \emph {et~al.}(2015)\citenamefont
  {Hermanns}, \citenamefont {O'Brien},\ and\ \citenamefont
  {Trebst}}]{Hermanns}%
  \BibitemOpen
  \bibfield  {author} {\bibinfo {author} {\bibfnamefont {M.}~\bibnamefont
  {Hermanns}}, \bibinfo {author} {\bibfnamefont {K.}~\bibnamefont {O'Brien}}, \
  and\ \bibinfo {author} {\bibfnamefont {S.}~\bibnamefont {Trebst}},\ }\href
  {\doibase 10.1103/PhysRevLett.114.157202} {\bibfield  {journal} {\bibinfo
  {journal} {Phys. Rev. Lett.}\ }\textbf {\bibinfo {volume} {114}},\ \bibinfo
  {pages} {157202} (\bibinfo {year} {2015})}\BibitemShut {NoStop}%
\bibitem [{\citenamefont {Jackeli}\ and\ \citenamefont
  {Khaliullin}(2009)}]{Jackeli}%
  \BibitemOpen
  \bibfield  {author} {\bibinfo {author} {\bibfnamefont {G.}~\bibnamefont
  {Jackeli}}\ and\ \bibinfo {author} {\bibfnamefont {G.}~\bibnamefont
  {Khaliullin}},\ }\href {\doibase 10.1103/PhysRevLett.102.017205} {\bibfield
  {journal} {\bibinfo  {journal} {Phys. Rev. Lett.}\ }\textbf {\bibinfo
  {volume} {102}},\ \bibinfo {pages} {017205} (\bibinfo {year}
  {2009})}\BibitemShut {NoStop}%
\bibitem [{\citenamefont {Chaloupka}\ \emph {et~al.}(2010)\citenamefont
  {Chaloupka}, \citenamefont {Jackeli},\ and\ \citenamefont
  {Khaliullin}}]{Chaloupka10}%
  \BibitemOpen
  \bibfield  {author} {\bibinfo {author} {\bibfnamefont {J.~c.~v.}\
  \bibnamefont {Chaloupka}}, \bibinfo {author} {\bibfnamefont {G.}~\bibnamefont
  {Jackeli}}, \ and\ \bibinfo {author} {\bibfnamefont {G.}~\bibnamefont
  {Khaliullin}},\ }\href {\doibase 10.1103/PhysRevLett.105.027204} {\bibfield
  {journal} {\bibinfo  {journal} {Phys. Rev. Lett.}\ }\textbf {\bibinfo
  {volume} {105}},\ \bibinfo {pages} {027204} (\bibinfo {year}
  {2010})}\BibitemShut {NoStop}%
\bibitem [{\citenamefont {Plumb}\ \emph {et~al.}(2014)\citenamefont {Plumb},
  \citenamefont {Clancy}, \citenamefont {Sandilands}, \citenamefont {Shankar},
  \citenamefont {Hu}, \citenamefont {Burch}, \citenamefont {Kee},\ and\
  \citenamefont {Kim}}]{Plumb14}%
  \BibitemOpen
  \bibfield  {author} {\bibinfo {author} {\bibfnamefont {K.~W.}\ \bibnamefont
  {Plumb}}, \bibinfo {author} {\bibfnamefont {J.~P.}\ \bibnamefont {Clancy}},
  \bibinfo {author} {\bibfnamefont {L.~J.}\ \bibnamefont {Sandilands}},
  \bibinfo {author} {\bibfnamefont {V.~V.}\ \bibnamefont {Shankar}}, \bibinfo
  {author} {\bibfnamefont {Y.~F.}\ \bibnamefont {Hu}}, \bibinfo {author}
  {\bibfnamefont {K.~S.}\ \bibnamefont {Burch}}, \bibinfo {author}
  {\bibfnamefont {H.-Y.}\ \bibnamefont {Kee}}, \ and\ \bibinfo {author}
  {\bibfnamefont {Y.-J.}\ \bibnamefont {Kim}},\ }\href {\doibase
  10.1103/PhysRevB.90.041112} {\bibfield  {journal} {\bibinfo  {journal} {Phys.
  Rev. B}\ }\textbf {\bibinfo {volume} {90}},\ \bibinfo {pages} {041112}
  (\bibinfo {year} {2014})}\BibitemShut {NoStop}%
\bibitem [{\citenamefont {Kim}\ \emph {et~al.}(2015{\natexlab{a}})\citenamefont
  {Kim}, \citenamefont {V.}, \citenamefont {Catuneanu},\ and\ \citenamefont
  {Kee}}]{Kim15-1}%
  \BibitemOpen
  \bibfield  {author} {\bibinfo {author} {\bibfnamefont {H.-S.}\ \bibnamefont
  {Kim}}, \bibinfo {author} {\bibfnamefont {V.~S.}\ \bibnamefont {V.}},
  \bibinfo {author} {\bibfnamefont {A.}~\bibnamefont {Catuneanu}}, \ and\
  \bibinfo {author} {\bibfnamefont {H.-Y.}\ \bibnamefont {Kee}},\ }\href
  {\doibase 10.1103/PhysRevB.91.241110} {\bibfield  {journal} {\bibinfo
  {journal} {Phys. Rev. B}\ }\textbf {\bibinfo {volume} {91}},\ \bibinfo
  {pages} {241110} (\bibinfo {year} {2015}{\natexlab{a}})}\BibitemShut
  {NoStop}%
\bibitem [{\citenamefont {Modic}\ \emph {et~al.}(2014)\citenamefont {Modic},
  \citenamefont {Smidt}, \citenamefont {Kimchi}, \citenamefont {Breznay},
  \citenamefont {Biffin}, \citenamefont {Choi}, \citenamefont {Johnson},
  \citenamefont {Coldea}, \citenamefont {Watkins-Curry}, \citenamefont
  {McCandess} \emph {et~al.}}]{Modic}%
  \BibitemOpen
  \bibfield  {author} {\bibinfo {author} {\bibfnamefont {K.}~\bibnamefont
  {Modic}}, \bibinfo {author} {\bibfnamefont {T.~E.}\ \bibnamefont {Smidt}},
  \bibinfo {author} {\bibfnamefont {I.}~\bibnamefont {Kimchi}}, \bibinfo
  {author} {\bibfnamefont {N.~P.}\ \bibnamefont {Breznay}}, \bibinfo {author}
  {\bibfnamefont {A.}~\bibnamefont {Biffin}}, \bibinfo {author} {\bibfnamefont
  {S.}~\bibnamefont {Choi}}, \bibinfo {author} {\bibfnamefont {R.~D.}\
  \bibnamefont {Johnson}}, \bibinfo {author} {\bibfnamefont {R.}~\bibnamefont
  {Coldea}}, \bibinfo {author} {\bibfnamefont {P.}~\bibnamefont
  {Watkins-Curry}}, \bibinfo {author} {\bibfnamefont {G.~T.}\ \bibnamefont
  {McCandess}},  \emph {et~al.},\ }\href {\doibase 10.1038/ncomms5203}
  {\bibfield  {journal} {\bibinfo  {journal} {Nature communications}\ }\textbf
  {\bibinfo {volume} {5}} (\bibinfo {year} {2014}),\
  10.1038/ncomms5203}\BibitemShut {NoStop}%
\bibitem [{\citenamefont {Kimchi}\ \emph {et~al.}(2014)\citenamefont {Kimchi},
  \citenamefont {Analytis},\ and\ \citenamefont {Vishwanath}}]{Kimchi14}%
  \BibitemOpen
  \bibfield  {author} {\bibinfo {author} {\bibfnamefont {I.}~\bibnamefont
  {Kimchi}}, \bibinfo {author} {\bibfnamefont {J.~G.}\ \bibnamefont
  {Analytis}}, \ and\ \bibinfo {author} {\bibfnamefont {A.}~\bibnamefont
  {Vishwanath}},\ }\href {\doibase 10.1103/PhysRevB.90.205126} {\bibfield
  {journal} {\bibinfo  {journal} {Phys. Rev. B}\ }\textbf {\bibinfo {volume}
  {90}},\ \bibinfo {pages} {205126} (\bibinfo {year} {2014})}\BibitemShut
  {NoStop}%
\bibitem [{\citenamefont {Takayama}\ \emph {et~al.}(2015)\citenamefont
  {Takayama}, \citenamefont {Kato}, \citenamefont {Dinnebier}, \citenamefont
  {Nuss}, \citenamefont {Kono}, \citenamefont {Veiga}, \citenamefont {Fabbris},
  \citenamefont {Haskel},\ and\ \citenamefont {Takagi}}]{Takayama}%
  \BibitemOpen
  \bibfield  {author} {\bibinfo {author} {\bibfnamefont {T.}~\bibnamefont
  {Takayama}}, \bibinfo {author} {\bibfnamefont {A.}~\bibnamefont {Kato}},
  \bibinfo {author} {\bibfnamefont {R.}~\bibnamefont {Dinnebier}}, \bibinfo
  {author} {\bibfnamefont {J.}~\bibnamefont {Nuss}}, \bibinfo {author}
  {\bibfnamefont {H.}~\bibnamefont {Kono}}, \bibinfo {author} {\bibfnamefont
  {L.~S.~I.}\ \bibnamefont {Veiga}}, \bibinfo {author} {\bibfnamefont
  {G.}~\bibnamefont {Fabbris}}, \bibinfo {author} {\bibfnamefont
  {D.}~\bibnamefont {Haskel}}, \ and\ \bibinfo {author} {\bibfnamefont
  {H.}~\bibnamefont {Takagi}},\ }\href {\doibase
  10.1103/PhysRevLett.114.077202} {\bibfield  {journal} {\bibinfo  {journal}
  {Phys. Rev. Lett.}\ }\textbf {\bibinfo {volume} {114}},\ \bibinfo {pages}
  {077202} (\bibinfo {year} {2015})}\BibitemShut {NoStop}%
\bibitem [{\citenamefont {Hwan~Chun}\ \emph {et~al.}(2015)\citenamefont
  {Hwan~Chun}, \citenamefont {Kim}, \citenamefont {Kim}, \citenamefont {Zheng},
  \citenamefont {Stoumpos}, \citenamefont {Malliakas}, \citenamefont
  {Mitchell}, \citenamefont {Mehlawat}, \citenamefont {Singh}, \citenamefont
  {Choi}, \citenamefont {Gog}, \citenamefont {Al-Zein}, \citenamefont {Sala},
  \citenamefont {Krisch}, \citenamefont {Chaloupka}, \citenamefont {Jackeli},
  \citenamefont {Khaliullin},\ and\ \citenamefont {Kim}}]{Chun2015}%
  \BibitemOpen
  \bibfield  {author} {\bibinfo {author} {\bibfnamefont {S.}~\bibnamefont
  {Hwan~Chun}}, \bibinfo {author} {\bibfnamefont {J.-W.}\ \bibnamefont {Kim}},
  \bibinfo {author} {\bibfnamefont {J.}~\bibnamefont {Kim}}, \bibinfo {author}
  {\bibfnamefont {H.}~\bibnamefont {Zheng}}, \bibinfo {author} {\bibfnamefont
  {C.~C.}\ \bibnamefont {Stoumpos}}, \bibinfo {author} {\bibfnamefont {C.~D.}\
  \bibnamefont {Malliakas}}, \bibinfo {author} {\bibfnamefont {J.~F.}\
  \bibnamefont {Mitchell}}, \bibinfo {author} {\bibfnamefont {K.}~\bibnamefont
  {Mehlawat}}, \bibinfo {author} {\bibfnamefont {Y.}~\bibnamefont {Singh}},
  \bibinfo {author} {\bibfnamefont {Y.}~\bibnamefont {Choi}}, \bibinfo {author}
  {\bibfnamefont {T.}~\bibnamefont {Gog}}, \bibinfo {author} {\bibfnamefont
  {A.}~\bibnamefont {Al-Zein}}, \bibinfo {author} {\bibfnamefont {M.~M.}\
  \bibnamefont {Sala}}, \bibinfo {author} {\bibfnamefont {M.}~\bibnamefont
  {Krisch}}, \bibinfo {author} {\bibfnamefont {J.}~\bibnamefont {Chaloupka}},
  \bibinfo {author} {\bibfnamefont {G.}~\bibnamefont {Jackeli}}, \bibinfo
  {author} {\bibfnamefont {G.}~\bibnamefont {Khaliullin}}, \ and\ \bibinfo
  {author} {\bibfnamefont {B.~J.}\ \bibnamefont {Kim}},\ }\href {\doibase
  10.1038/nphys3322} {\bibfield  {journal} {\bibinfo  {journal} {Nature
  Physics}\ }\textbf {\bibinfo {volume} {11}},\ \bibinfo {pages} {462}
  (\bibinfo {year} {2015})}\BibitemShut {NoStop}%
\bibitem [{\citenamefont {Banerjee}\ \emph {et~al.}(2016)\citenamefont
  {Banerjee}, \citenamefont {Bridges}, \citenamefont {Yan}, \citenamefont
  {Aczel}, \citenamefont {Li}, \citenamefont {Stone}, \citenamefont {Granroth},
  \citenamefont {Lumsden}, \citenamefont {Yiu}, \citenamefont {Knolle} \emph
  {et~al.}}]{Banerjee16}%
  \BibitemOpen
  \bibfield  {author} {\bibinfo {author} {\bibfnamefont {A.}~\bibnamefont
  {Banerjee}}, \bibinfo {author} {\bibfnamefont {C.}~\bibnamefont {Bridges}},
  \bibinfo {author} {\bibfnamefont {J.-Q.}\ \bibnamefont {Yan}}, \bibinfo
  {author} {\bibfnamefont {A.}~\bibnamefont {Aczel}}, \bibinfo {author}
  {\bibfnamefont {L.}~\bibnamefont {Li}}, \bibinfo {author} {\bibfnamefont
  {M.}~\bibnamefont {Stone}}, \bibinfo {author} {\bibfnamefont
  {G.}~\bibnamefont {Granroth}}, \bibinfo {author} {\bibfnamefont
  {M.}~\bibnamefont {Lumsden}}, \bibinfo {author} {\bibfnamefont
  {Y.}~\bibnamefont {Yiu}}, \bibinfo {author} {\bibfnamefont {J.}~\bibnamefont
  {Knolle}},  \emph {et~al.},\ }\href {\doibase 10.1038/nmat4604} {\bibfield
  {journal} {\bibinfo  {journal} {Nature materials}\ } (\bibinfo {year}
  {2016}),\ 10.1038/nmat4604}\BibitemShut {NoStop}%
\bibitem [{\citenamefont {Knolle}\ \emph
  {et~al.}(2014{\natexlab{b}})\citenamefont {Knolle}, \citenamefont
  {Kovrizhin}, \citenamefont {Chalker},\ and\ \citenamefont
  {Moessner}}]{Knolle14-1}%
  \BibitemOpen
  \bibfield  {author} {\bibinfo {author} {\bibfnamefont {J.}~\bibnamefont
  {Knolle}}, \bibinfo {author} {\bibfnamefont {D.~L.}\ \bibnamefont
  {Kovrizhin}}, \bibinfo {author} {\bibfnamefont {J.~T.}\ \bibnamefont
  {Chalker}}, \ and\ \bibinfo {author} {\bibfnamefont {R.}~\bibnamefont
  {Moessner}},\ }\href {\doibase 10.1103/PhysRevLett.112.207203} {\bibfield
  {journal} {\bibinfo  {journal} {Phys. Rev. Lett.}\ }\textbf {\bibinfo
  {volume} {112}},\ \bibinfo {pages} {207203} (\bibinfo {year}
  {2014}{\natexlab{b}})}\BibitemShut {NoStop}%
\bibitem [{\citenamefont {Sizyuk}\ \emph {et~al.}(2014)\citenamefont {Sizyuk},
  \citenamefont {Price}, \citenamefont {W\"olfle},\ and\ \citenamefont
  {Perkins}}]{Sizyuk}%
  \BibitemOpen
  \bibfield  {author} {\bibinfo {author} {\bibfnamefont {Y.}~\bibnamefont
  {Sizyuk}}, \bibinfo {author} {\bibfnamefont {C.}~\bibnamefont {Price}},
  \bibinfo {author} {\bibfnamefont {P.}~\bibnamefont {W\"olfle}}, \ and\
  \bibinfo {author} {\bibfnamefont {N.~B.}\ \bibnamefont {Perkins}},\ }\href
  {\doibase 10.1103/PhysRevB.90.155126} {\bibfield  {journal} {\bibinfo
  {journal} {Phys. Rev. B}\ }\textbf {\bibinfo {volume} {90}},\ \bibinfo
  {pages} {155126} (\bibinfo {year} {2014})}\BibitemShut {NoStop}%
\bibitem [{lat()}]{lattice_foot}%
  \BibitemOpen
  \href@noop {} {}\bibinfo {note} {There is one additional known trivalent 3D
  lattice with 120 degree bond angles: the (10,3)a, or hyperoctagon
  lattice.\cite{Obrien} However, the Fermi-surface that is realized by Majorana
  spinons on this lattice is thought not to be stable to interactions making it
  more difficult to treat quantitatively and is therefore not treated
  here.}\BibitemShut {Stop}%
\bibitem [{\citenamefont {Lieb}(1994)}]{Lieb}%
  \BibitemOpen
  \bibfield  {author} {\bibinfo {author} {\bibfnamefont {E.~H.}\ \bibnamefont
  {Lieb}},\ }\href {\doibase 10.1103/PhysRevLett.73.2158} {\bibfield  {journal}
  {\bibinfo  {journal} {Phys. Rev. Lett.}\ }\textbf {\bibinfo {volume} {73}},\
  \bibinfo {pages} {2158} (\bibinfo {year} {1994})}\BibitemShut {NoStop}%
\bibitem [{\citenamefont {Kimchi}\ and\ \citenamefont {You}(2011)}]{Kimchi}%
  \BibitemOpen
  \bibfield  {author} {\bibinfo {author} {\bibfnamefont {I.}~\bibnamefont
  {Kimchi}}\ and\ \bibinfo {author} {\bibfnamefont {Y.-Z.}\ \bibnamefont
  {You}},\ }\href {\doibase 10.1103/PhysRevB.84.180407} {\bibfield  {journal}
  {\bibinfo  {journal} {Phys. Rev. B}\ }\textbf {\bibinfo {volume} {84}},\
  \bibinfo {pages} {180407} (\bibinfo {year} {2011})}\BibitemShut {NoStop}%
\bibitem [{\citenamefont {Kim}\ \emph {et~al.}(2009)\citenamefont {Kim},
  \citenamefont {Ohsumi}, \citenamefont {Komesu}, \citenamefont {Sakai},
  \citenamefont {Morita}, \citenamefont {Takagi},\ and\ \citenamefont
  {Arima}}]{Kim}%
  \BibitemOpen
  \bibfield  {author} {\bibinfo {author} {\bibfnamefont {B.~J.}\ \bibnamefont
  {Kim}}, \bibinfo {author} {\bibfnamefont {H.}~\bibnamefont {Ohsumi}},
  \bibinfo {author} {\bibfnamefont {T.}~\bibnamefont {Komesu}}, \bibinfo
  {author} {\bibfnamefont {S.}~\bibnamefont {Sakai}}, \bibinfo {author}
  {\bibfnamefont {T.}~\bibnamefont {Morita}}, \bibinfo {author} {\bibfnamefont
  {H.}~\bibnamefont {Takagi}}, \ and\ \bibinfo {author} {\bibfnamefont
  {T.}~\bibnamefont {Arima}},\ }\href {\doibase 10.1126/science.1167106}
  {\bibfield  {journal} {\bibinfo  {journal} {Science}\ }\textbf {\bibinfo
  {volume} {323}},\ \bibinfo {pages} {1329} (\bibinfo {year}
  {2009})}\BibitemShut {NoStop}%
\bibitem [{chi()}]{chiral_foot}%
  \BibitemOpen
  \href@noop {} {}\bibinfo {note} {This use of the word 'chiral' is in the
  sense of Ref. \onlinecite{Ryu10}. Below we use the same word in a separate
  way to refer to direction-polarized boundary modes.}\BibitemShut {Stop}%
\bibitem [{\citenamefont {Schaffer}\ \emph {et~al.}(2015)\citenamefont
  {Schaffer}, \citenamefont {Lee}, \citenamefont {Lu},\ and\ \citenamefont
  {Kim}}]{Schaffer}%
  \BibitemOpen
  \bibfield  {author} {\bibinfo {author} {\bibfnamefont {R.}~\bibnamefont
  {Schaffer}}, \bibinfo {author} {\bibfnamefont {E.~K.-H.}\ \bibnamefont
  {Lee}}, \bibinfo {author} {\bibfnamefont {Y.-M.}\ \bibnamefont {Lu}}, \ and\
  \bibinfo {author} {\bibfnamefont {Y.~B.}\ \bibnamefont {Kim}},\ }\href
  {\doibase 10.1103/PhysRevLett.114.116803} {\bibfield  {journal} {\bibinfo
  {journal} {Phys. Rev. Lett.}\ }\textbf {\bibinfo {volume} {114}},\ \bibinfo
  {pages} {116803} (\bibinfo {year} {2015})}\BibitemShut {NoStop}%
\bibitem [{\citenamefont {Ryu}\ \emph {et~al.}(2010)\citenamefont {Ryu},
  \citenamefont {Schnyder}, \citenamefont {Furusaki},\ and\ \citenamefont
  {Ludwig}}]{Ryu10}%
  \BibitemOpen
  \bibfield  {author} {\bibinfo {author} {\bibfnamefont {S.}~\bibnamefont
  {Ryu}}, \bibinfo {author} {\bibfnamefont {A.~P.}\ \bibnamefont {Schnyder}},
  \bibinfo {author} {\bibfnamefont {A.}~\bibnamefont {Furusaki}}, \ and\
  \bibinfo {author} {\bibfnamefont {A.~W.~W.}\ \bibnamefont {Ludwig}},\ }\href
  {\doibase 10.1088/1367-2630/12/6/065010} {\bibfield  {journal} {\bibinfo
  {journal} {New Journal of Physics}\ }\textbf {\bibinfo {volume} {12}},\
  \bibinfo {pages} {065010} (\bibinfo {year} {2010})}\BibitemShut {NoStop}%
\bibitem [{Sch()}]{Schaffer_supp}%
  \BibitemOpen
  \href@noop {} {}\bibinfo {note} {Ref. \onlinecite{Schaffer} supplementary
  material}\BibitemShut {NoStop}%
\bibitem [{\citenamefont {Kitaev}(2001)}]{Kitaev01}%
  \BibitemOpen
  \bibfield  {author} {\bibinfo {author} {\bibfnamefont {A.~Y.}\ \bibnamefont
  {Kitaev}},\ }\href {\doibase 10.1070/1063-7869/44/10S/S29} {\bibfield
  {journal} {\bibinfo  {journal} {Physics-Uspekhi}\ }\textbf {\bibinfo {volume}
  {44}},\ \bibinfo {pages} {131} (\bibinfo {year} {2001})}\BibitemShut
  {NoStop}%
\bibitem [{\citenamefont {Fidkowski}\ and\ \citenamefont
  {Kitaev}(2011)}]{Fidkowski11}%
  \BibitemOpen
  \bibfield  {author} {\bibinfo {author} {\bibfnamefont {L.}~\bibnamefont
  {Fidkowski}}\ and\ \bibinfo {author} {\bibfnamefont {A.}~\bibnamefont
  {Kitaev}},\ }\href {\doibase 10.1103/PhysRevB.83.075103} {\bibfield
  {journal} {\bibinfo  {journal} {Phys. Rev. B}\ }\textbf {\bibinfo {volume}
  {83}},\ \bibinfo {pages} {075103} (\bibinfo {year} {2011})}\BibitemShut
  {NoStop}%
\bibitem [{\citenamefont {Wan}\ \emph {et~al.}(2011{\natexlab{b}})\citenamefont
  {Wan}, \citenamefont {Turner}, \citenamefont {Vishwanath},\ and\
  \citenamefont {Savrasov}}]{Wan}%
  \BibitemOpen
  \bibfield  {author} {\bibinfo {author} {\bibfnamefont {X.}~\bibnamefont
  {Wan}}, \bibinfo {author} {\bibfnamefont {A.}~\bibnamefont {Turner}},
  \bibinfo {author} {\bibfnamefont {A.}~\bibnamefont {Vishwanath}}, \ and\
  \bibinfo {author} {\bibfnamefont {S.}~\bibnamefont {Savrasov}},\ }\href
  {\doibase 10.1103/PhysRevB.83.205101} {\bibfield  {journal} {\bibinfo
  {journal} {Phys. Rev. B}\ }\textbf {\bibinfo {volume} {83}},\ \bibinfo
  {pages} {205101} (\bibinfo {year} {2011}{\natexlab{b}})}\BibitemShut
  {NoStop}%
\bibitem [{\citenamefont {Rau}\ \emph {et~al.}(2014)\citenamefont {Rau},
  \citenamefont {Lee},\ and\ \citenamefont {Kee}}]{Rau14}%
  \BibitemOpen
  \bibfield  {author} {\bibinfo {author} {\bibfnamefont {J.~G.}\ \bibnamefont
  {Rau}}, \bibinfo {author} {\bibfnamefont {E.~K.-H.}\ \bibnamefont {Lee}}, \
  and\ \bibinfo {author} {\bibfnamefont {H.-Y.}\ \bibnamefont {Kee}},\ }\href
  {\doibase 10.1103/PhysRevLett.112.077204} {\bibfield  {journal} {\bibinfo
  {journal} {Phys. Rev. Lett.}\ }\textbf {\bibinfo {volume} {112}},\ \bibinfo
  {pages} {077204} (\bibinfo {year} {2014})}\BibitemShut {NoStop}%
\bibitem [{\citenamefont {Rau}\ \emph {et~al.}(2016)\citenamefont {Rau},
  \citenamefont {Lee},\ and\ \citenamefont {Kee}}]{Rau16}%
  \BibitemOpen
  \bibfield  {author} {\bibinfo {author} {\bibfnamefont {J.~G.}\ \bibnamefont
  {Rau}}, \bibinfo {author} {\bibfnamefont {E.~K.-H.}\ \bibnamefont {Lee}}, \
  and\ \bibinfo {author} {\bibfnamefont {H.-Y.}\ \bibnamefont {Kee}},\ }\href
  {\doibase 10.1146/annurev-conmatphys-031115-011319} {\bibfield  {journal}
  {\bibinfo  {journal} {Annual Review of Condensed Matter Physics}\ }\textbf
  {\bibinfo {volume} {7}},\ \bibinfo {pages} {195} (\bibinfo {year}
  {2016})}\BibitemShut {NoStop}%
\bibitem [{tri()}]{trigfoot}%
  \BibitemOpen
  \href@noop {} {}\bibinfo {note} {In the case of trigonal distortion, one also
  needs to revise the wave function of the doublet states.}\BibitemShut {Stop}%
\bibitem [{\citenamefont {Yamaji}\ \emph {et~al.}(2014)\citenamefont {Yamaji},
  \citenamefont {Nomura}, \citenamefont {Kurita}, \citenamefont {Arita},\ and\
  \citenamefont {Imada}}]{Yamaji14}%
  \BibitemOpen
  \bibfield  {author} {\bibinfo {author} {\bibfnamefont {Y.}~\bibnamefont
  {Yamaji}}, \bibinfo {author} {\bibfnamefont {Y.}~\bibnamefont {Nomura}},
  \bibinfo {author} {\bibfnamefont {M.}~\bibnamefont {Kurita}}, \bibinfo
  {author} {\bibfnamefont {R.}~\bibnamefont {Arita}}, \ and\ \bibinfo {author}
  {\bibfnamefont {M.}~\bibnamefont {Imada}},\ }\href {\doibase
  10.1103/PhysRevLett.113.107201} {\bibfield  {journal} {\bibinfo  {journal}
  {Phys. Rev. Lett.}\ }\textbf {\bibinfo {volume} {113}},\ \bibinfo {pages}
  {107201} (\bibinfo {year} {2014})}\BibitemShut {NoStop}%
\bibitem [{\citenamefont {Foyevtsova}\ \emph {et~al.}(2013)\citenamefont
  {Foyevtsova}, \citenamefont {Jeschke}, \citenamefont {Mazin}, \citenamefont
  {Khomskii},\ and\ \citenamefont {Valent\'{\i}}}]{Foyevtsova13}%
  \BibitemOpen
  \bibfield  {author} {\bibinfo {author} {\bibfnamefont {K.}~\bibnamefont
  {Foyevtsova}}, \bibinfo {author} {\bibfnamefont {H.~O.}\ \bibnamefont
  {Jeschke}}, \bibinfo {author} {\bibfnamefont {I.~I.}\ \bibnamefont {Mazin}},
  \bibinfo {author} {\bibfnamefont {D.~I.}\ \bibnamefont {Khomskii}}, \ and\
  \bibinfo {author} {\bibfnamefont {R.}~\bibnamefont {Valent\'{\i}}},\ }\href
  {\doibase 10.1103/PhysRevB.88.035107} {\bibfield  {journal} {\bibinfo
  {journal} {Phys. Rev. B}\ }\textbf {\bibinfo {volume} {88}},\ \bibinfo
  {pages} {035107} (\bibinfo {year} {2013})}\BibitemShut {NoStop}%
\bibitem [{\citenamefont {Kim}\ \emph {et~al.}(2015{\natexlab{b}})\citenamefont
  {Kim}, \citenamefont {Lee},\ and\ \citenamefont {Kim}}]{kim15}%
  \BibitemOpen
  \bibfield  {author} {\bibinfo {author} {\bibfnamefont {H.-S.}\ \bibnamefont
  {Kim}}, \bibinfo {author} {\bibfnamefont {E.~K.-H.}\ \bibnamefont {Lee}}, \
  and\ \bibinfo {author} {\bibfnamefont {Y.~B.}\ \bibnamefont {Kim}},\ }\href
  {\doibase 10.1209/0295-5075/112/67004} {\bibfield  {journal} {\bibinfo
  {journal} {Europhys. Lett.}\ }\textbf {\bibinfo {volume} {112}},\ \bibinfo
  {pages} {67004} (\bibinfo {year} {2015}{\natexlab{b}})}\BibitemShut {NoStop}%
\bibitem [{\citenamefont {Winter}\ \emph {et~al.}(2016)\citenamefont {Winter},
  \citenamefont {Li}, \citenamefont {Jeschke},\ and\ \citenamefont
  {Valenti}}]{Winter16}%
  \BibitemOpen
  \bibfield  {author} {\bibinfo {author} {\bibfnamefont {S.~M.}\ \bibnamefont
  {Winter}}, \bibinfo {author} {\bibfnamefont {Y.}~\bibnamefont {Li}}, \bibinfo
  {author} {\bibfnamefont {H.~O.}\ \bibnamefont {Jeschke}}, \ and\ \bibinfo
  {author} {\bibfnamefont {R.}~\bibnamefont {Valenti}},\ }\href
  {http://arxiv.org/abs/1603.02548} {\bibfield  {journal} {\bibinfo  {journal}
  {arXiv:1603.02548}\ } (\bibinfo {year} {2016})}\BibitemShut {NoStop}%
\bibitem [{\citenamefont {Loudon}(1963)}]{Loudon}%
  \BibitemOpen
  \bibfield  {author} {\bibinfo {author} {\bibfnamefont {R.}~\bibnamefont
  {Loudon}},\ }\href {\doibase 10.1098/rspa.1963.0166} {\bibfield  {journal}
  {\bibinfo  {journal} {Proceedings of the Royal Society of London A:
  Mathematical, Physical and Engineering Sciences}\ }\textbf {\bibinfo {volume}
  {275}},\ \bibinfo {pages} {218} (\bibinfo {year} {1963})}\BibitemShut
  {NoStop}%
\bibitem [{\citenamefont {Fleury}\ and\ \citenamefont
  {Loudon}(1968)}]{fleury68}%
  \BibitemOpen
  \bibfield  {author} {\bibinfo {author} {\bibfnamefont {P.~A.}\ \bibnamefont
  {Fleury}}\ and\ \bibinfo {author} {\bibfnamefont {R.}~\bibnamefont
  {Loudon}},\ }\href {\doibase 10.1103/PhysRev.166.514} {\bibfield  {journal}
  {\bibinfo  {journal} {Phys. Rev.}\ }\textbf {\bibinfo {volume} {166}},\
  \bibinfo {pages} {514} (\bibinfo {year} {1968})}\BibitemShut {NoStop}%
\bibitem [{\citenamefont {Shastry}\ and\ \citenamefont
  {Shraiman}(1990)}]{Shastry}%
  \BibitemOpen
  \bibfield  {author} {\bibinfo {author} {\bibfnamefont {B.~S.}\ \bibnamefont
  {Shastry}}\ and\ \bibinfo {author} {\bibfnamefont {B.~I.}\ \bibnamefont
  {Shraiman}},\ }\href {\doibase 10.1103/PhysRevLett.65.1068} {\bibfield
  {journal} {\bibinfo  {journal} {Phys. Rev. Lett.}\ }\textbf {\bibinfo
  {volume} {65}},\ \bibinfo {pages} {1068} (\bibinfo {year}
  {1990})}\BibitemShut {NoStop}%
\bibitem [{\citenamefont {Perkins}\ and\ \citenamefont
  {Brenig}(2008)}]{Perkins08}%
  \BibitemOpen
  \bibfield  {author} {\bibinfo {author} {\bibfnamefont {N.}~\bibnamefont
  {Perkins}}\ and\ \bibinfo {author} {\bibfnamefont {W.}~\bibnamefont
  {Brenig}},\ }\href {\doibase 10.1103/PhysRevB.77.174412} {\bibfield
  {journal} {\bibinfo  {journal} {Phys. Rev. B}\ }\textbf {\bibinfo {volume}
  {77}},\ \bibinfo {pages} {174412} (\bibinfo {year} {2008})}\BibitemShut
  {NoStop}%
\bibitem [{\citenamefont {Ko}\ \emph {et~al.}(2010)\citenamefont {Ko},
  \citenamefont {Liu}, \citenamefont {Ng},\ and\ \citenamefont {Lee}}]{Ko}%
  \BibitemOpen
  \bibfield  {author} {\bibinfo {author} {\bibfnamefont {W.-H.}\ \bibnamefont
  {Ko}}, \bibinfo {author} {\bibfnamefont {Z.-X.}\ \bibnamefont {Liu}},
  \bibinfo {author} {\bibfnamefont {T.-K.}\ \bibnamefont {Ng}}, \ and\ \bibinfo
  {author} {\bibfnamefont {P.~A.}\ \bibnamefont {Lee}},\ }\href {\doibase
  10.1103/PhysRevB.81.024414} {\bibfield  {journal} {\bibinfo  {journal} {Phys.
  Rev. B}\ }\textbf {\bibinfo {volume} {81}},\ \bibinfo {pages} {024414}
  (\bibinfo {year} {2010})}\BibitemShut {NoStop}%
\bibitem [{\citenamefont {Perkins}\ \emph {et~al.}(2013)\citenamefont
  {Perkins}, \citenamefont {Chern},\ and\ \citenamefont {Brenig}}]{Perkins13}%
  \BibitemOpen
  \bibfield  {author} {\bibinfo {author} {\bibfnamefont {N.~B.}\ \bibnamefont
  {Perkins}}, \bibinfo {author} {\bibfnamefont {G.-W.}\ \bibnamefont {Chern}},
  \ and\ \bibinfo {author} {\bibfnamefont {W.}~\bibnamefont {Brenig}},\ }\href
  {\doibase 10.1103/PhysRevB.87.174423} {\bibfield  {journal} {\bibinfo
  {journal} {Phys. Rev. B}\ }\textbf {\bibinfo {volume} {87}},\ \bibinfo
  {pages} {174423} (\bibinfo {year} {2013})}\BibitemShut {NoStop}%
\bibitem [{\citenamefont {J.~Nasu}\ \emph {et~al.}(2016)\citenamefont
  {J.~Nasu}, \citenamefont {Knolle}, \citenamefont {Kovrizhin}, \citenamefont
  {Motome},\ and\ \citenamefont {Moessner}}]{Nasu2016}%
  \BibitemOpen
  \bibfield  {author} {\bibinfo {author} {\bibfnamefont {J.}~\bibnamefont
  {J.~Nasu}}, \bibinfo {author} {\bibfnamefont {J.}~\bibnamefont {Knolle}},
  \bibinfo {author} {\bibfnamefont {D.~L.}\ \bibnamefont {Kovrizhin}}, \bibinfo
  {author} {\bibfnamefont {Y.}~\bibnamefont {Motome}}, \ and\ \bibinfo {author}
  {\bibfnamefont {R.}~\bibnamefont {Moessner}},\ }\href@noop {} {\  (\bibinfo
  {year} {2016})},\ \Eprint {http://arxiv.org/abs/arXiv:1602.05277}
  {arXiv:1602.05277} \BibitemShut {NoStop}%
\bibitem [{\citenamefont {Bruus}\ and\ \citenamefont
  {Flensberg}(2004)}]{Bruus04}%
  \BibitemOpen
  \bibfield  {author} {\bibinfo {author} {\bibfnamefont {H.}~\bibnamefont
  {Bruus}}\ and\ \bibinfo {author} {\bibfnamefont {K.}~\bibnamefont
  {Flensberg}},\ }\href@noop {} {\emph {\bibinfo {title} {Many-body quantum
  theory in condensed matter physics: an introduction}}}\ (\bibinfo
  {publisher} {OUP Oxford},\ \bibinfo {year} {2004})\BibitemShut {NoStop}%
\bibitem [{\citenamefont {Shastry}\ and\ \citenamefont
  {Shraiman}(1991)}]{Shastry2}%
  \BibitemOpen
  \bibfield  {author} {\bibinfo {author} {\bibfnamefont {B.~S.}\ \bibnamefont
  {Shastry}}\ and\ \bibinfo {author} {\bibfnamefont {B.~I.}\ \bibnamefont
  {Shraiman}},\ }\href {\doibase 10.1142/S0217979291000237} {\bibfield
  {journal} {\bibinfo  {journal} {International Journal of Modern Physics B}\
  }\textbf {\bibinfo {volume} {5}},\ \bibinfo {pages} {365} (\bibinfo {year}
  {1991})}\BibitemShut {NoStop}%
\bibitem [{\citenamefont {Devereaux}\ and\ \citenamefont
  {Hackl}(2007)}]{Devereaux07}%
  \BibitemOpen
  \bibfield  {author} {\bibinfo {author} {\bibfnamefont {T.~P.}\ \bibnamefont
  {Devereaux}}\ and\ \bibinfo {author} {\bibfnamefont {R.}~\bibnamefont
  {Hackl}},\ }\href {\doibase 10.1103/RevModPhys.79.175} {\bibfield  {journal}
  {\bibinfo  {journal} {Rev. Mod. Phys.}\ }\textbf {\bibinfo {volume} {79}},\
  \bibinfo {pages} {175} (\bibinfo {year} {2007})}\BibitemShut {NoStop}%
\bibitem [{\citenamefont {Smith}\ \emph {et~al.}(2015)\citenamefont {Smith},
  \citenamefont {Knolle}, \citenamefont {Kovrizhin}, \citenamefont {Chalker},\
  and\ \citenamefont {Moessner}}]{Smith}%
  \BibitemOpen
  \bibfield  {author} {\bibinfo {author} {\bibfnamefont {A.}~\bibnamefont
  {Smith}}, \bibinfo {author} {\bibfnamefont {J.}~\bibnamefont {Knolle}},
  \bibinfo {author} {\bibfnamefont {D.~L.}\ \bibnamefont {Kovrizhin}}, \bibinfo
  {author} {\bibfnamefont {J.~T.}\ \bibnamefont {Chalker}}, \ and\ \bibinfo
  {author} {\bibfnamefont {R.}~\bibnamefont {Moessner}},\ }\href {\doibase
  10.1103/PhysRevB.92.180408} {\bibfield  {journal} {\bibinfo  {journal} {Phys.
  Rev. B}\ }\textbf {\bibinfo {volume} {92}},\ \bibinfo {pages} {180408}
  (\bibinfo {year} {2015})}\BibitemShut {NoStop}%
\bibitem [{\citenamefont {Sandilands}\ \emph {et~al.}(2015)\citenamefont
  {Sandilands}, \citenamefont {Tian}, \citenamefont {Plumb}, \citenamefont
  {Kim},\ and\ \citenamefont {Burch}}]{Sandilands}%
  \BibitemOpen
  \bibfield  {author} {\bibinfo {author} {\bibfnamefont {L.~J.}\ \bibnamefont
  {Sandilands}}, \bibinfo {author} {\bibfnamefont {Y.}~\bibnamefont {Tian}},
  \bibinfo {author} {\bibfnamefont {K.~W.}\ \bibnamefont {Plumb}}, \bibinfo
  {author} {\bibfnamefont {Y.-J.}\ \bibnamefont {Kim}}, \ and\ \bibinfo
  {author} {\bibfnamefont {K.~S.}\ \bibnamefont {Burch}},\ }\href {\doibase
  10.1103/PhysRevLett.114.147201} {\bibfield  {journal} {\bibinfo  {journal}
  {Phys. Rev. Lett.}\ }\textbf {\bibinfo {volume} {114}},\ \bibinfo {pages}
  {147201} (\bibinfo {year} {2015})}\BibitemShut {NoStop}%
\bibitem [{\citenamefont {Rousseau}\ \emph {et~al.}(1981)\citenamefont
  {Rousseau}, \citenamefont {Bauman},\ and\ \citenamefont {Porto}}]{Rousseau}%
  \BibitemOpen
  \bibfield  {author} {\bibinfo {author} {\bibfnamefont {D.~L.}\ \bibnamefont
  {Rousseau}}, \bibinfo {author} {\bibfnamefont {R.~P.}\ \bibnamefont
  {Bauman}}, \ and\ \bibinfo {author} {\bibfnamefont {S.~P.~S.}\ \bibnamefont
  {Porto}},\ }\href {\doibase 10.1002/jrs.1250100152} {\bibfield  {journal}
  {\bibinfo  {journal} {Journal of Raman Spectroscopy}\ }\textbf {\bibinfo
  {volume} {10}},\ \bibinfo {pages} {253} (\bibinfo {year} {1981})}\BibitemShut
  {NoStop}%
\bibitem [{\citenamefont {Klein}\ and\ \citenamefont {Porto}(1969)}]{Klein}%
  \BibitemOpen
  \bibfield  {author} {\bibinfo {author} {\bibfnamefont {M.~V.}\ \bibnamefont
  {Klein}}\ and\ \bibinfo {author} {\bibfnamefont {S.~P.~S.}\ \bibnamefont
  {Porto}},\ }\href {\doibase 10.1103/PhysRevLett.22.782} {\bibfield  {journal}
  {\bibinfo  {journal} {Phys. Rev. Lett.}\ }\textbf {\bibinfo {volume} {22}},\
  \bibinfo {pages} {782} (\bibinfo {year} {1969})}\BibitemShut {NoStop}%
\bibitem [{\citenamefont {Niu}\ \emph {et~al.}(2012)\citenamefont {Niu},
  \citenamefont {Chung}, \citenamefont {Hsu}, \citenamefont {Mandal},
  \citenamefont {Raghu},\ and\ \citenamefont {Chakravarty}}]{Niu12}%
  \BibitemOpen
  \bibfield  {author} {\bibinfo {author} {\bibfnamefont {Y.}~\bibnamefont
  {Niu}}, \bibinfo {author} {\bibfnamefont {S.~B.}\ \bibnamefont {Chung}},
  \bibinfo {author} {\bibfnamefont {C.-H.}\ \bibnamefont {Hsu}}, \bibinfo
  {author} {\bibfnamefont {I.}~\bibnamefont {Mandal}}, \bibinfo {author}
  {\bibfnamefont {S.}~\bibnamefont {Raghu}}, \ and\ \bibinfo {author}
  {\bibfnamefont {S.}~\bibnamefont {Chakravarty}},\ }\href {\doibase
  10.1103/PhysRevB.85.035110} {\bibfield  {journal} {\bibinfo  {journal} {Phys.
  Rev. B}\ }\textbf {\bibinfo {volume} {85}},\ \bibinfo {pages} {035110}
  (\bibinfo {year} {2012})}\BibitemShut {NoStop}%
\bibitem [{\citenamefont {Shankar}(1994)}]{ShankarRG}%
  \BibitemOpen
  \bibfield  {author} {\bibinfo {author} {\bibfnamefont {R.}~\bibnamefont
  {Shankar}},\ }\href {\doibase 10.1103/RevModPhys.66.129} {\bibfield
  {journal} {\bibinfo  {journal} {Rev. Mod. Phys.}\ }\textbf {\bibinfo {volume}
  {66}},\ \bibinfo {pages} {129} (\bibinfo {year} {1994})}\BibitemShut
  {NoStop}%
\end{thebibliography}%

\end{document}